\documentclass[a4paper,fleqn,usenatbib]{mnras}
\usepackage{graphicx}
\usepackage[normalem]{ulem}
\usepackage[T1]{fontenc}
\usepackage{ae,aecompl}

\usepackage{amsmath}	
\usepackage{amssymb}	

\usepackage{url}

\hypersetup{draft}

\usepackage{upgreek}
\usepackage[usenames]{color} 
\usepackage{appendix}
\usepackage{booktabs}
\usepackage{color}
\usepackage{float}

\title[The star that stopped pulsating]{HD~60435: The star that stopped pulsating }

\author[D. W. Kurtz et al.]{Donald W. Kurtz$^{1,2}$\thanks{E-mail: kurtzdw@gmail.com}, Gerald Handler$^{3}$, Daniel L. Holdsworth$^{4,2,5}$, Margarida S. Cunha$^{6}$,  \newauthor Hideyuki Saio$^{7}$, Thebe Medupe$^{1}$, Simon J. Murphy$^{8}$, Joachim Kr\"uger$^{8}$, \newauthor E. Brunsden$^5$, Victoria Antoci$^9$, Daniel R. Hey$^{10}$,  Noi Shitrit$^{11}$, Jaymie M. Matthews$^{12}$   \\
$^{1}$Centre for Space Research, North-West University, Dr Albert Luthuli Drive, Mahikeng 2735, South Africa\\
$^{2}$Jeremiah Horrocks Institute, University of Central Lancashire, Preston PR1 2HE, UK\\
$^{3}$Nicolaus Copernicus Astronomical Center, Polish Academy of Sciences, ul. Bartycka 18, 00-716, Warszawa, Poland\\
$^{4}$South African Astronomical Observatory, P.O. Box 9, Observatory 7935, Cape Town, South Africa\\
$^5$School of Physics, Engineering and Technology, University of York, Heslington, York, YO10 5DD, UK\\
$^{6}$Instituto de Astrof\'isica e Ci\^encias do Espa\c{c}o, Universidade do Porto CAUP, Rua das Estrelas, P-PT4150-762 Porto, Portugal\\
$^7$Astronomical Institute, Graduate School of Science, Tohoku University, Sendai 980-8578, Japan\\
$^{8}$Centre for Astrophysics, University of Southern Queensland, Toowoomba, QLD 4350, Australia\\
$^{9}$DTU Space, Technical University of Denmark, Elektrovej 327, Kgs. Lyngby, 2800, Denmark\\
$^{10}$Institute for Astronomy,University of Hawai'i, Honolulu, HI 96822,USA\\
$^{11}$The School of Physics and Astronomy, Tel Aviv University, Tel Aviv 69978, Israel\\
$^{12}$Department of Physics and Astronomy, University of British Columbia, 6224 Agricultural Road, Vancouver, BC V6T 1Z1, Canada
}

\begin{document}
\date{\today}
\pagerange{\pageref{firstpage}--\pageref{lastpage}} \pubyear{2024}
\maketitle
\label{firstpage}

\begin{abstract}
HD\,60435 is a well-known rapidly oscillating (roAp) Ap star with a series of alternating even and odd degree modes, making it a prime asteroseismic target. It is also an oblique pulsator with rotational inclination, $i$, and magnetic/pulsation obliquity, $\beta$, such that both magnetic/pulsation poles are viewed over the rotation period, $P_{\rm rot} = 7.679696$\,d, determined from rotational light variations. While some roAp stars have stable pulsation mode amplitudes over decades, HD\,60435 is known to have amplitude variations on time scales as short as 1\,d. We show from 5\,yr of {\it TESS} observations that there is strong amplitude modulation on this short time scale with possible mode interactions. Most remarkably, HD\,60435 stopped pulsating during the time span of the {\it TESS} observations. This is the first time that any pulsating star has been observed to cease pulsating entirely. That has implications for mode interaction, excitation and damping, and is relevant to the problem of why only some stars in many pulsation instability strips pulsate, while others do not. During a 24.45-d time span of the {\it TESS} data when there was mode stability for a dipole mode and a quadrupole mode, the oblique pulsator model constrained $i$ and $\beta$, which we used to model those modes with a magnetic pulsation model from which we determined a polar field strength of 4\,kG, in good agreement with a known magnetic measurement. We modelled the frequency separations showing that they can constrain the global metallicity, something that is not possible from spectroscopy of the highly peculiar Ap atmosphere. 
\end{abstract}  


\section{Introduction}
\label{intro}

Stellar pulsation is ubiquitous across the Hertzsprung-Russell (HR) diagram, occurring in many classes of stars. In recent years precise astrophysical inference of stellar structure has become possible through asteroseismology (e.g., \citealt{2010aste.book.....A}, \citealt{2021RvMP...93a5001A}, \citealt{2022ARA&A..60...31K}). A given star may pulsate as different kinds of variables during its lifetime, as it evolves into and out of various pulsational instability strips or regions in the HR~diagram (e.g., see figs 1 of \citealt{2021RvMP...93a5001A} or \citealt{2022ARA&A..60...31K}). Those regions are defined by boundaries in effective temperature and luminosity that depend on a competition between pulsational driving in certain layers of the star and damping throughout the star. 

For example, a 5-M$_\odot$ star may begin life on the main-sequence as a Slowly Pulsating B (SPB) star, move out of the SPB instability strip and stop pulsating, only then to evolve through the Cepheid instability strip where it becomes a $\delta$\,Cep star (Cepheid), evolve out of that instability strip with Cepheid pulsation stopping, then become a red giant where it becomes a stochastic pulsator and even a Mira variable at various stages. After shedding much of its mass in winds and as a planetary nebula, the typically 0.6-M$_\odot$ remnant core then evolves through several instability strips as a DOV, DBV, and DAV white dwarf, crossing regions of stability along the way. Stars of other masses can pass through other instability zones.

Thus the pulsating stars that are the subjects of asteroseismology have various eigenmodes excited to observable amplitudes as they evolve into instability strips, then extinguished by the dominance of damping as they evolve out at various stages in their evolutionary lives. There is considerable theoretical understanding of pulsation driving mechanisms for the various classes. This is usually the $\kappa$-mechanism operating in H, He, or Fe ionisation zones for hotter stars and stochastic driving for the cooler stars (see \citealt{2010aste.book.....A}, section 3.7, for discussion of these mechanisms). But theoretically predicting pulsation amplitude, or explaining observed amplitude in a particular star remains a complex and unsolved problem. 

Some stars within a particular instability region may pulsate while others do not; some may have high amplitude while others low amplitude, some may have many eigenmodes excited, while others only one or a few. The problem of mode selection is also complex and unsolved. It is clear that mode excitation and selection are finely tuned to many fundamental parameters, such as $T_{\rm eff}$, $\log g$, metallicity, magnetic field, rotation, binarity, and tides from companions, but our understanding of how these parameters govern observed mode amplitude is weak.

Thus it is of interest when pulsation modes in a star are observed to change amplitude, to appear or disappear, on a human time-scale. This is much shorter than the usual evolutionary time scales we model. The observations of mode amplitude growth and decay indicate energy exchange among modes --  mode interaction --  or pulsational energy gains or losses to the driving and damping layers in the star. When a new mode is observed to appear, or a known mode is observed to disappear, this has the potential to provide information on possibly small changes in a star's structure. These amplitude changes, along with frequency changes, can provide detailed understanding of changes in a star on a human time-scale. Such short time-scale changes in stellar structure may be far less smooth than the long time-scale evolution tracks we typically produce with models.

Pulsation amplitude and frequency changes in $\delta$~Sct stars are well-documented, beginning with 40 years of pioneering work on 4\,CVn  (e.g., \citealt{2000MNRAS.313..129B, 2017A&A...599A.116B}). \citet{2012MNRAS.427.1418M} showed that three pulsation modes in the $\delta$~Sct star KIC\,3429637 had linear changes in their pulsation amplitudes over a 2-yr time-span, suggesting evolutionary changes. \citet{2016MNRAS.460.1970B} studied 983 $\delta$~Sct stars with {\it Kepler} main-mission data showing how pulsation amplitude and phase varied over the 4-yr data set for some modes, while others remained at constant amplitude. They showed that the amplitude variations were secular, and not the result of beating among unresolved modes. A spectacular case was found in the $\delta$~Sct star KIC\,7106205 \citep{2014MNRAS.444.1909B} where one mode changed in amplitude from 5\,mmag to 0.5\,mmag in 650\,d, then remained stable for the remaining 820\,d of the 4-yr data set, while other modes in this star had constant amplitude over the full 1470\,d.  \citet{2022ApJ...936...48Y} studied KIC\,2857323, a high-amplitude $\delta$~Sct (HADS) star pulsating in the fundamental and first-overtone radial modes. Such stars usually show much more stable pulsation amplitudes than other, lower-amplitude $\delta$~Sct stars that pulsate in many modes, yet \citeauthor{2022ApJ...936...48Y} found a slow, secular decay of the first-overtone amplitude over the 4\,yr of the Kepler data set. So even `simple' $\delta$~Sct stars with few pulsation modes can show amplitude changes on a time-scale less than years. Similar to this $\delta$~Sct star, \citet{2024ApJ...973..157C} have shown in {\it TESS} and AAVSO data, over a time-scale of 100\,d, the disappearance, then reappearance of the first-overtone mode in the double-mode RR\,Lyr star V338\,Boo.

Changes in the pulsation spectra are also common for white dwarf stars. Among the most spectacular events observed was the appearance of a new pulsation mode and the vanishing of all others in the prototype pulsating DB white dwarf GD\,358 (V777\,Her) within a single day followed by a slow disappearance of this mode \citep{2003A&A...401..639K}. On the other hand, most of the pulsation modes in the pulsating DB white dwarf PG\,1456+103 disappeared at some point and were observed to grow back on a time scale of a just few weeks \citep{2013ASPC..469...53H}.

\citet{2023PASP..135l4201A} found that the well-known A supergiant star Deneb ($\alpha$\,Cyg, HR\,7924, HD\,197345) ceases (or nearly ceases) pulsating at times, then resumes on perhaps a $\sim$70-d time-scale. They demonstrated this in both radial velocity data going back a century and with recent {\it TESS} (Transiting Exoplanet Survey Satellite, \citealt{2015JATIS...1a4003R}) photometric data. \citet{2024arXiv241023985G} extended this discussion with further data from {\it TESS} and the American Association of Variable Star Observers (AAVSO). \citeauthor{2023PASP..135l4201A} speculated that the cause of the resumption of pulsation may lie in `the microvariations produced in convective layers below their atmospheres, pulsation-driven shocks and rarefactions, or pulsation-convection interactions.' These are interesting conjectures, but we do not know the causes of these amplitude variations. At the moment, no other star is known to show this pulsation behaviour seen in Deneb.

Polaris ($\alpha$\,UMi, HR\,424, HD\,8890) is another famous star that shows amplitude modulation of its 4-d Cepheid pulsation. \citet{2005PASP..117..207T} studied pulsation period changes in Polaris over 160\,yr from $1844 - 2004$ and found a steady increase for most of the time span, but with a more rapid decrease from $1963 - 1966$. They found an amplitude greater than 0.1\,mag in $V$ up to the $1963 - 1966$ hiatus, followed by a sharp decline to below $0.05$\,mag with erratic behaviour from cycle-to-cycle. \citet{2008ApJ...683..433B} documented a drop in pulsation amplitude in Polaris from 120\,mmag to 30\,mmag over a century up to the early 2000s. \citet{1989AJ.....98.2249D} detected a steady period increase in 1987 -- 1988 from radial velocity variations. From these studies it appeared that Polaris might have been evolving redward out of the instability strip and stopping pulsation. But \citeauthor{2008ApJ...683..433B} showed a recovery in amplitude from 2003 to 2006. We have examined the Sector 60 {\it TESS} data from 2022 December -- 2023 January that show clear pulsation with a period 3.973\,d and a peak-to-peak amplitude of 76\,mmag with the {\it TESS} red bandpass. So Polaris is not stopping pulsating at present.

Another famous bright star that showed significant decrease in pulsation amplitude, and was thought to be stopping pulsating, is Spica \citep{1978MNRAS.185..325L}, a spectroscopic binary with a $\beta$\,Cep primary that has been observed to pulsate in several modes, with a primary frequency of 5.75\,d$^{-1}$ \citep{2016MNRAS.458.1964T}. Subsequent observations -- including space-based observations from the MOST (Microvariability and Oscillations of Stars) satellite \citep{2009CoAst.158..303D} -- have shown that the 5.75-d$^{-1}$ frequency has continued, although at decreased amplitude that has at times been difficult to detect with ground-based observations (see \citealt{1985A&A...143..466C}).

We have discovered a star that has been observed since the 1980s to have a rich pulsation spectrum, and it has now stopped pulsating completely.  To our knowledge, this is the first pulsating star to have been observed to completely cease pulsating. This star is HD\,60435, a rapidly oscillating Ap (roAp) star that is the principal subject of this paper. 

\subsection{Frequency and amplitude variations in \MakeLowercase{ro}A\MakeLowercase{p} stars}

The rapidly oscillating Ap (roAp) stars are A to early-F spectroscopically peculiar stars with strong magnetic fields. They typically show stable, long-lived rotational light variations caused by spots associated with roughly dipolar magnetic fields that are inclined to the rotation axis. They pulsate in high radial overtone p~modes with the pulsation axis inclined to the rotation axis and close to the magnetic axis \citep[e.g.][]{2011A&A...536A..73B}. Detailed introductions to these stars can be found in ensemble studies of them in the {\it TESS} data sets by \citet{2019MNRAS.487.3523C} and \citet{2021MNRAS.506.1073H, 2024MNRAS.527.9548H}, and in a review by \citet[][section 2.15.12]{2022ARA&A..60...31K}.

Some roAp stars have pulsation modes that have stable frequencies and amplitudes over decades, while others show measurable changes in these quantities on times scales as short as one day (see \citealt{2021MNRAS.506.1073H, 2024MNRAS.527.9548H} for examples). An interesting and  illustrative case is that of HD\,217522. This star was discovered to have pulsations with frequencies near to 105\,d$^{-1}$ ($1.22$\,mHz) in 1981. Later observations in 1989 showed those frequencies plus a new one near 174\,d$^{-1}$ ($2.01$\,mHz) that was not present in 1981. The modes of these frequencies are separated by many radial overtones, with the intervening modes not excited. This excitation of non-consecutive radial overtone modes remains a more general problem in understanding mode selection in these stars. 

\citet{2015MNRAS.446.1347M} studied the pulsations in HD\,217522 spectroscopically, examining pulsation amplitude line-by-line and as a function of depth in individual lines. Because elements are radially stratified in roAp star atmospheres, this allows the examination of pulsation amplitude as a function of atmospheric depth. The complex modes are magneto-acoustic, as has been examined theoretically in detail by \citet{2018MNRAS.480.1676Q}. \citeauthor{2015MNRAS.446.1347M} found that the higher frequency seen in the 1989 photometric data of HD\,217522 was present at some depths in the atmosphere,  but not seen in others. They also pointed out, as have other authors, that this star shows amplitude variability on a time-scale as short as a day. 

Thus we understand that some pulsation modes may be visible at particular atmospheric depths and others not; \citet{2018MNRAS.480.1676Q} provide theoretical understanding of this. It is important when  studying the problem of mode excitation and damping, and in looking for counterparts of stars like Deneb, Spica, Polaris, the $\delta$~Sct and roAp stars, and others, that we consider whether different data sets are probing different parts of the stellar atmosphere, therefore giving different results, although that is far less likely for low radial overtone pulsators like Deneb, Spica, and Polaris, than for high radial overtone pulsators like the roAp stars. 

\section{HD\,60435, an  \MakeLowercase{ro}A\MakeLowercase{p} star that has stopped pulsating} 

HD\,60435 is a magnetic Ap star that was discovered to be an roAp star by \citet{1984MNRAS.209..841K} from observations obtained during 1983 February to April. It was studied extensively by \citet{1987ApJ...313..782M} using ground-based photometric data in Johnson $B$ obtained contemporaneously from observatories in Chile and South Africa from 1984 November through 1985 March. They found a series of p~modes of consecutive radial overtone and alternating even and odd degrees ($\ell = 1, 2$) for radial overtones ranging from $n = 13 - 28$, making the star a prime asteroseismic target. 

From a study of the rotational light variation caused by stable spots, \citet{1990MNRAS.243..289K} determined the rotation period to be $P_{\rm rot} = 7.6793 \pm 0.0006$\,d$^{-1}$ and showed that spots on opposite hemispheres of the star are seen, presumably at, or near to, both magnetic poles. They confirmed the result of  \citeauthor{1987ApJ...313..782M} that the pulsation amplitudes vary with the rotation of the star -- as expected for an oblique pulsator -- but also non-periodically on time scales shorter than the rotation period. 

\citet{1998CoSka..28..109Z} carried out a spectroscopic study and found variability in the equivalent width of lines of Li, Ca, Fe, and Pr with the rotation period. Their abundance analysis found typical Ap abundance anomalies with overabundances in some rare earth elements of up to 300 times solar. 

\citet{2019MNRAS.487.2117B}  and \citet{2021MNRAS.506.1073H} both reported studies of {\it TESS} data for HD\,60435 that show pulsation amplitude variations on a time scale as short as a day. \citeauthor{2019MNRAS.487.2117B} presented a set of plots (their fig.\,9) showing amplitude spectra for the data on  a 1-d time scale. While their plots did not account for the 7.68-d rotational variations, the amplitude variations are so strong that the case was clear for rapid changes in mode amplitude. 

HD\,60435 was observed extensively by the {\it TESS} mission over a time-span of 5\,yr. The star showed significant, albeit variable, pulsation amplitude in a series of even and odd degree modes for the first two years of {\it TESS} observations, then no pulsation at all, except for a brief resurgence of a single mode during one rotation cycle. Uniquely, at present, this star has stopped pulsating.

\subsection{The {\it TESS} data} 
\label{data}

HD\,60435 was observed in the years $2018 - 2023$ by {\it TESS} in some sectors (hereafter abbreviated `S') from S3--69, while other sectors were not observed. Pre-search Data Conditioning Simple Aperture Photometry (PDCSAP) data are available at 120-s cadence for S3, 6, 7-10, 13, 27, 30, 33-35, 37, 68-69, which cover a time span from 2018 September 20 to 2023 September 20 -- 5 years. An additional four sectors of Full Frame Image (FFI) 200-s data are available for S61, 62, 63, and 67.

Because we are examining rapid changes in pulsation amplitude, and the total cessation of pulsation, and because we are concerned with the time scales for these, it is important to show a record of when the star was observed. We also must distinguish between secular pulsation amplitude changes, and those caused by oblique pulsation where the observed amplitude varies with rotation and viewing aspect. 

The {\it TESS} elliptical orbit of Earth is $\sim$13.5\,d with data downloads at each perigee passage. The sectors comprise two orbits, hence there are usually continuous data strings of one orbit, or half a sector, minus time for the data downloads. There are also changes in the satellite physical conditions with each re-acquisition of the observed field. These lead to some data being removed by the PDCSAP pipeline at the beginnings and ends of each orbit. Fortuitously for the study of HD\,60435, the observing strategy gives continuous data sets of up to  $14$\,d, which is close to 2 rotation cycles. This is sufficient to resolve the frequency multiplets (split by the rotation frequency) generated by oblique pulsation.

For our analysis, we therefore divided the data into what we call `half-sectors'. These are tabulated in an example Table\,\ref{table:1a} (with the full Table\,\ref{table:1b} in the Appendix) where they are labelled with the half-sector names (e.g., S3.1, S3.2) and with a time-stamped name giving the truncated BJD (${\rm BJD} - 2400000.0$) of the central time of each half-sector (e.g., JD58390.03 and JD58401.42 for S3.1 and S3.2, respectively). These time-stamped names are used in plots of amplitude spectra in this paper so that the gaps between those can be judged.  

\begin{figure}
\begin{center}
\includegraphics[width=1.0\linewidth,angle=0]{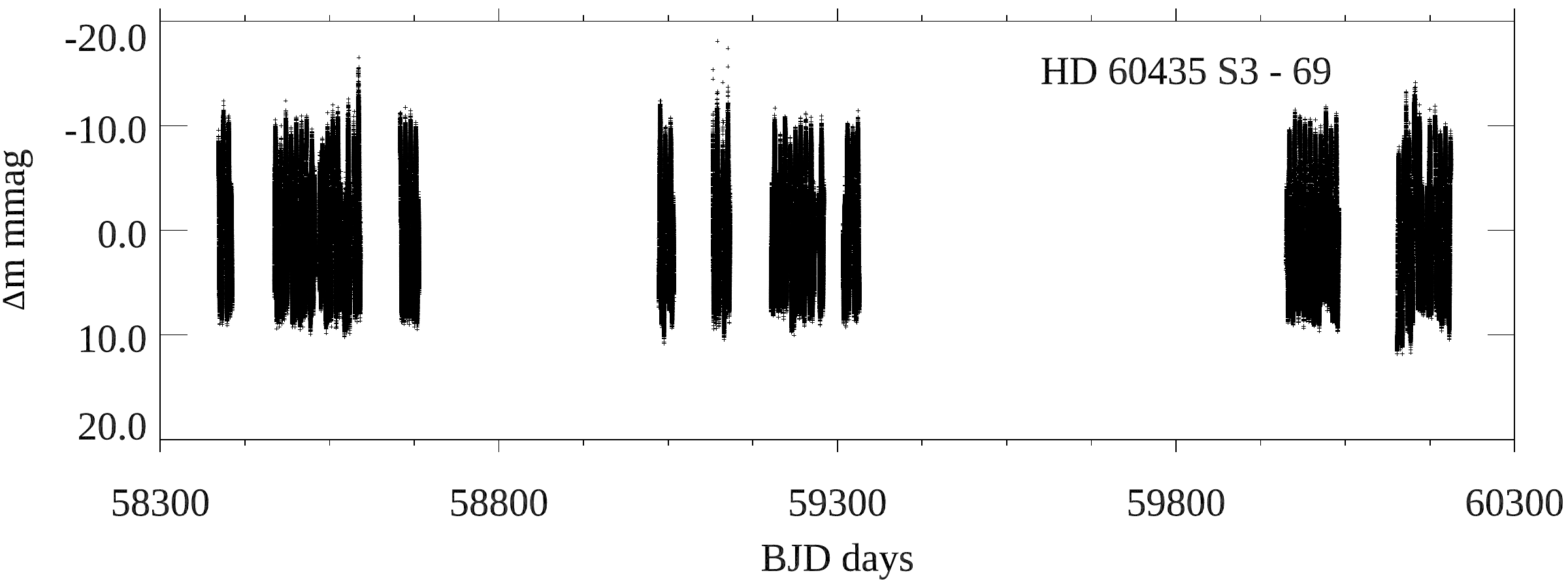}\\
\caption{A light curve of the S3--69 {\it TESS} PDCSAP data. This includes the rotational variations as well as the pulsation, although neither can be discerned at the resolution of the 5-yr span of the data. This plot accompanies Table\,\ref{table:1b} to show the time span and gaps in the observations. The time shown is ${\rm BJD} - 2400000.$}
\label{fig:lc1} 
\end{center}
\end{figure}

Fig.\,\ref{fig:lc1} shows a light curve of all of the {\it TESS} data. The data in the light curve include the rotation and pulsation variations, along with some low-frequency instrumental artefacts, all of which are so compacted at the scale of the figure that they cannot be seen. The purpose of this light curve is to show the time distribution of the data in accompaniment with Tables\,\ref{table:1a} and \ref{table:1b}.

\begin{table*}
\centering
\caption{Times and durations of the {\it TESS} half-sectors and full sectors. The data files are named with the mid-time of each half-sector for additional identification of the individual amplitude spectra in Figs\,\ref{fig:as} and \ref{fig:as2}. Columns 5 and 6 give, respectively, the duration of the half-sectors and their combined sector in days. This table is a sample; the full table is Table\,\ref{table:1b} the Appendix.} 
\begin{tabular}{ccrrrr}
\hline 
\multicolumn{1}{c}{{\it TESS} Sector} & \multicolumn{1}{c}{data set name} & \multicolumn{1}{c}{BJD time start} & \multicolumn{1}{c}{BJD time end} & \multicolumn{2}{c}{duration days}  \\
\multicolumn{1}{c}{} & \multicolumn{1}{c}{} & \multicolumn{2}{c}{$2400000+$} & \multicolumn{1}{r}{half sector} &  \multicolumn{1}{r}{full sector}  \\
\hline
\hline
3.1 & JD58390.03 & 58385.934815 & 58394.154367 & 8.22 &   \\ 
3.2 & JD58401.42 & 58396.639116 & 58406.212895 & 9.57 & 20.28  \\ 
6.1 & JD58472.65 & 58468.272565 & 58477.021355 & 8.75 &   \\ 
6.2 & JD58484.18 & 58478.243596 & 58490.045204 & 11.80 & 21.77  \\ 
7.1 & JD58497.34 & 58491.634114 & 58503.038448 & 11.40 &   \\ 
7.2 & JD58510.40 & 58504.710685 & 58516.087190 & 11.38 & 24.45  \\ 
\hline
\hline
\end{tabular}
\label{table:1a}
\end{table*}

\citet{1978Ap&SS..56..285L} showed that a time base of $T=1.5/\Delta f$ is necessary to fully resolve two frequencies separated by $\Delta f$ in a given data set, and to determine their amplitudes and phases without systematic error. It can be seen in Table\,\ref{table:1b} that most of the half-sectors have sufficient duration to resolve the rotational modulation of the pulsation amplitudes by this criterion. 

\section{The rotation and magnetic field}

\begin{figure}
\begin{center}
\includegraphics[width=1.0\linewidth,angle=0]{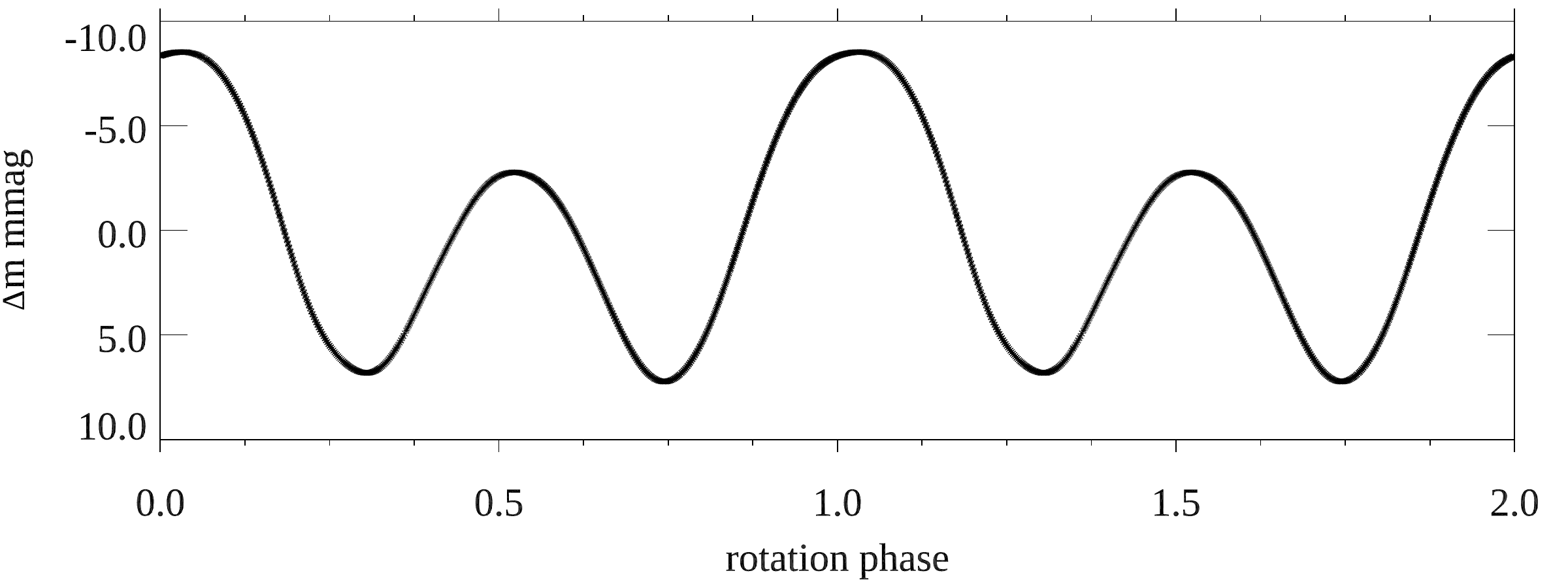}\\
\caption{The rotation curve. Two cycles are shown. $P_{\rm rot} = 7.679696 \pm 0.000005$\,d. The rotational curve was generated from a 10-harmonic fit of the rotation frequency to the entire data set, but then sampled at the same times as the S7 observations for illustrative purposes.}
\label{fig:rotation1} 
\end{center}
\end{figure}

Fig.\,\ref{fig:rotation1} shows the light variability caused by the stable spots, as is typical of $\alpha^2$\,CVn stars. These stars are known to have stable rotational variations from the spots and stable magnetic fields for observations spanning nearly a century for some stars. Importantly for HD\,60435, that rotational variability is completely stable over the time-span of the data, implying that there is no change in the spots or magnetic field that might impact on the pulsations. The spot variations in the S3-69 data yield an improved rotation period of $P_{\rm rot} = 7.679696 \pm 0.000005$\,d, both more precise and also in agreement with previous determinations. 

It is worth noting that roAp stars observed in Johnson $B$ usually show pulsation maximum near to rotational light minimum; i.e., the spots that cause the rotational light variations lie close to the poles of the pulsation axis. Often for Ap stars the rotational variation is a maximum in the red when it is minimum in blue. That is caused by flux redistribution in the spots as a consequence of the increased temperature gradient caused by the line opacity of the rare earth elements. This relationship between the maxima in red and minima in blue light for HD\,60435 is shown clearly to be the case in fig.\,2 of \citet{1990MNRAS.243..289K}, who studied the rotational light variations of this star through $UBVRI$ filters.

We assume here that the axes of the magnetic field and pulsation coincide, or nearly coincide \citep{2011A&A...536A..73B}, and that the spots are close to the magnetic poles. We also assume that the epoch of pulsation  maximum, ${\rm BJD}\,2458500.53070$, as derived from the oblique pulsator model in section\,\ref{aopm} below, also nearly coincides with the time of rotational light maximum, caused by the spots. These assumptions are justified in the results to come below. We then find an ephemeris for the times of pulsation, magnetic field, and rotational light maxima to be:
\begin{equation}
t_{\rm max} =  {\rm BJD}\,2458500.53070 + 7.679696 \pm 0.000005 \times E \, .
\label{ephemeris}
\end{equation}

\citet{2006AN....327..289H} measured a longitudinal magnetic field of $B_\ell = -296 \pm 52$\,G at ${\rm BJD} = 2453000.072$. Using equation\,\ref{ephemeris} we find that the fractional rotational phase of the \citeauthor{2006AN....327..289H} longitudinal magnetic field measurement is $\Phi = 0.234 \pm 0.005$, which is essentially in quadrature. In the standard oblique rotator model (e.g., \citealt{1950MNRAS.110..395S,1967ApJ...150..547P}),  for a dipole field the longitudinal field measured varies as 
\begin{equation}
    B_\ell=\frac{1}{20}\frac{15+u}{3-u}B_p(\cos\beta\cos i +\sin\beta\sin i \cos (2\pi\Phi)) \, ,
\label{eq_b1}
\end{equation} 
\noindent where $u$ is a limb-darkening coefficient. In sections 6 and 7 below we derive the rotational inclination to be $i = 53^\circ$ and the magnetic obliquity to be $\beta = 63^\circ$ from the oblique pulsator model. Assuming the magnetic and pulsation axes to be closely aligned, and taking $u = 0.5$  yields a polar field strength of $B_p \approx 2.8$\,kG, which is acceptably close to the asteroseismic polar field strength of $B_p \approx 4$\,kG we derive in section\,\ref{HSmodels} below from modelling the quadrupole pulsation amplitude and phase variations with rotation.

Further measurements of the longitudinal magnetic field strength in HD\,60435 are desirable. 
 
\section{Amplitude spectra}
\label{sec:as}

\begin{figure}
\begin{center}
\includegraphics[width=1.0\linewidth,angle=0]{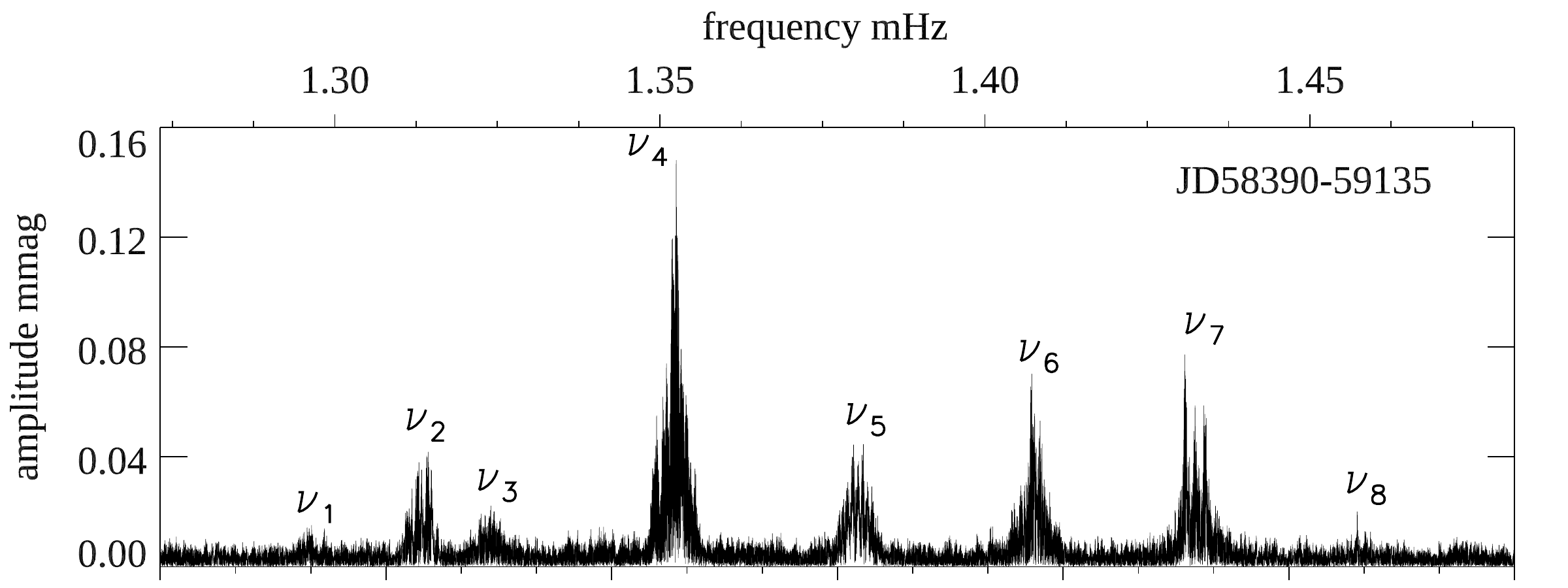}\\
\vspace{-0.6cm}
\includegraphics[width=1.0\linewidth,angle=0]{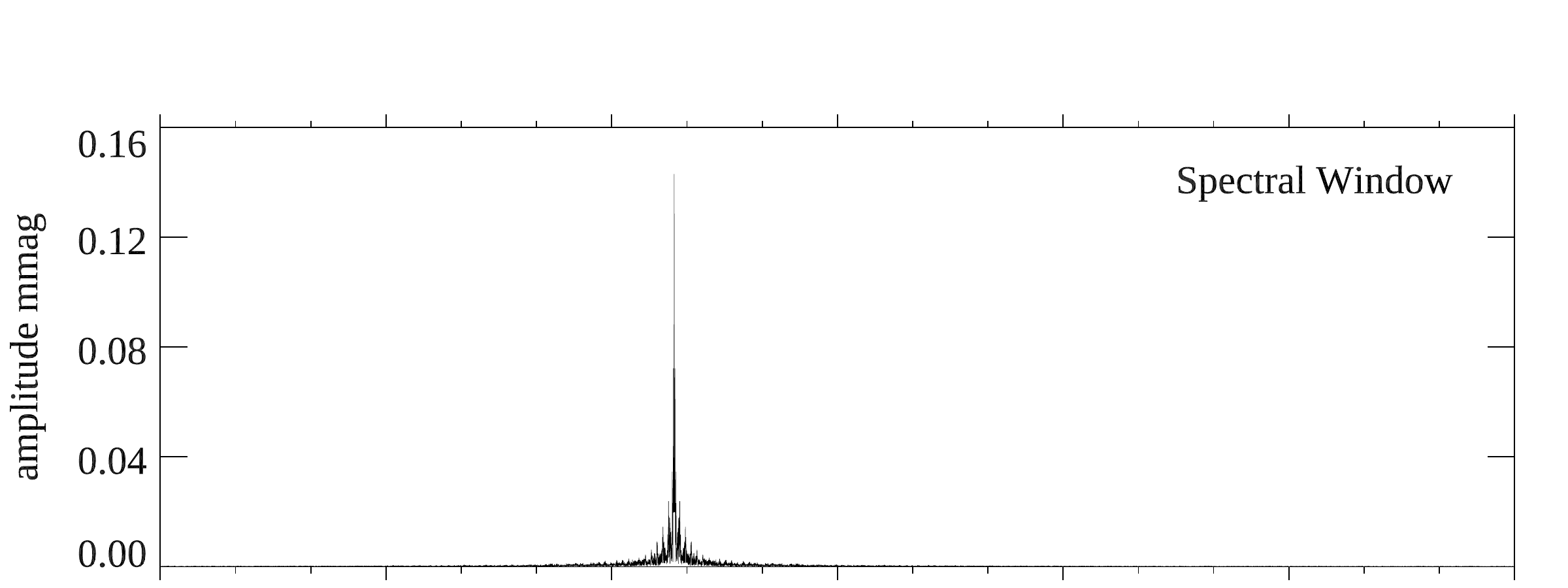}\\
\vspace{-0.6cm}
\includegraphics[width=1.00\linewidth,angle=0]{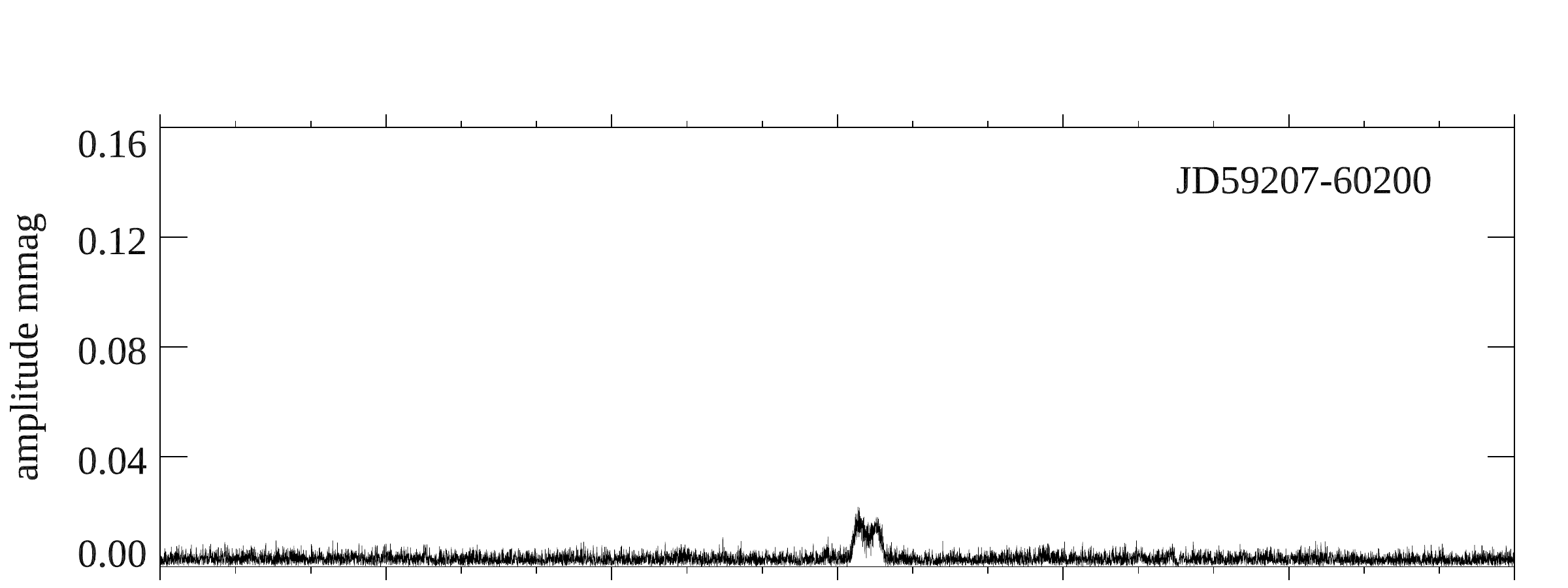}\\
\includegraphics[width=1.0\linewidth,angle=0]{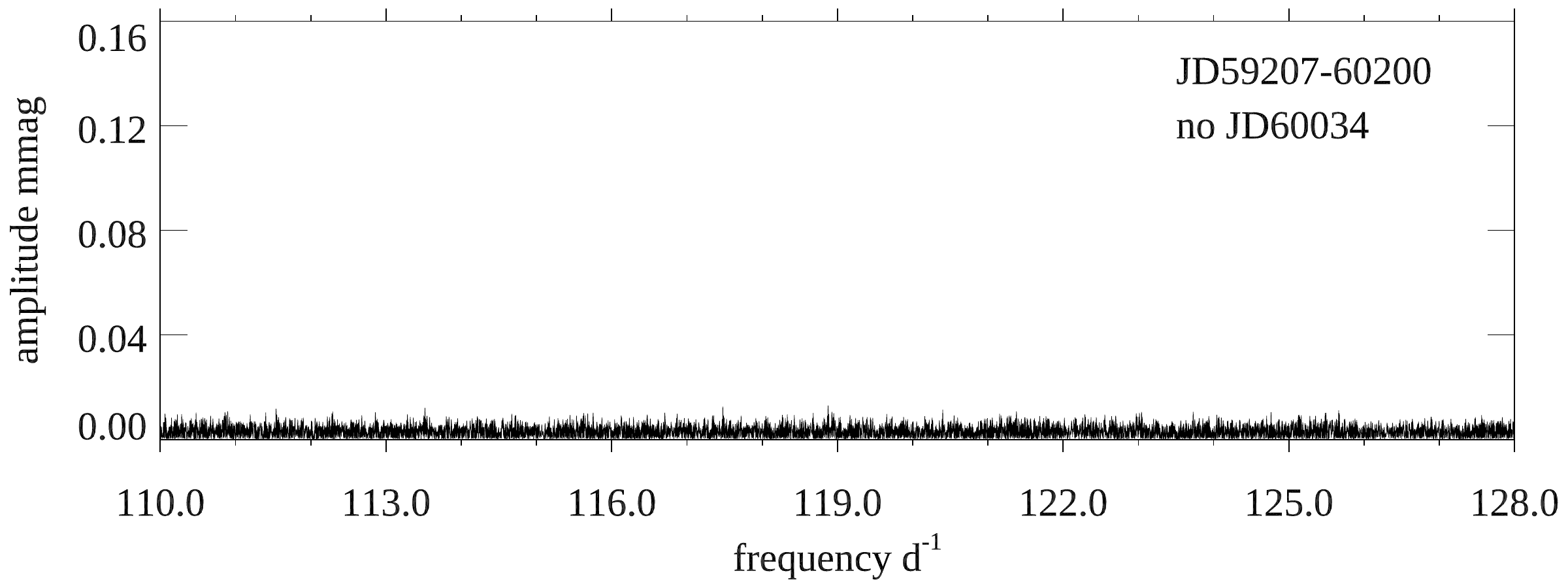}\\
\caption{Top: Amplitude spectrum over a time-span of 755.5\,d when the star was showing pulsations. The mode frequency peaks, which are broadened by amplitude modulation, are labelled as $\nu_1$ to $\nu_8$ for discussion. Second panel: The spectral window for $\nu_4$. Third panel: Amplitude spectrum over a time span of 1004\,d after the pulsations ceased, except for a brief reappearance of $\nu_5$ in the half-sector JD60034.25 (see Fig.\,\ref{fig:as2}). Bottom: The same as the third panel, but with the half sector JD60034.25 left out. Even with the nonperiodic amplitude modulation the rotational triplet structure of $\nu_7$ is evident in the top panel.}
\label{fig:as0} 
\end{center}
\end{figure}

To study the pulsation frequencies, amplitudes, and changes in those amplitudes we generated amplitude spectra using the rapid Fourier transform algorithm of \citet{1985MNRAS.213..773K}. The pulsation frequencies are well separated from the low-frequency rotation variations and instrumental artefacts, so we ran a high-pass filter to remove all the low frequency variation -- both astrophysical and instrumental -- so that the noise in the amplitude spectrum is white. This allows better determination of the uncertainties on frequency, amplitude, and phase by least-squares fitting, and it removes the high-frequency tail of the spectral window pattern of the low-frequency peaks, which have much higher amplitude than the pulsations. The high-pass filter was a simple sequential pre-whitening of low-frequency peaks up to 4\,d$^{-1}$ until the noise level at higher frequencies was reached. Throughout the rest of this paper the data analysed for the pulsation frequencies are those high-pass filtered data. 

To set the stage, the top panel of Fig.\,\ref{fig:as0} shows an amplitude spectrum for the combined S3--30 data where strong, amplitude-variable pulsation is seen in many modes. Because of the rotational modulation caused by oblique pulsation, and the amplitude modulation on a short time-scale, the peaks are broadened; they are labelled as $\nu_1$ to  $\nu_8$ in the top panel for discussion. To show the broadening, the spectral window for $\nu_4$ is shown in the second panel. The third panel shows the amplitude spectrum of the S33--69 data where there is no pulsational signal, except for $\nu_5$, which only appeared at the end of data set JD60034.25 (S63.2). When that half sector is left out the bottom panel shows a complete absence of pulsation over a time span of 1004\,d.

Because of the broadened peaks caused by amplitude modulation as shown in  Fig.\,\ref{fig:as0}, the frequencies were estimated from the broadened peaks in the amplitude spectra of the half-sectors of data when they were excited to observable amplitude. $\nu_1$, $\nu_3$, $\nu_4$, and $\nu_5$ were measured from the JD58390.03 data; $\nu_2$ was measured from JD58675.49 data; $\nu_6$ and $\nu_7$ were taken from the central frequencies of the rotational multiplets given in Table\,\ref{table:4} from the combined half-sectors JD58497.34 and JD58510.40; $\nu_8$ was measured from the JD58549.94 data.

\begin{table*}
\centering
\caption{Comparison between the ground-based $B$-data frequencies of \citet{1987ApJ...313..782M} and the {\it TESS} frequencies measured for $\nu_1$ to $\nu_8$.  The sixth column shows the frequency differences, which are used to determine the large frequency separation, $\Delta \nu_0 = 56$\,$\upmu$Hz. The final column shows the observed ratios discussed in section~\ref{sec:modeID_models}.
}
\begin{tabular}{ccccrcr}
\hline
\multicolumn{1}{c}{label} & \multicolumn{2}{c}{ground-based} & \multicolumn{2}{c}{{\it TESS}}  & \multicolumn{1}{c}{frequency } & \multicolumn{1}{c}{$r_{\rm obs}$}\\
 & \multicolumn{2}{c}{frequency} & \multicolumn{2}{c}{frequency}  & \multicolumn{1}{c}{difference} & \\
 & \multicolumn{1}{c}{mHz} &\multicolumn{1}{c}{d$^{-1}$} & \multicolumn{1}{c}{d$^{-1}$} &\multicolumn{1}{c}{mHz} & \multicolumn{1}{c}{$\upmu$Hz} &\multicolumn{1}{c}{$\upmu$Hz} \\
\hline
\hline
 & $0.7090$ & $61.2576$ &  & & &\\ 
 & $0.7614$ & $65.7850$ &  &  & &\\
 & $0.8428$ & $72.8179$ &  &  & &\\
 & $0.9397$ & $81.1901$ &  &  & &\\
 & $0.9906$ & $85.5878$ &  &  & &\\
 & $1.0433$ & $90.1411$ &  &  & &\\
 & $1.0990$ & $94.9536$ &  &  & &\\
 & $1.1482$ & $99.2045$ &  &  & &\\
 & $1.1734$ & $101.3818$ &  &  & &\\
 &$1.2250$  & $105.8400$ &  &  & &\\
 &$1.2848$  & $111.0067$ &  &  & &\\
 $\nu_1$ & &  & $112.16 \pm 0.14$ &  $1.29814 \pm 0.00162$  &$\nu_2 - \nu_1 = 14.2$  &\\
 $\nu_2$ & &  & $113.39 \pm 0.04$ & $1.31235 \pm 0.00046$ & $\nu_3 - \nu_2 = 12.1$ &\\
$\nu_3$ & $1.3281$ & $114.7478$ & $114.43 \pm 0.05$ & $1.32447 \pm 0.00058$ & $\nu_3 - \nu_1 = 26.3$ & $-0.01 \pm 0.02$\\ 
 $\nu_4$ & $1.3525$ & $116.8560$ & $116.83 \pm 0.02$ & $1.35221 \pm 0.00023$ & $\nu_4 - \nu_3 = 27.7$ &\\
 $\nu_5$ & $1.3810$ & $119.3184$ & $119.35 \pm 0.04$ & $1.38137 \pm 0.00046$ & $\nu_5 - \nu_4 = 29.2$ & $0.02 \pm 0.01$\\
 $\nu_6$ & $1.4073$ & $121.5907$ & $121.62 \pm 0.01$ & $1.40760 \pm 0.00012$ & $\nu_6 - \nu_5 = 26.2$ &\\
 $\nu_7$  & $1.4334$ & $123.8458$ & $123.77 \pm 0.01$ & $1.43248 \pm 0.00012$ & $\nu_7 - \nu_6 = 24.9$ & $0.00 \pm 0.01$\\
 $\nu_8$ & $1.4572$ & $125.9021$ & $125.91 \pm 0.09$ & $1.45728 \pm 0.00104$ &  $\nu_8 - \nu_7 = 24.8$ &\\
\hline
\hline
\end{tabular}
\label{table:3}
\end{table*}
 
\begin{figure}
\begin{center}
\includegraphics[width=1.0\linewidth,angle=0]{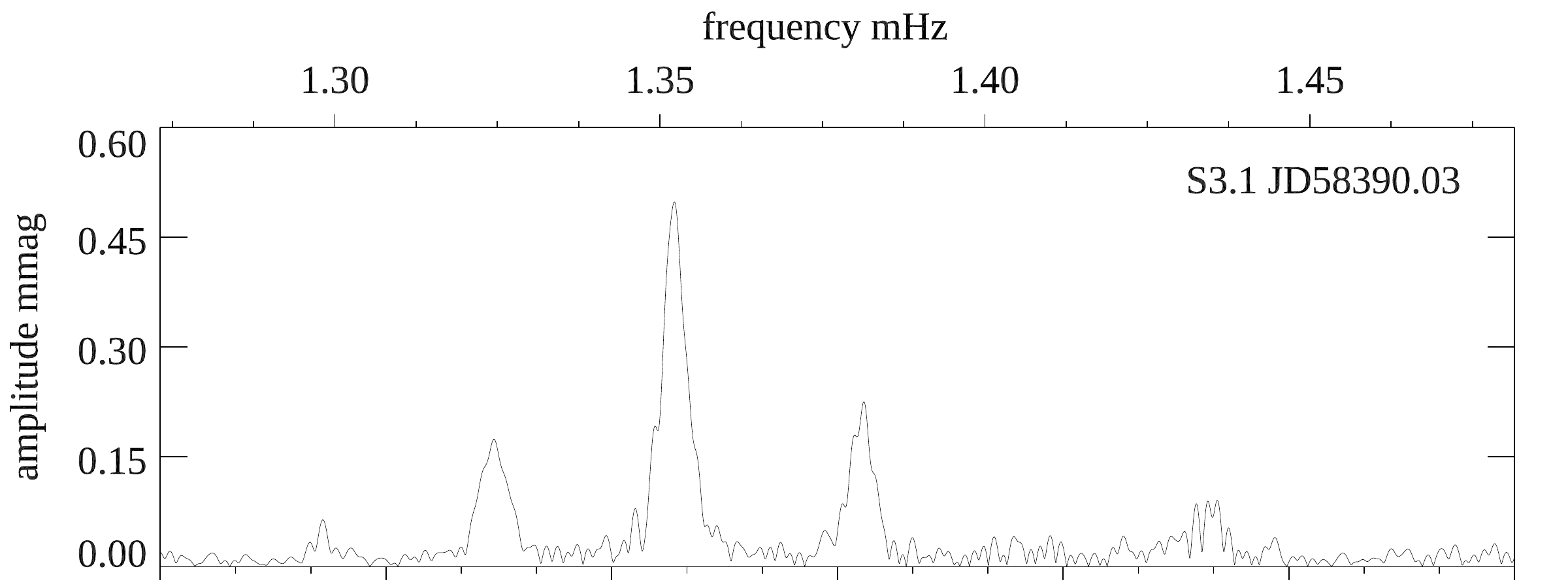}\\
\vspace{-0.6cm}
\includegraphics[width=1.0\linewidth,angle=0]{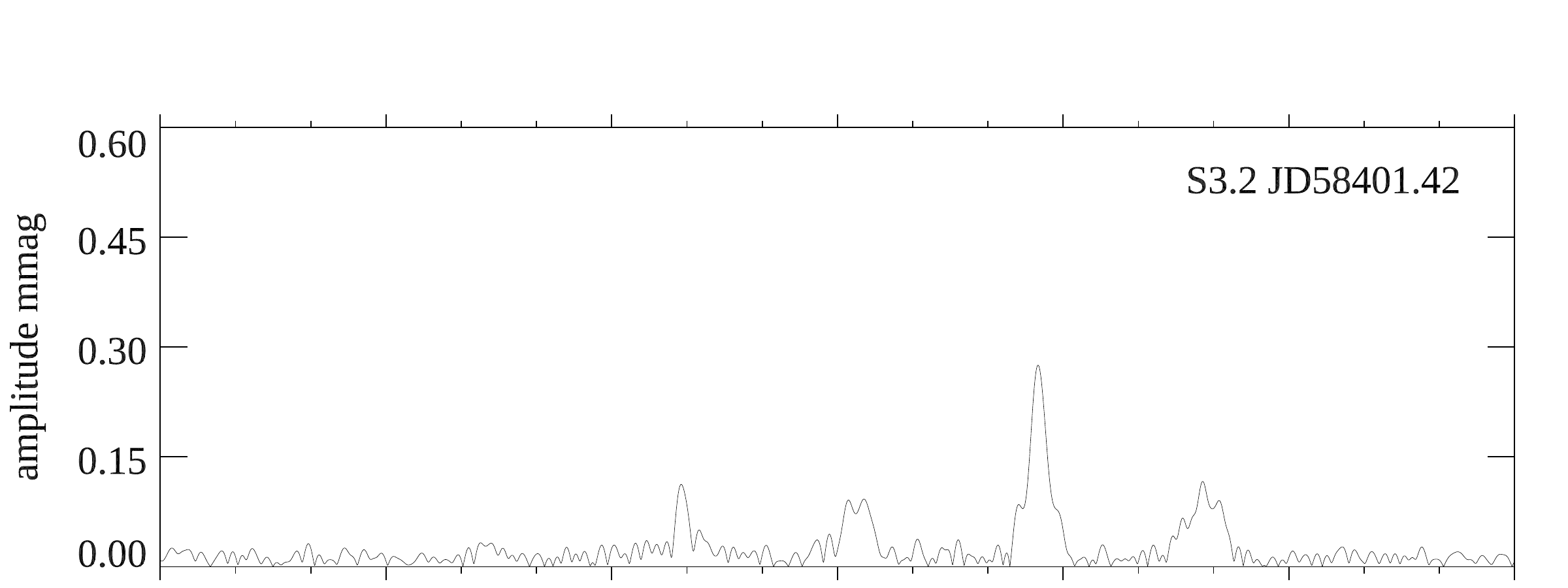}\\
\vspace{-0.6cm}
\includegraphics[width=1.0\linewidth,angle=0]{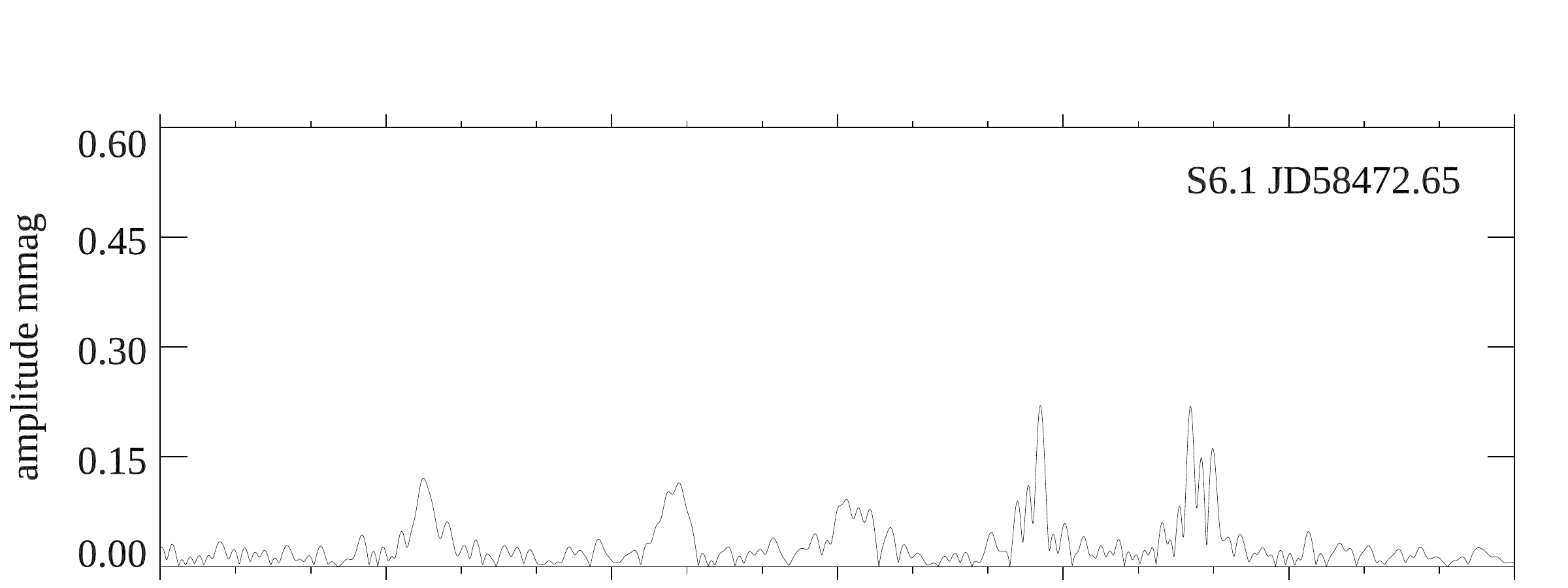}\\
\vspace{-0.6cm}
\includegraphics[width=1.0\linewidth,angle=0]{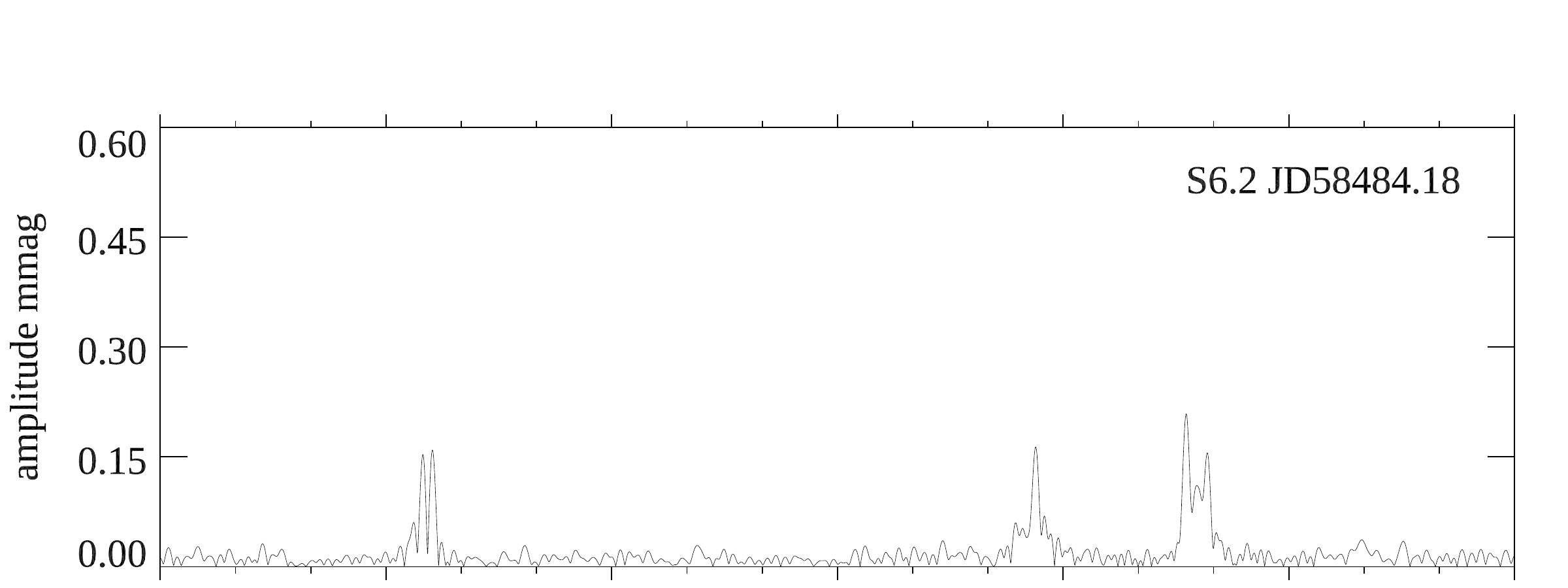}\\
\vspace{-0.6cm}
\includegraphics[width=1.0\linewidth,angle=0]{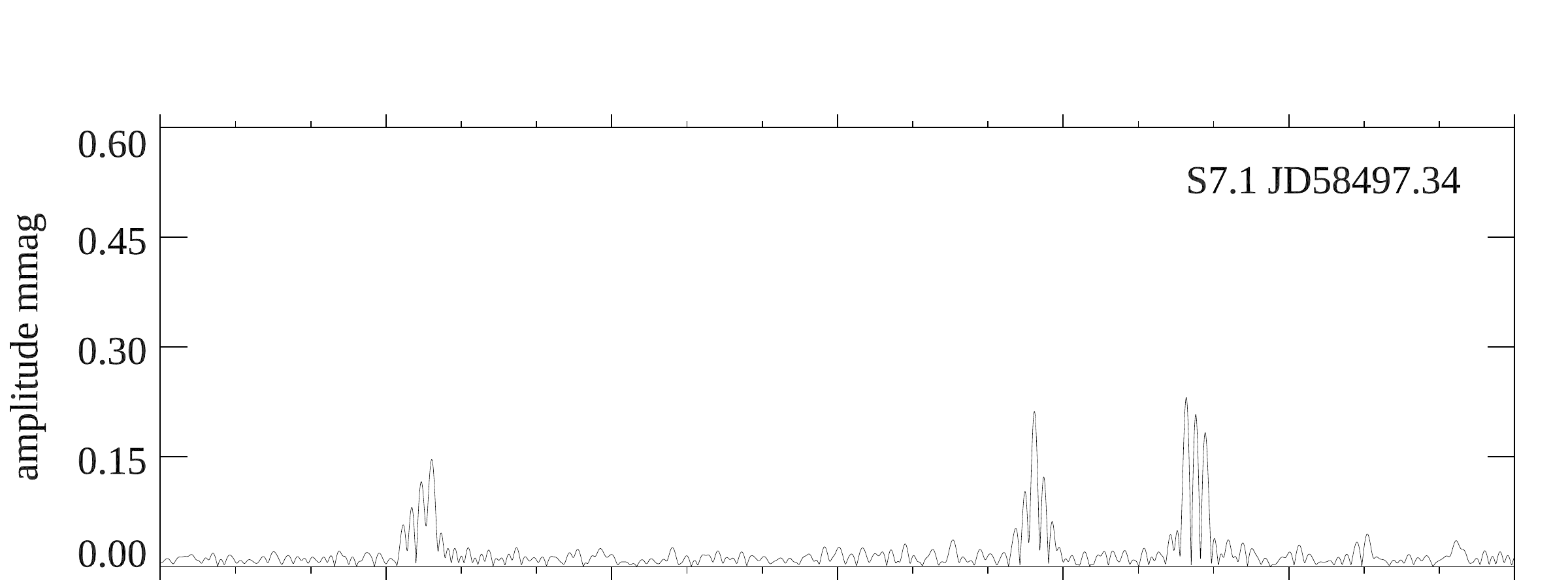}\\
\includegraphics[width=1.0\linewidth,angle=0]{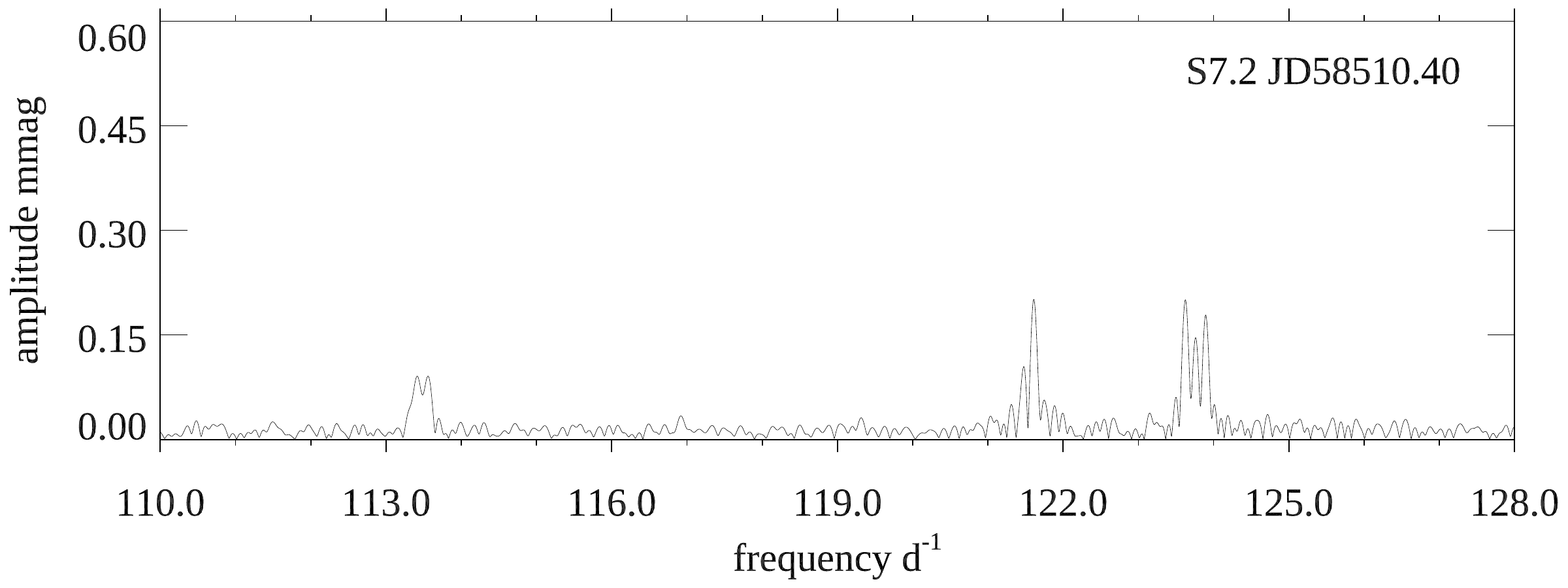}\\
\caption{Example amplitude spectra of the half-sectors showing the amplitude modulation and mode changes. The full set of amplitude spectra for all of the half-sectors is shown in Fig.\,\ref{fig:as2} in the Appendix. }
\label{fig:as} 
\end{center}
\end{figure}

Table\,\ref{table:3} shows a comparison between the ground-based frequencies found by \citet{1987ApJ...313..782M} and the {\it TESS} frequencies found in this study.  The uncertainties given are the formal errors determined from the data subsets (chosen as explained above) because of the amplitude modulation broadening of the peaks in the amplitude spectra, as seen in Figs\,\ref{fig:as} and \ref{fig:as2}. These uncertainties appear reasonable, given the good agreement between the {\it TESS} and ground-based values for six of the frequencies in data taken decades apart in time. 

Other than those 6 frequencies, most of the rest of the ground-based frequencies are outside of the frequency range of detected peaks in the {\it TESS} data. \citet{1993ApJ...413L.125T} obtained 5.5\,hr of high-speed photometric observations of HD\,60435 in the ultraviolet with the Hubble Space Telescope. They found one certain pulsation mode frequency at $123.70$\,d$^{-1}$, which corresponds to the $\nu_7$ dipole frequency seen in Fig.\,\ref{fig:as0}, showing that HD\,60435 was pulsating in 1991 August. The other possible mode frequencies they suggested do not have counterparts in Table\,\ref{table:3}. 

For the {\it TESS} data Table\,\ref{table:3} also shows that there is a series of alternating even-and-odd degree modes with separation of $\sim$$28$\,$\upmu$Hz, hence the large separation is $56$\,$\upmu$Hz. Many of the ground-based frequencies that have no counterpart in the {\it TESS} data show separations around $52$\,$\upmu$Hz, suggesting that those may be from consecutive modes of the same degree. Because of the challenges caused by the complex spectral windows for the ground-based data, we do not use those further in this analysis. We return to identifying the degrees of the modes for the {\it TESS} data from models in section\,\ref{sec:modeID_models} below.

Fig.\,\ref{fig:as} shows a sample of the amplitude spectra for each of the half-sector data strings given in Table\,\ref{table:1a} (the full set of amplitude spectra are in Fig.\,\ref{fig:as2} of the Appendix), where extreme changes in pulsation amplitude for various modes can be seen. It is notable that when the $\nu_4$ pulsation is present, the $\nu_6$ and $\nu_7$ pulsations are not, and vice versa. This gives the appearance of mode switching via energy transfer between these modes, but the complete disappearance of all pulsation in the S33-69 data shows the situation is more complex. Clearly, the pulsation energy in the observed modes is not conserved.

\subsection{Tracing the amplitude changes during the {\it {\it TESS}} observations}

The amplitude changes in HD\,60435 and the cessation of pulsation are the most important results of this paper. It is useful to have complementary ways of viewing these changes, as we try to understand their causes and consequences; thus we present the amplitude changes in three separate ways. The first method is the traditional amplitude spectrum views in Figs.\,\ref{fig:as} and \ref{fig:as2} presented in the last section. The other two methods we present in this section are least-squares fits of the five most prominent pulsation frequencies to short sections of the data, and a wavelet analysis.

\subsubsection{Amplitude changes seen with least-squares fits}

\begin{figure*}
\begin{center}
\includegraphics[width=0.85\linewidth,angle=0]{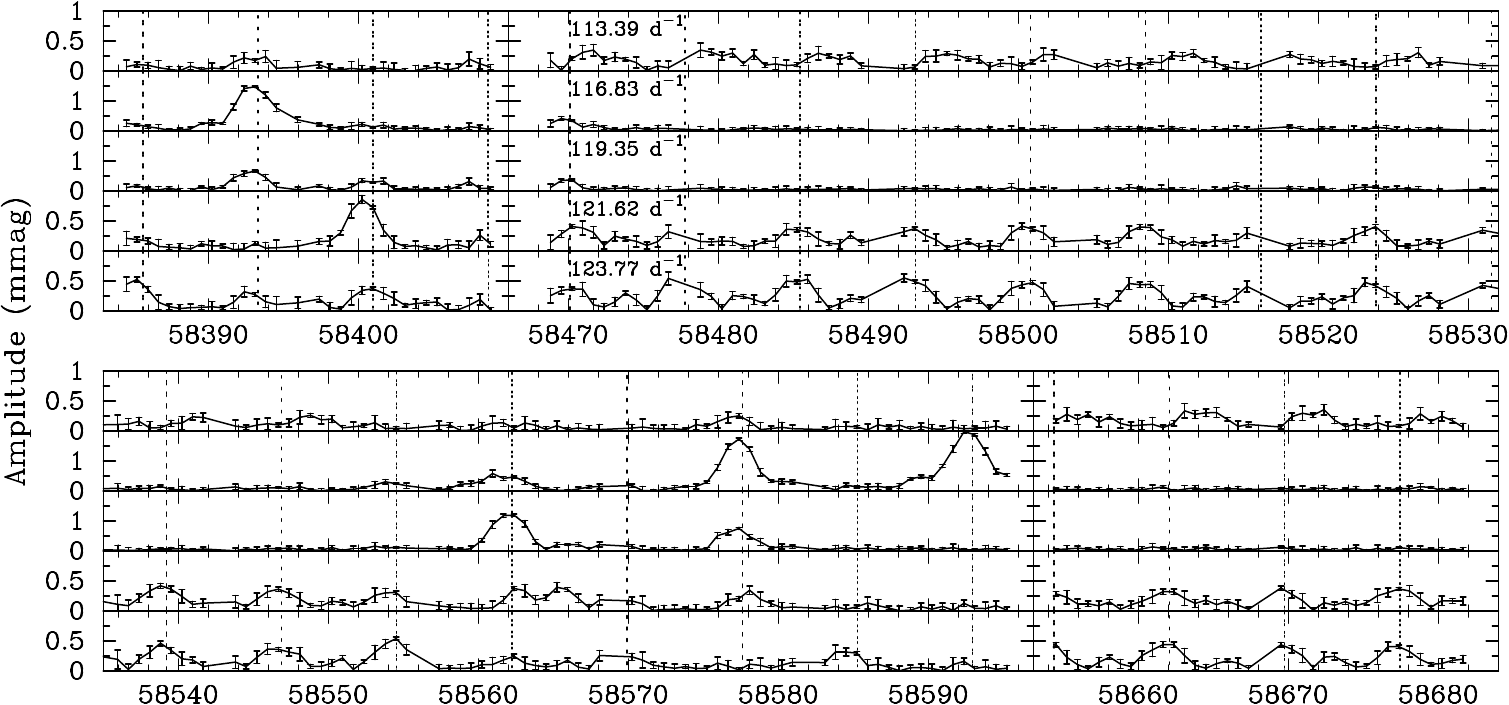} \\
\vspace{1mm}
\includegraphics[width=0.85\linewidth,angle=0]{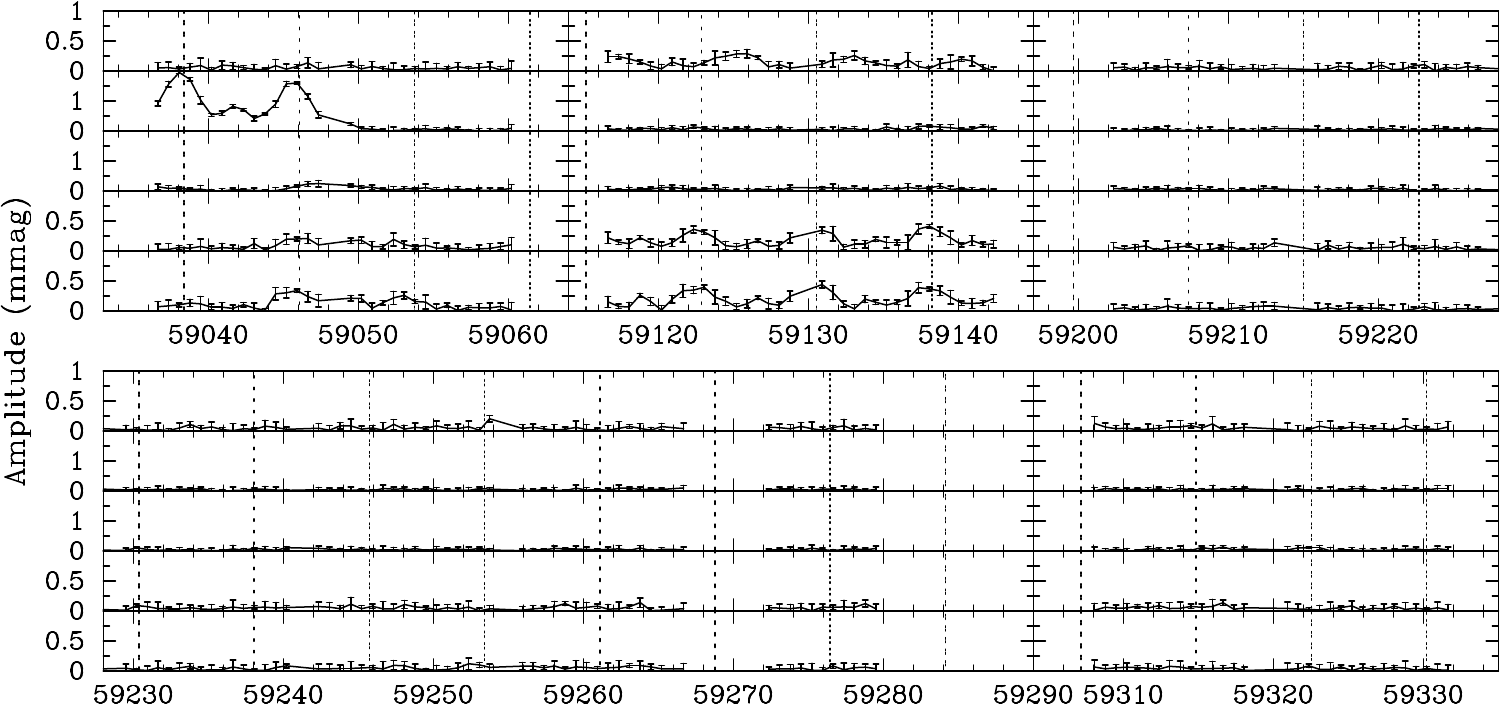} \\
\vspace{1mm}
\includegraphics[width=0.85\linewidth,angle=0]{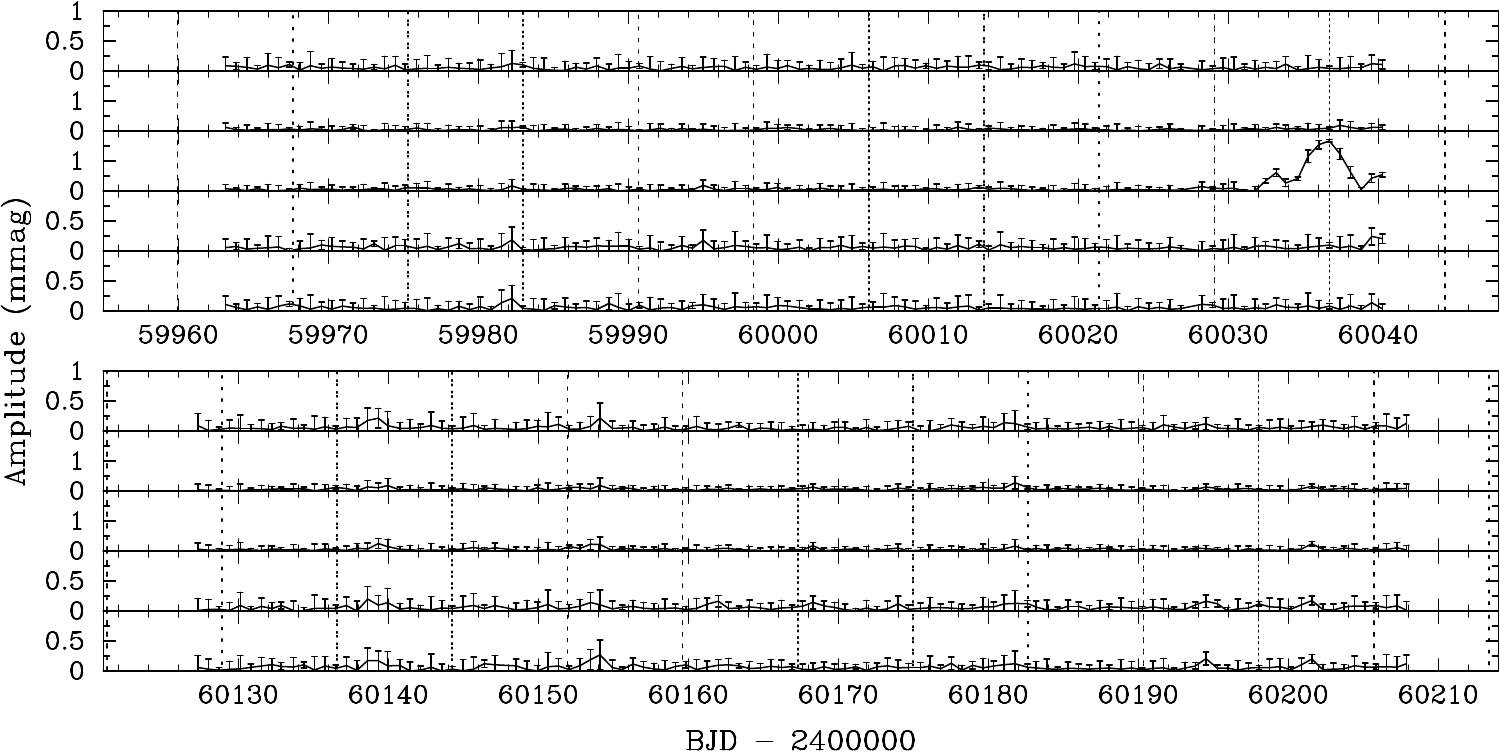} \\
\caption{The run of the amplitudes of the five strongest pulsation modes -- $\nu_2$, $\nu_4$, $\nu_5$, $\nu_6$, and $\nu_7$ -- of HD 60435 over time. The error bars are formal values computed according to \citet{1999DSSN...13...28M}. Vertical dashed lines denote times of rotational light maximum in the TESS bandpass.}
\label{fig:amps} 
\end{center}
\end{figure*} 

To describe the changes in amplitude of the five strongest pulsation modes of HD 60435 in more detail, we subdivided the {\it TESS} data into small subsets. To assure that the frequencies are resolved, we took the frequency separation to be $28$\,$\upmu$Hz which requires data time spans of 0.6\,d to fulfil the \citet{1978Ap&SS..56..285L} criterion, $T=1.5/\Delta f$. Conservatively, we chose data time spans of 0.7\,d. The run of the pulsation amplitudes of  five modes is shown in Fig.\,\ref{fig:amps}. They are $\nu_2$, $\nu_4$, $\nu_5$, $\nu_6$, and $\nu_7$ as listed in Table\,\ref{table:3}.  The amplitudes of $\nu_1$, $\nu_3$, and $\nu_8$ are too small to be traced meaningfully in this way. The short time-scale amplitude modulations of the modes are evident, as is especially the cessation of pulsation for all modes from ${\rm BJD}\,2459200$, with the exception of a brief return of $\nu_5$ around ${\rm BJD}\,2460037$. Fig.\,\ref{fig:amps} demonstrates that amplitude variations of individual modes occur on time scales shorter than a single rotation period.

\subsubsection{Amplitude changes with a wavelet analysis}

\begin{figure}
\begin{center}
\includegraphics[width=1.0\linewidth,angle=0]{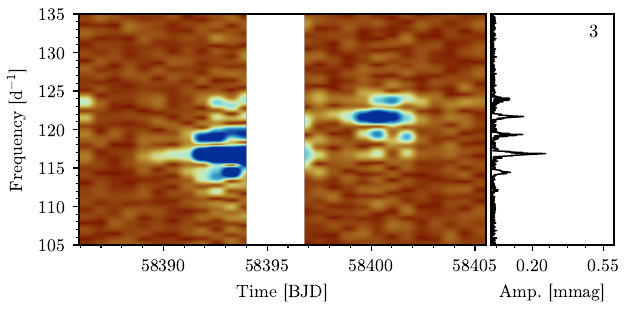}\\
\includegraphics[width=1.0\linewidth,angle=0]{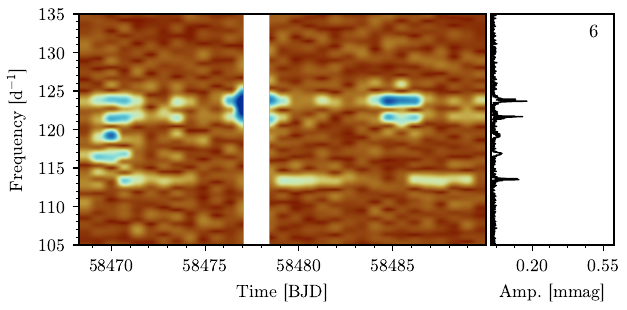}\\
\includegraphics[width=1.0\linewidth,angle=0]{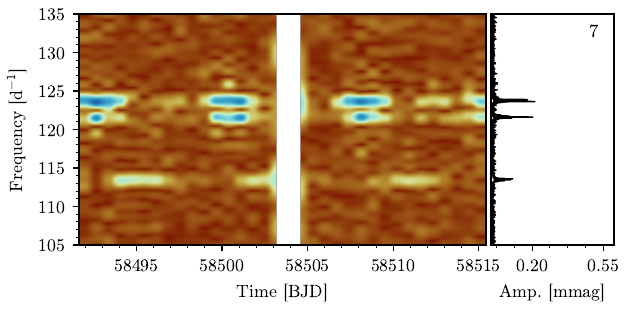}\\
\caption{Example wavelet plots by sector. The time is ${\rm BJD} - 2400000.$  The amplitude modulations from both the changing view of oblique pulsation with the $7.679696$-d rotation and the secular changes can be seen. The colours range linearly from amplitude zero (red) to amplitude 0.6\,mmag (blue). The sectors are numbered in the top right corner of each plot. The full set of plots are in Fig.\,\ref{fig:wav2} in the Appendix.}
\label{fig:wav1} 
\end{center}
\end{figure}

The third method of viewing the amplitude changes with time is with a wavelet analysis.  Using the Astropy\footnote{http://www.astropy.org} implementation of the Lomb-Scargle periodogram \citep{2022ApJ...935..167A} we calculated a sliding window amplitude spectrum to illustrate the mode evolution over time. To do this, we convolved the time series with a sliding rectangular window of width 0.8\,d, at steps of 0.5\,d. By testing, we found that smaller window sizes result in moderately decreased frequency resolution, whereas larger sizes show the changes in the modes less well. We also experimented with using a Gaussian window, but found only minor differences with the rectangular window approach. In each segment of the light curve, we calculated the amplitude spectrum between 105 and 135\,d$^{-1}$ with an oversampling factor of 5. We performed the analysis for each sector individually and show the results in Figs\,\ref{fig:wav1} and \ref{fig:wav2}.

\subsubsection{The harmonic and the 20-s data}

There is a second harmonic observable when $\nu_4$ is near its greatest amplitude. That is visible in the 120-s data. There are 20-s cadence data available in S27, 30, 33-35, 37, but given the amplitude modulation, we make no astrophysical inference from this second harmonic, except to mention that it is observed to be present, which is common for many roAp stars;  see \citet{2021MNRAS.506.1073H,2024MNRAS.527.9548H} for examples.

\subsection{ Further ground-based observations}

\begin{table}
\centering
\caption{A journal of the LCOGT Johnson $B$ observations. The BJD is the mid-point of the nightly run. The third  column is the rotation phase at that BJD; the fourth column is the duration of the observing run. The integration time was $10.5$\,s. The final column gives the height of the highest noise peaks in the amplitude spectrum of the observations each night in the frequency range of the previously known pulsation of HD\,60435.} 
\begin{tabular}{ccccrc}
\hline
Date & BJD & rot.& $\Delta$t &  \multicolumn{1}{c}{N} & A$_{\rm max}$ \\
	& $-2400000.0$ &phase & hr & & mmag\\
\hline
$2023-12-25$ & 60303.99055 & 0.83 & 0.68 & 89 & 3.5 \\
$2023-12-25$ & 60304.20083 & 0.86 & 0.49 & 138 & 2.7 \\
$2023-12-26$ & 60305.15951 & 0.99 & 0.99 & 127 & 2.5 \\
$2023-12-27$ & 60306.02235 & 0.10& 0.49 & 130 & 1.8\\
$2023-12-28$ & 60307.14724 & 0.25& 0.98 & 124  & 11.3\\ 
$2023-12-30$ & 60309.02237 & 0.49& 0.98 &126 & 2.4\\
$2024-01-01$ & 60311.10257 & 0.76& 0.98 &138 & 1.7\\
$2024-01-02$ & 60312.04046 & 0.88& 0.98 & 134 & 2.1\\ 
$2024-01-03$ & 60313.03587 & 0.01& 0.98 & 133 & 2.8\\
$2024-01-05$ & 60315.02172 & 0.27& 0.95 & 128 & 1.4 \\
$2024-01-06$ & 60316.01748 & 0.40& 0.74 & \phantom{0}94 & 8.1\\
\hline
\end{tabular}
\label{table:lcogt}
\end{table}

To test whether pulsations had restarted, we used the Las Cumbres Observatory Global Telescope (LCOGT) network of telescopes \citep{2013PASP..125.1031B} to observe HD\,60435 through a Johnson $B$ filter for $0.5 - 1.0$\,hr over 10 nights in 2023 December and 2024 January. All observations were collected with one of the 0.4-m telescopes at Siding Spring Observatory, Australia, using the QHY camera in full frame mode. The data were reduced and aperture photometry performed with an adapted version of the TEA-Phot reduction code \citep{2019A&A...629A..21B}.

A journal of the observations is given in Table\,\ref{table:lcogt}. The LCOGT observations span $98 - 110$\,d after the last of the {\it TESS} observations, and they cover near two rotations of HD\,60435. From equation 1 we calculate the rotation phase of the mid-point of each nightly run, as seen in column 3 of Table\,\ref{table:lcogt}. Several nights were near to rotation phase zero, which is rotational light maximum, hence pulsation maximum. In the past HD\,60435 has had peaks in the amplitude spectrum as great as 6\,mmag in Johnson $B$ observations \citep{1984MNRAS.209..841K}, although \citet{1987ApJ...313..782M} generally found peaks near $3$\,mmag during pulsation maximum. As can be seen from Table\,\ref{table:lcogt} the noise peaks range between $2.1 - 3.5$\,mmag on the nights where observations were near rotation and pulsation maximum, suggesting that HD\,60435 had not returned to pulsating during the time span of our new $B$ observations.

\subsection{Pulsational radial velocity variations}

\begin{table}
\centering
\caption{A journal of the SALT HRS observations. The BJD is the mid-point of the nightly run. The third  column is the rotation phase at that BJD; the fourth column, N, is the number of exposures. The integration time was $70$\,s and the cadence was $113$\,s. } 
\begin{tabular}{ccccc}
\hline
\multicolumn{1}{c}{date} & \multicolumn{1}{c}{BJD} & rot.  &
\multicolumn{1}{c}{N} & \multicolumn{1}{c}{duration}  \\
\multicolumn{1}{c}{} & \multicolumn{1}{c}{$-2400000.0$} & phase  &
\multicolumn{1}{c}{} & \multicolumn{1}{c}{s}  \\\hline
\hline
$ 2014-11-08 $ & $56969.501843$ &$0.64$ &$23 $ & $  2556 $ \\
$ 2014-11-09 $ & $56970.506103$ & $0.77$ &$23 $ & $  2556 $ \\
$ 2014-12-23 $ & $57014.611339$ &$0.51$ &$24 $ & $  2669 $ \\
$ 2015-01-03 $ & $ 57026.345736$ &$0.04$ &$23 $ & $  2556 $ \\
\hline
\hline
\end{tabular}
\label{table:hrs}
\end{table}

\begin{figure}
\begin{center}
\includegraphics[width=1.0\linewidth,angle=0]{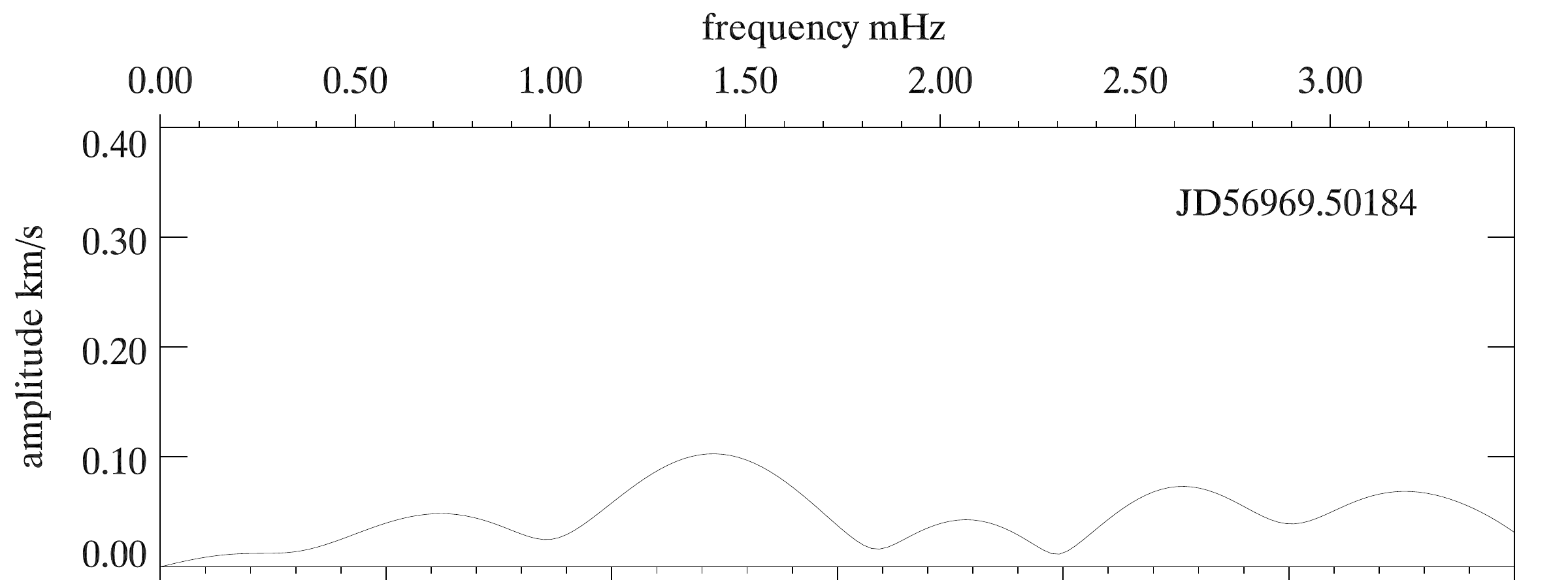}\\
\includegraphics[width=1.0\linewidth,angle=0]{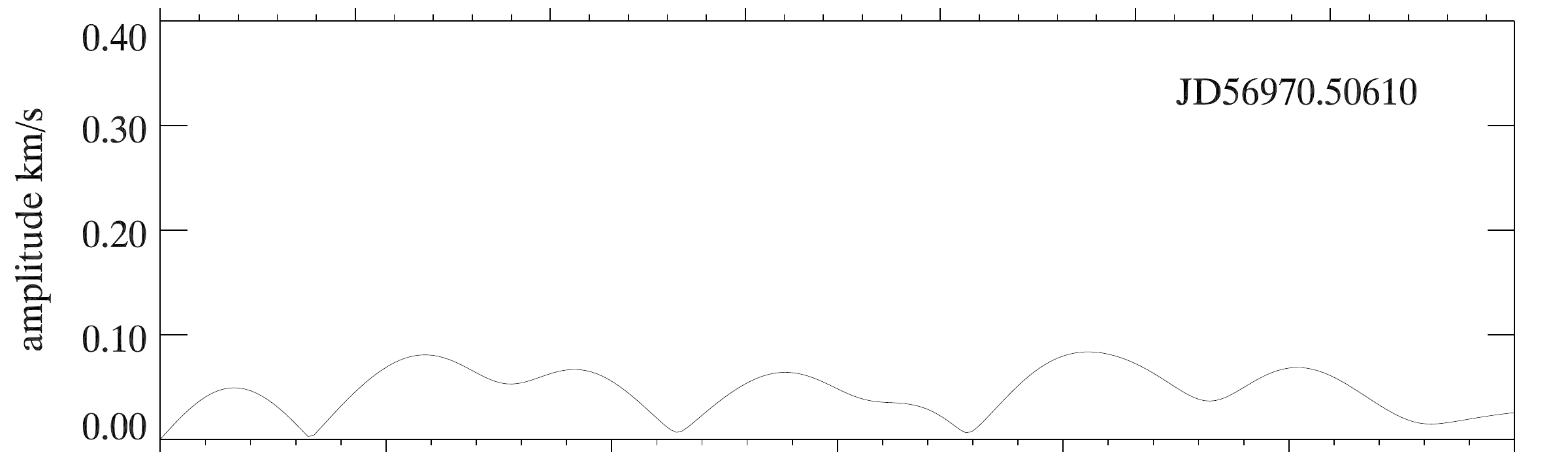}\\
\includegraphics[width=1.0\linewidth,angle=0]{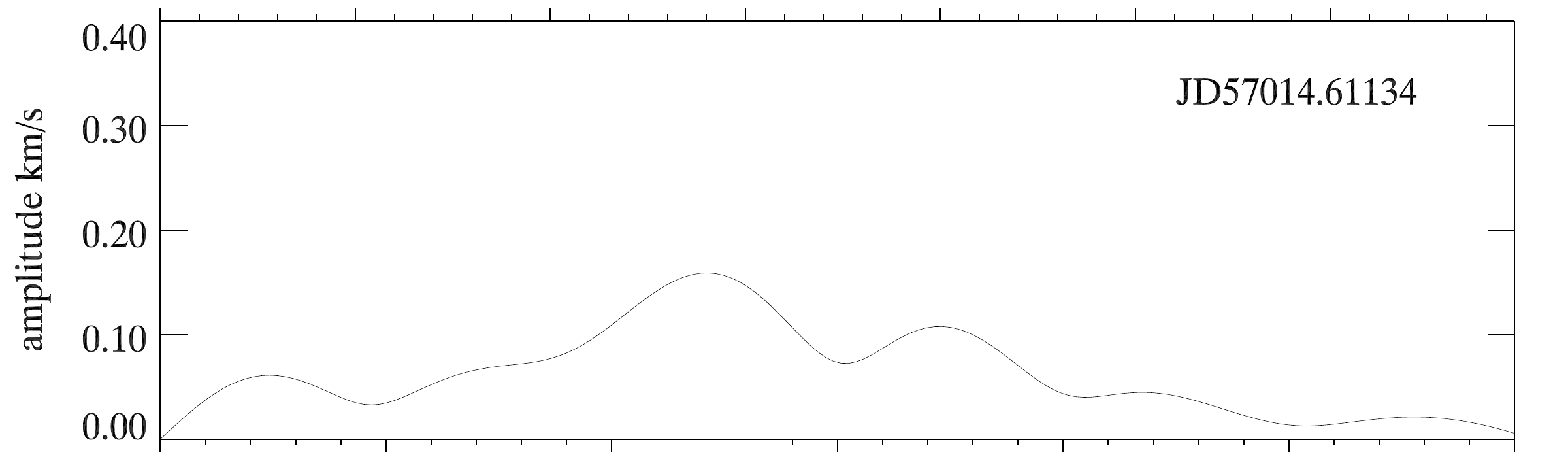}\\
\includegraphics[width=1.0\linewidth,angle=0]{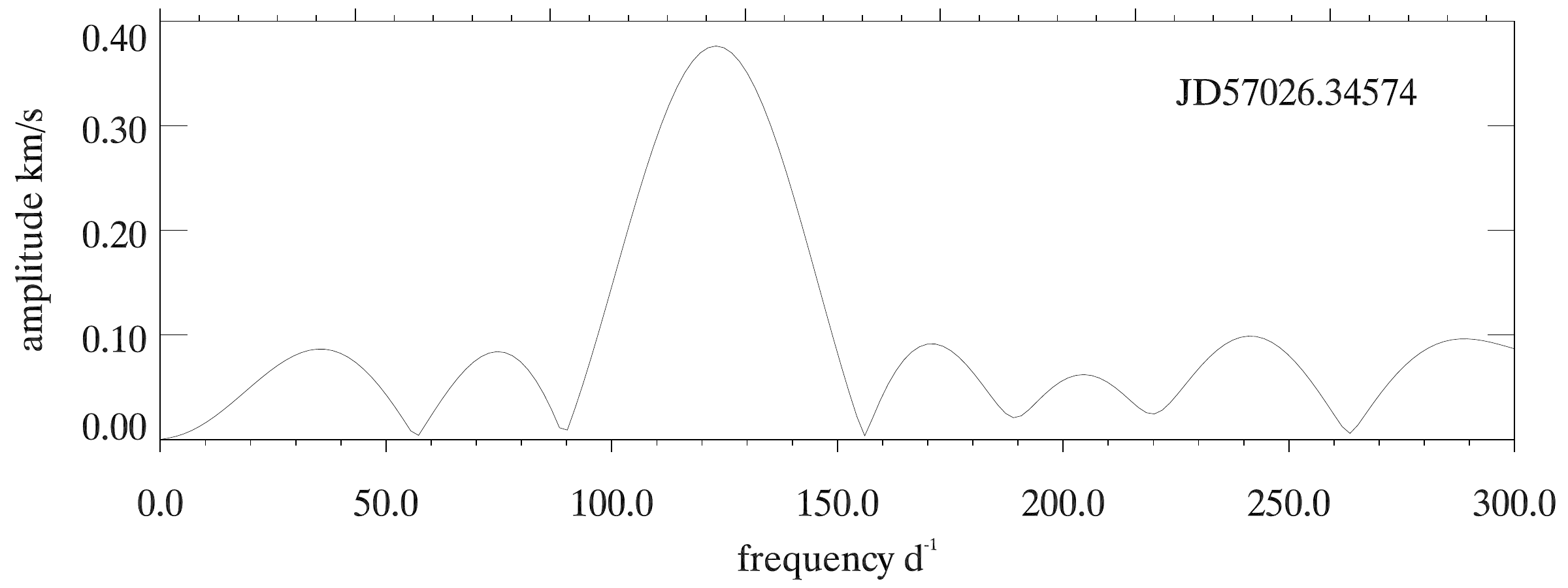}\\
\caption{Amplitude spectra of the pulsational radial velocity results from the four nights of SALT observations. It is clear that on the fourth (bottom panel) night that the star is pulsating within the expected frequency range.}
\label{fig:rvs} 
\end{center}
\end{figure}

HD\,60435 was observed with the High Resolution Spectrograph \citep[HRS;][]{bramall2010,crause2014} on the Southern African Large Telescope \citep[SALT;][]{buckley2006} on four separate nights in $2014 - 2015$. HRS is a dual-beam spectrograph with wavelength coverage of $3700-5500$\,\AA\ and $5500-8900$\,\AA. The observations were automatically reduced using the SALT custom pipeline, which is based on ESO's {\sc{midas}} pipeline \citep{pyhrs2,pyhrs3}. Observations were made under the proposal code 2014-2-SCI-068 (PI: Medupe), and were collected in the Medium Resolution mode ($R\sim 45000$) with exposure times of 70\,s resulting in a cadence of 113\,s. Table\,\ref{table:hrs} gives a journal of the observations.

For the pulsation analysis, we normalised the spectra using the SUPPN{\sc{et}} code \citep{2022A&A...659A.199R} and treated each night separately. Ions of the rare earth elements are typically stratified in the atmospheres of roAp stars, and there is often detectable pulsation amplitude seen in lines of Nd\,{\sc{iii}} (see, e.g., \citealt{2007A&A...473..907R}, \citealt{2005MNRAS.364..864E}).
We identified lines of Nd\,{\sc{iii}} that were locally isolated and had sufficient equivalent width for cross-correlation. This resulted in the use of seven lines at $\lambda\lambda$: 5802.53, 5845.02, 5851.54, 5987.68, 6145.07, 6327.26, and 6550.23\,\AA. The central wavelength and equivalent width (0$^{\rm th}$ moment) of each line was measured by the moment method \citep{1992AandA...266..294A}. These were constructed into a $\delta$-function template using the wavelengths as positions and the equivalent widths as the depth to weight the lines \citep[see][]{Wright2008}. Cross-correlation with this template produced line profiles that combine the information of the variation from all the lines chosen. From the cross-correlation profiles the radial velocity measurements were made using the first moment \citep{1992AandA...266..294A}.

Fourier analysis of the resultant radial velocities showed a clear detection of pulsation on night 4 with a semi-amplitude of $385\pm35$\,m\,s$^{-1}$, when the rotation phase of $0.04$ is at the expected time of pulsation amplitude maximum. There are tentative detections on two of the other nights (Fig.\ref{fig:rvs}). While the frequency resolution and limited number of observations do not allow for a detailed study, they do allow us to confirm that the star was pulsating during the observation period in $2014 - 2015$, 30 years after the seminal observations of \citet{1987ApJ...313..782M}, and $4$ years prior to the beginning of the {\it TESS} data analysed in this paper.

\section{Fundamental data}
\label{fund}

From {\it Gaia} DR3 \citep{2016A&A...595A...1G, 2023A&A...674A...1G, 2023A&A...674A..26C} we find $T_{\rm eff} = 7800$\,K, $\log g = 4.06$ (cgs), ${\rm [Fe/H]} = -0.08$. Using the Infrared Flux Method (IRFM), Smalley (private communication) found $T_{\rm eff} = 7900 \pm 200$\,K. The {\it TESS} Input catalogue (TIC) gives $T_{\rm eff} = 8150$\,K, $\log g = 4.13$. There are two published spectral types for HD\,60435. One is ApSr(Eu) from the Michigan Spectral Catalog 
\citep{1975mcts.book.....H}, which does not give a spectral subtype from which $T_{\rm eff}$ could be inferred. This is not unusual for Ap stars, as their abnormal line strengths confound giving a temperature-related subclass. The other spectral classification is A3 from the original Henry Draper (HD) Catalog from over a century ago with an updated reference of \citet{1993yCat.3135....0C}. That would imply $T_{\rm eff} \sim 8700$\,K, but we give that low weight as a consequence of the low resolution of the HD catalog. The other values for $T_{\rm eff}$ are consistent within the uncertainties. We adopt $T_{\rm eff} = 7900\pm200$\,K based on the result from the IRFM.  The Gaia and TIC surface gravities are consistent.

To infer the luminosity, equatorial rotational velocity, and the inclination of the rotation axis we ran a Monte Carlo simulation with $10^{6}$ draws. The uncertainties on each parameter were calculated as percentiles 15.86 and 84.14 of the Monte Carlo simulations. From Gaia DR3 we adopted the parallax and its uncertainty, $\pi = 4.077\pm0.014$\,mas, mean apparent $g$ magnitude $g=8.87\pm0.01$\,mag with an assumed uncertainty. The interstellar extinction in the Gaia g band may be as large as $A_g=0.14$ as derived by the Gaia DR3 pipeline for hot stars \citep{2023A&A...674A..28F}. With that extinction the absolute G magnitude was calculated as
\begin{eqnarray}
G = g -5\left[\log_{10} \left( \frac{1000}{\pi} \right)-1\right]-a_g \, .
\end{eqnarray}
We used a bolometric correction of zero (which is appropriate for $T_{\rm eff} = 7900$\,K) and calculated the luminosity,
\begin{eqnarray}
L/{\rm L}_{\odot}  = 10^{-(G - {\rm M}_{{\rm bol,}\odot})/2.5} \, ,
\end{eqnarray}
where ${\rm M}_{{\rm bol,}\odot} = 4.74$ is the bolometric luminosity of the Sun. We thereby arrived at a stellar luminosity of $L = 15.25\pm0.21$\,L$_{\odot}$. Taking the solar $T_{\rm eff} = 5772$\,K, we used our luminosity and adopted $T_{\rm eff}$ to calculate a radius, $R=2.08\pm0.10$\,R$_{\odot}$. With our measured rotation period of $P_{\rm rot} = 7.679696\pm0.000005$\,d, this allowed us to calculate the equatorial rotation velocity, assuming zero flattening, via
\begin{eqnarray}
v_{\rm eq} = \frac{2 \uppi R}{P_{\rm rot}}
\end{eqnarray}
to be $v_{\rm eq} = 13.8\pm0.7$\,km\,s$^{-1}$. We will use this to calculate the stellar inclination angle in section\,\ref{aopm}.

\subsection{Spectroscopic monitoring}
\label{spec}

We conducted spectroscopic observations of HD 60435 using the \textit{Veloce} spectrograph at the Anglo-Australian Telescope, using the \textit{Verde} arm (4420 -- 5960\,\AA), which is an extension to the established \textit{Veloce Rosso} \citep{2018SPIE10702E..0YG} instrument. \textit{Veloce} is a high-resolution ($R = 80\,000$) \'echelle spectrograph using a 19-fibre Integral Field Unit (IFU) that is rearranged into a pseudo-slit. We conducted the observations without simultaneous calibration and performed a simple data extraction that co-adds flux from all object fibres.

HD 60435 was observed five times during three nights in 2023 December (JD~2460304.12 -- 2460306.13) and six times on 2024 May 23 (JD~2460435.86 -- 2460435.94). Table \ref{tab:aat_log_lsd} provides a log of these observations. The spectra were normalised on an order-by-order basis using iSpec  \citep{2014A&A...569A.111B, 2019MNRAS.486.2075B}. After heliocentric correction, we derived stellar line profiles from each \'echelle order via a least-squares deconvolution (LSD) analysis \citep{Donati_1997} using LSDpy package\footnote{\url{https://github.com/folsomcp/LSDpy}} against a $\delta$-function atomic line mask derived from the Vienna Atomic Line Database (VALD, \citealt{VALD}). A master profile was created for each spectrum by taking an average of LSD profiles weighted by flux uncertainties and omitting regions with strong lines (Na\,{\sc i} D lines, H$_\beta$, and H$_\gamma$ red wing). Resulting profiles were fitted with a profile created as a convolution of intrinsic profile ($T_{\rm eff}$  and micro-turbulent velocity), instrumental profile, and broadening kernel, yielding a measurement of radial velocity and projected rotational velocity, $v \sin i$. For spectroscopic analysis, we assumed atmospheric parameters $T_{\rm eff}$, $\log g$, and $[\rm M/\rm H]$ from {\it Gaia} DR3 \citep{Gaia_2016, Gaia_2023, Creevey_2023}.

Individual results of radial velocity and $v \sin i$ are presented in Table \ref{tab:aat_log_lsd}. There was no radial velocity value in the {\it Gaia} archive, but our values, measured 0.4-yr apart, are consistent at 19.22\,km\,s$^{-1}$. Combining measurements from 11 independent observations, we measured $v \sin i = 11.0 \pm0.6$ km\,s$^{-1}$, with the uncertainties conservatively drawn from MCMC samples added in quadrature with a systematic noise floor estimated to be $0.5\,\rm km\,s^{-1}$ following \citet{Houdebine_2011}. We use that measured $v \sin i$  to constrain the rotation inclination in section\,\ref{aopm} below. We also find no indication of binarity in our 0.4-yr of radial velocity measurements. This is unsurprising, as the incidence of binarity in Ap stars on a time scale shorter than that is low \citep{2012A&A...545A..38S}.

\begin{table}
\caption{Log of observations from Veloce at AAT. Uncertainties are $1 \sigma$ level values drawn from MCMC samples.}
        \centering
    \begin{tabular}{cccc}
        \hline
   date &     JD & radial velocity & $v \sin i$ \\
  &      $-2400000.0$ & km\,s$^{-1}$ & km\,s$^{-1}$ \\
        \hline
$2023-12-25$ &      60304.11552 & $19.22\pm0.09$ & $10.97 \pm 0.30$ \\
 $2023-12-25$ &     60304.13029 & $19.22\pm0.09$ & $10.96 \pm 0.29$ \\
 $2023-12-26$ &     60305.14378 & $19.22\pm0.10$ & $10.98 \pm 0.29$ \\ 
 $2023-12-26$&     60305.24039 & $19.22\pm0.11$ & $10.97 \pm 0.29$ \\ 
 $2023-12-27$ &     60306.12876 & $19.22\pm0.10$ & $10.97 \pm 0.29$ \\ 
 $2024-05-23$&     60453.85750 & $19.22\pm0.08$ & $10.94 \pm 0.30$ \\
 $2024-05-23$&     60453.87216 & $19.22\pm0.09$ & $10.93 \pm 0.30$ \\
 $2024-05-23$&     60453.89224 & $19.22\pm0.08$ & $10.95 \pm 0.29$ \\
 $2024-05-23$&     60453.90690 & $19.22\pm0.09$ & $10.94 \pm 0.29$ \\ 
 $2024-05-23$&     60453.92210 & $19.22\pm0.09$ & $10.94 \pm 0.29$ \\ 
 $2024-05-23$&     60453.93676 & $19.22\pm0.09$ & $10.95 \pm 0.29$ \\
        \hline
    \end{tabular}
    \label{tab:aat_log_lsd}
\end{table}

\section{Application of the oblique pulsator model to two S7 multiplets}
\label{aopm}

\begin{figure}
\begin{center}
\includegraphics[width=1.00\linewidth,angle=0]{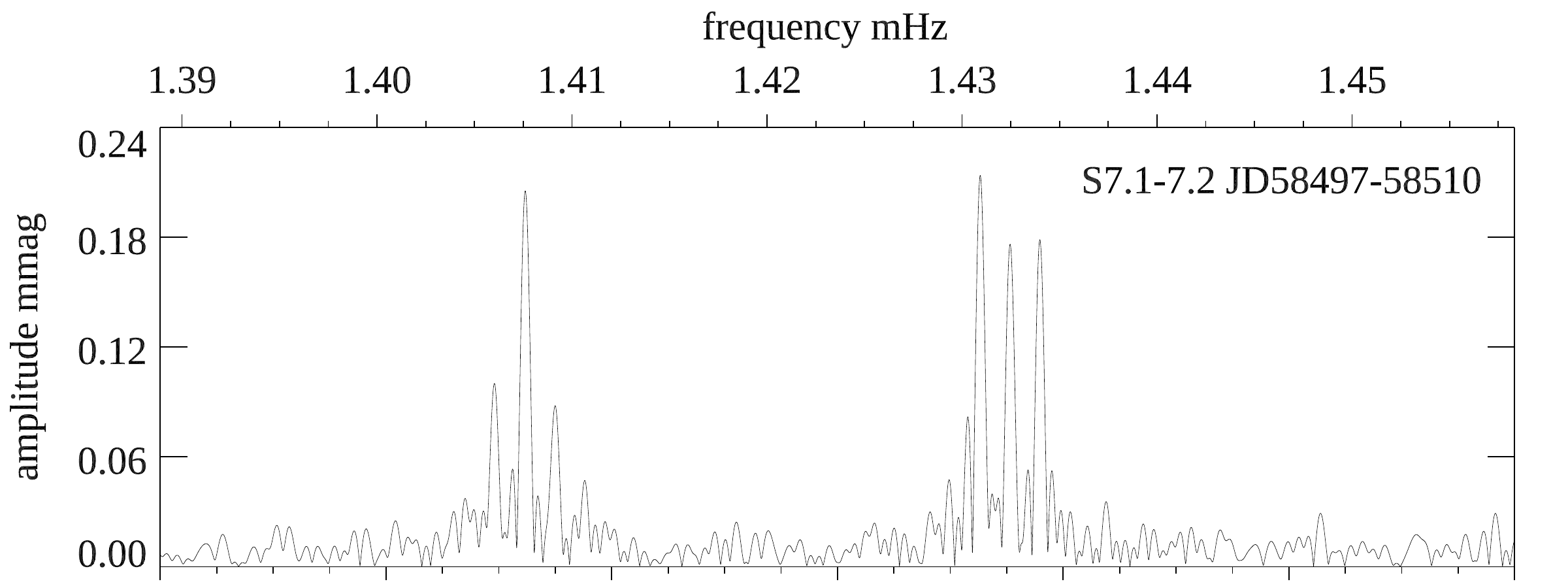}\\
\includegraphics[width=1.00\linewidth,angle=0]{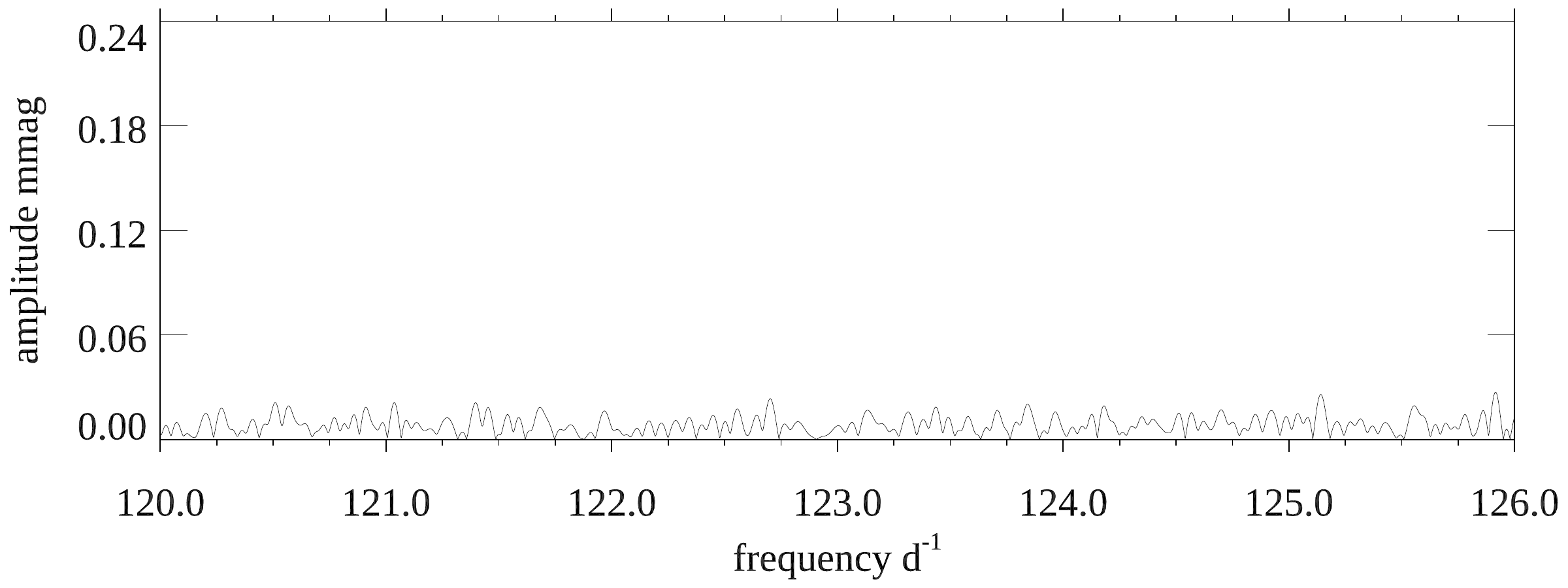}
\caption{Top: Amplitude spectrum for the S7.1 and S7.2 data over a time-span of 24.45\,d showing two multiplets, the one on the left for a quadrupole mode, the one on the right for a dipole mode. Bottom: Amplitude spectrum of the residuals after pre-whitening by a quintuplet and a triplet exactly split by the rotation frequency, as given in Table\,\ref{table:4}. These frequencies are stable during this time-span. There is no significant signal left. The highest peaks in the residuals are 27\,$\upmu$mag. }
\label{fig:as6} 
\end{center}
\end{figure}

Some of the half-sectors seen in Figs\,\ref{fig:as} and \ref{fig:as2} show stable pulsation amplitude and rotationally split multiplets from oblique pulsation. The best example of that is for the S7.1 JD58497.34 and S7.2 JD58510.40 data sets where there are two multiplets with no amplitude modulation -- other than the oblique pulsator rotational modulation -- during the time span of the full sector. Fig.\,\ref{fig:as6} shows an oblique pulsation dipole mode triplet for $\nu_7$ and a distorted quadrupole mode quintuplet for $\nu_6$.

Table\,\ref{table:4} gives the values for the frequencies of these multiplets where the splitting has been chosen to be exactly the rotation frequency. The phases of the dipole $\nu_7$  rotational sidelobes were set equal at $t_0 = {\rm BJD}\,2458500.53070$, which is very close to rotational (spot) maximum, hence, by inference, magnetic maximum. This is typical of oblique pulsators. Note that the phase of the central peak is close to that of the sidelobes, indicating that the mode is nearly a pure dipole. Note also that the phases of the $\nu_6$ quintuplet components are also nearly equal, showing that this mode also has maximum amplitude at rotational light maximum, as does the dipole mode.

\begin{figure}
\begin{center}
\includegraphics[width=1.0\linewidth,angle=0]{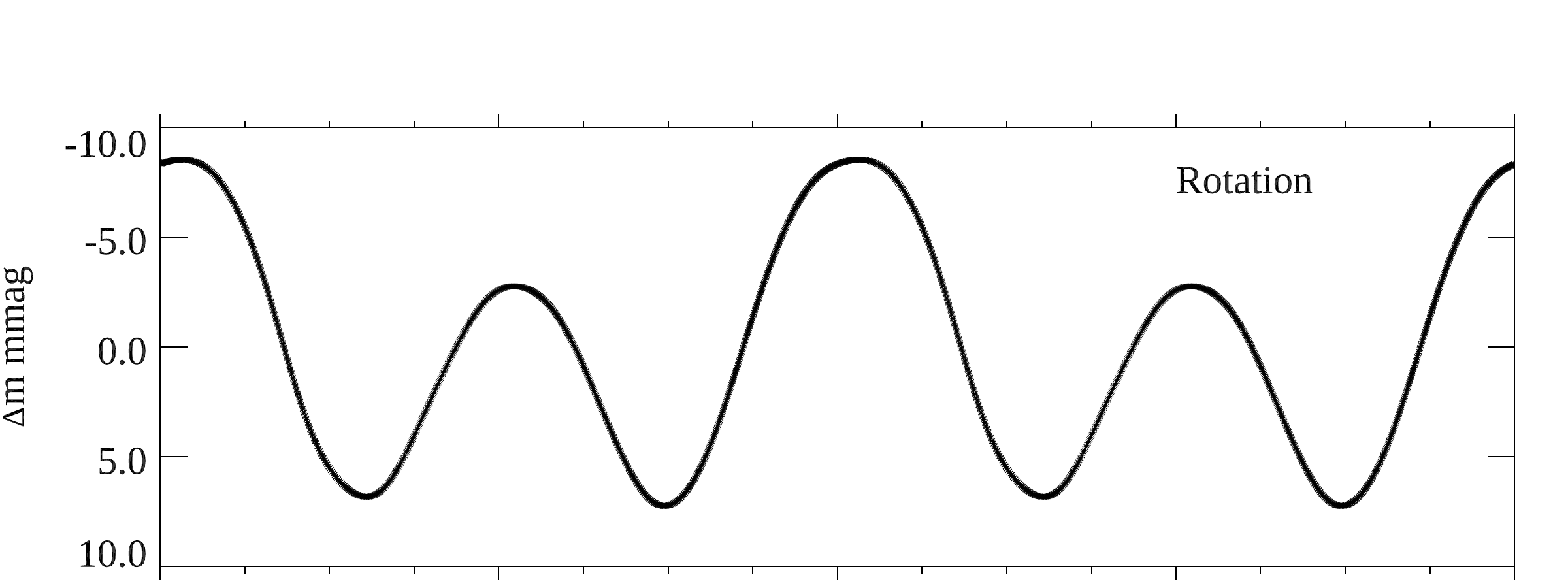}\\
\vspace{-0.6cm}
\includegraphics[width=1.0\linewidth,angle=0]{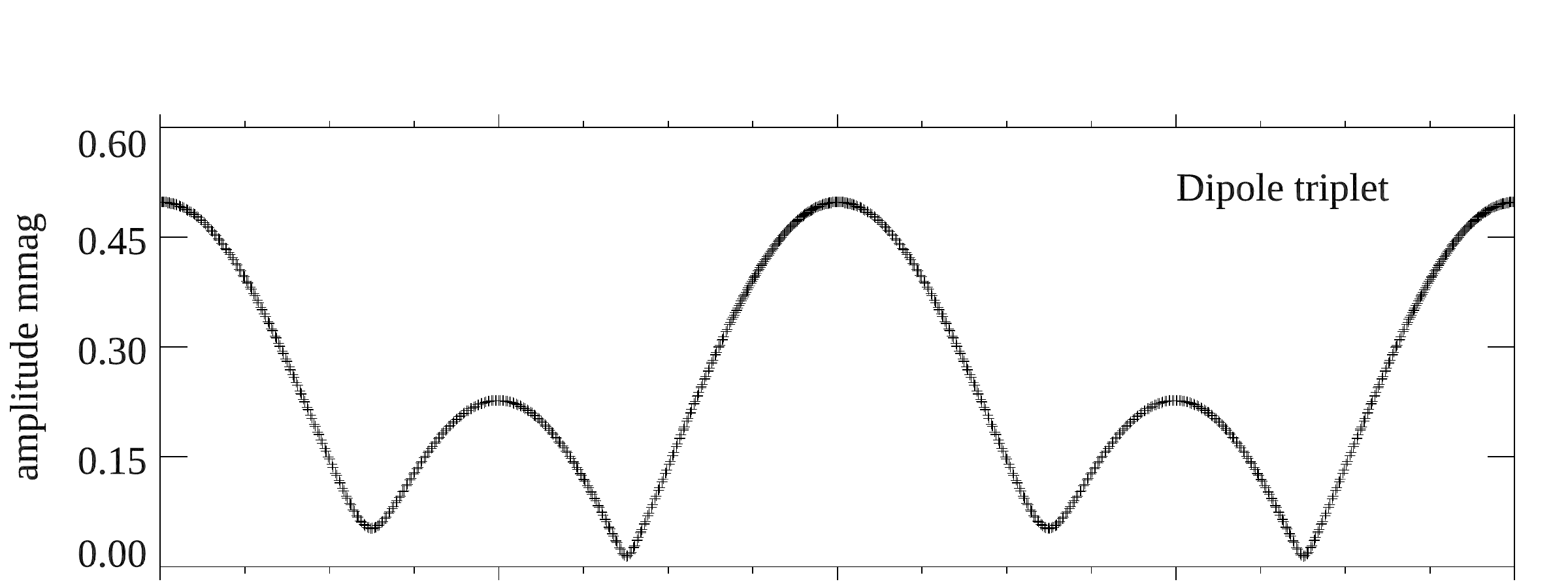}\\
\vspace{-0.6cm}
\includegraphics[width=1.0\linewidth,angle=0]{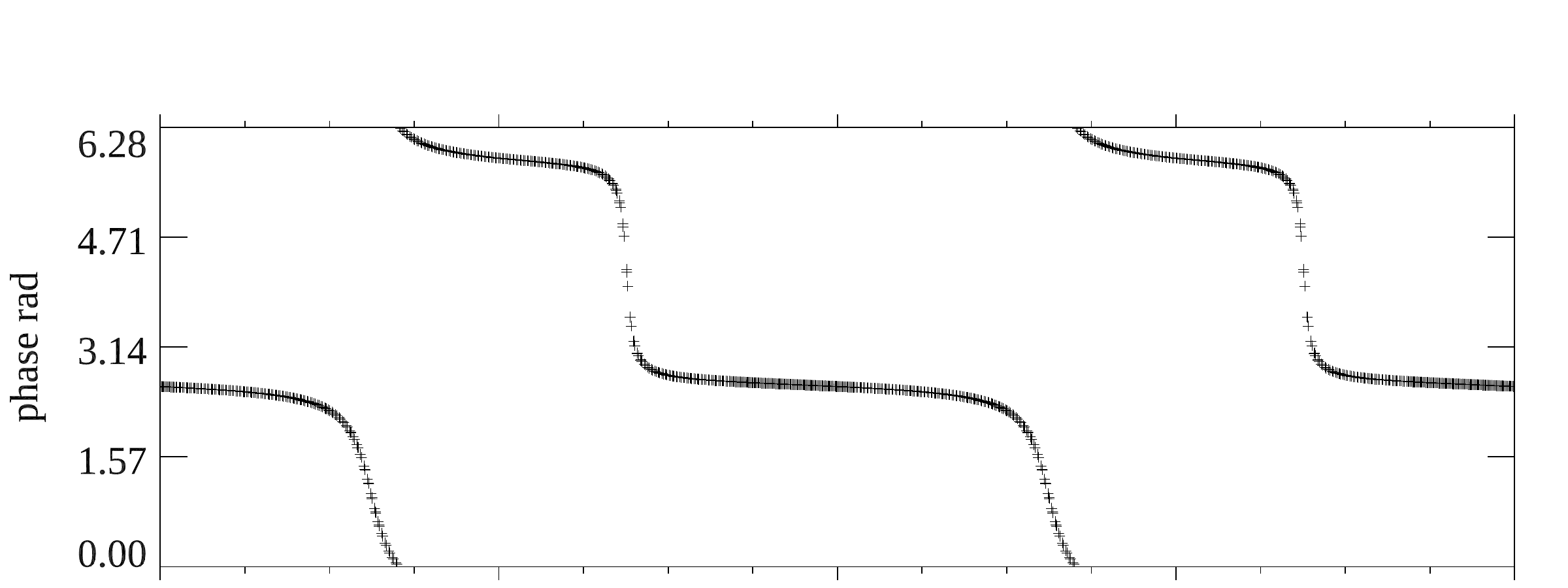}\\
\vspace{-0.6cm}
\includegraphics[width=1.0\linewidth,angle=0]{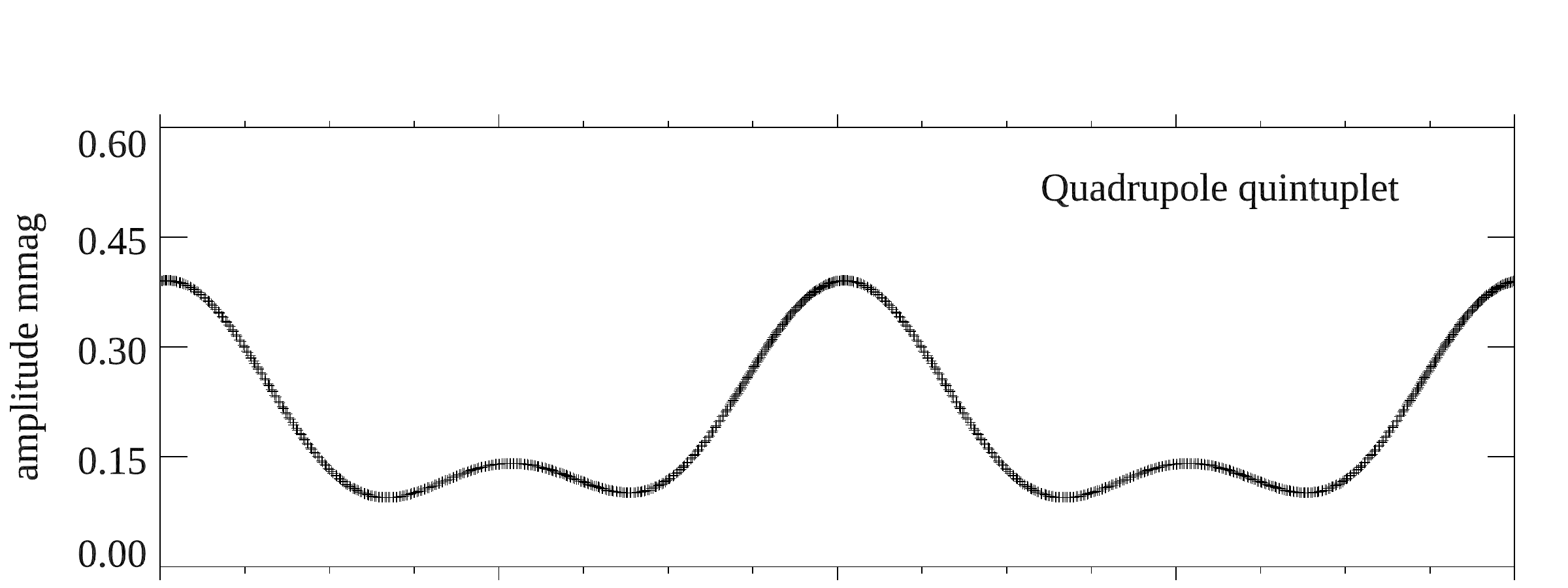}\\
\includegraphics[width=1.0\linewidth,angle=0]{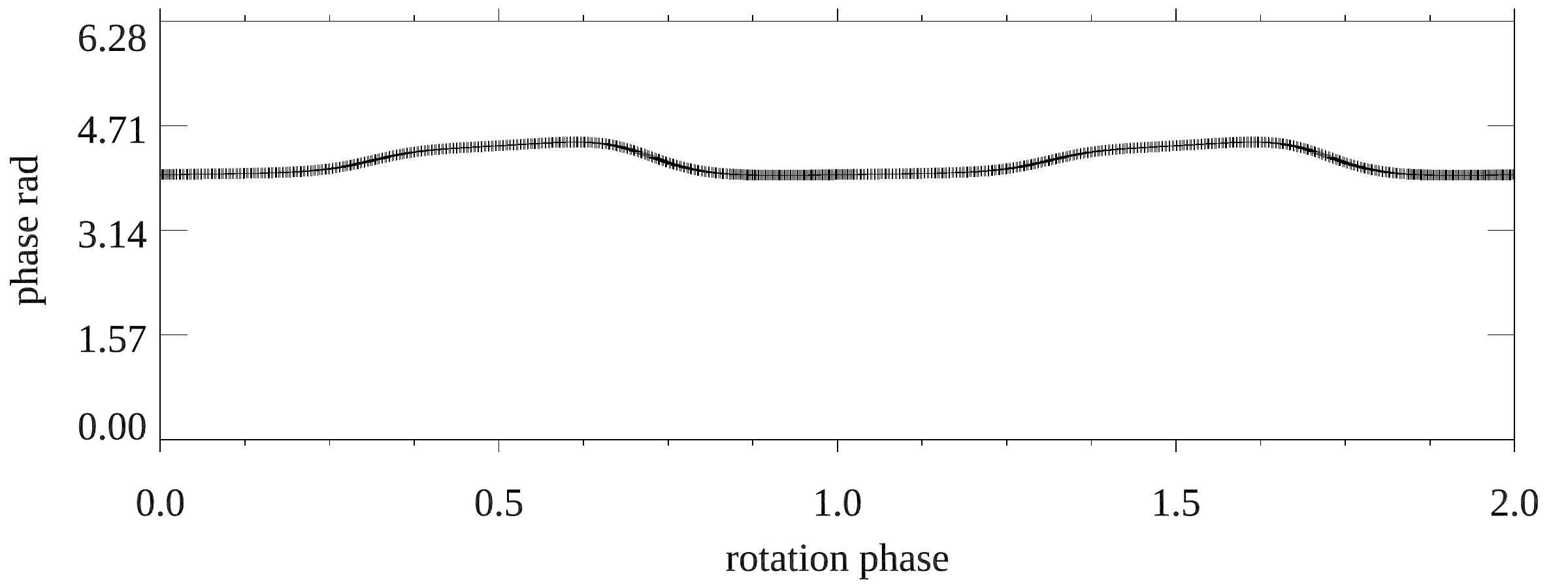}\\
\caption{Top panel: The rotation curve. $P_{\rm rot} = 7.679696$\,d; $t_0 = {\rm BJD}\,2458500.53070$. Two rotation cycles are shown for visibility across rotation phase zero. Second panel: The pulsation amplitude of the dipole triplet for the S7.1-7.2 JD58497-58510 data. Third panel: The pulsation phase for the dipole triplet. Fourth panel: The pulsation amplitude of the quadrupole quintuplet for the S7.1-7.2 JD58497-58510 data. Bottom panel: The pulsation phase for the quadrupole quintuplet; this has been shifted down by 2\,rad for visibility because it wraps around at $2 \uppi$\,rad. The phase is nearly constant.}
\label{fig:rotation} 
\end{center}
\end{figure}

Fig.\,\ref{fig:rotation} shows the rotational light curve and the pulsation amplitudes and pulsation phases as a function of rotation phase for the dipole triplet and quadrupole quintuplet. As explained in the caption to Fig.\,\ref{fig:rotation1} above, the rotational curve was generated from a 10-harmonic fit of the rotation frequency to the entire data set, but then sampled at the same times as the S7 observations. Similarly, the pulsation amplitudes and phases were derived from data generated at the times of observations using the frequencies, amplitudes and phases from Table\,\ref{table:4}. Thus, the noise has been removed from these data for better clarity in the figure. 

The rotation curve is slightly non-sinusoidal, which is described by the significant harmonics of the rotation frequency that were input to generate the curve. In particular, the principal maximum at rotation phase zero (or 1 or 2, since the curve is shown through two rotations) is broadened and slightly offset from the pulsation mode maxima. This shows that the abundance spots that cause the rotational variation are spread somewhat away from the pulsation poles, which are themselves most likely very near the magnetic poles. It is  not uncommon  for spots in magnetic Ap stars ($\alpha^2$\,CVn stars) to have a rough symmetry about the magnetic poles, but to be more complex than that. 

The rotational variations show that spots at both magnetic poles are seen, as was deduced by \citet{1987ApJ...313..782M} in their original study of this star. The dipole amplitude and phase variations are representative of a nearly pure dipole mode. The larger and smaller maxima show that neither the rotational inclination, $i$, nor the obliquity of the pulsation axis, $\beta$, is $90^\circ$. If either of those were $90^\circ$, we would see equal pulsation maxima at half-rotation intervals. Furthermore, the pulsation amplitude goes nearly to zero at rotation quadrature and the pulsation phase reverses by $\uppi$\,rad, as expected for a pure dipole pulsation. We can, therefore, put a constraint on $i$ and $\beta$.

For a dipole oblique pulsation \citep{1982MNRAS.200..807K} and using the values in Table\,\ref{table:4}:
\begin{equation}
\frac{A_{-1}^{(1)} + A_{+1}^{(1)}}{A_0^{(1)}} = \tan i \tan \beta =2.64 \pm 0.15 \, .
\label{eqn:opmdipole}
\end{equation}
\noindent In section\,\ref{fund} above we found $v \sin i = 11.0\pm0.6$ km\,s$^{-1}$ observationally and we calculated that $v_{\rm eq} = 13.8 \pm 0.7$\,km\,s$^{-1}$ from values of $R$ and $P_{\rm rot}$. We thus determine that $\sin i = 0.80 \pm 0.06$ the rotational inclination is $i = 53^{\circ} \pm 6^{\circ}$. With $i = 53^\circ$ the constraint of equation\,\ref{eqn:opmdipole} then gives  $\beta = 63^\circ \pm 9^\circ$. These values are important inputs to our models in section\,\ref{HSmodels}.

The sum $i + \beta = 116^\circ > 90^\circ$, showing that we see both pulsation poles over the rotational cycle. This is consistent with the double-wave rotational light curve seen in the top panel of Fig.\,\ref{fig:rotation}. The pulsation maxima for the dipole mode occur when the pulsation pole comes  closest to the line of sight. For the most favourable view at rotation phase zero the pulsation pole is $|i - \beta| = \alpha = 10^\circ$ from the line of sight. The secondary pulsation maximum occurs when $|i + \beta -180| = \alpha = 64^\circ$. For a dipole mode the observed amplitude is proportional to $\cos \alpha$ times the intrinsic amplitude.  We therefore expect for a pure dipole pulsation that the two pulsation maxima should have an amplitude ratio of $\cos 10^\circ/\cos 64^\circ = 2.2$, very close to what is seen in the second panel of Fig.\,\ref{fig:rotation}.

The quintuplet variations are more complex; this mode is highly distorted. It has a main maximum at rotation phase zero, as does the dipole. With the angle between the pulsation pole and the line of sight being $\alpha = 10^\circ$ at this rotation phase, the quadrupole pulsation is dominated by the polar cap, but the equatorial band has some visibility, given that the nodal lines for a zonal quadrupole ($\ell =2, m=0$) mode are at co-latitude $54.7^\circ$ from the pulsation pole. At rotation phase 0.5 for the other quadrupole maximum the equatorial belt is much more prominent. In any case, the nearly constant pulsation phase seen for this mode over the rotation cycle in the bottom panel of  Fig.\,\ref{fig:rotation} indicates extreme distortion from a normal quadrupole mode, with the equatorial belt contributing little to the observed pulsation amplitude. This is seen better in the models below in section\,\ref{HSmodels}.

\begin{table}
\centering
\caption{A least squares fit to the S7.1 and S7.2 data of the frequency quintuplet for the distorted quadrupole mode (top five frequencies), and the frequency triplet for the dipole mode (bottom three frequencies) as shown in Fig.\,\ref{fig:as6} where the rotational sidelobes have been forced to be equally split from the central  mode frequency by exactly the rotation frequency, $\nu_{\rm rot} = 0.13021349$\,d$^{-1}$ ($P_{\rm rot} = 7.679696$\,d).  The zero point for the phases, $t_0 = {\rm BJD}\,2458500.53070$, was chosen to be a time when the phases of the two rotational sidelobes of the dipole triplet are equal. Note that the phase of the central frequency of the dipole, which is the mode frequency, is also equal (within the uncertainty) at this time, showing that the mode is essentially a pure dipole mode. We note that the phases for the quintuplet are all nearly equal at this $t_0$, also. This shows that the quadrupole and dipole reach maximum observed amplitude at the same time, and, as we show in section 4 this is close to the time of maximum rotational light, hence the pulsation poles are close to the spots, therefore most probably the magnetic poles. } 
\begin{tabular}{cccr}
\hline
\multicolumn{2}{c}{frequency} & \multicolumn{1}{c}{amplitude} &   
\multicolumn{1}{c}{phase}  \\
\multicolumn{1}{c}{d$^{-1}$} &\multicolumn{1}{c}{mHz} & \multicolumn{1}{c}{mmag} &   
\multicolumn{1}{c}{radians}   \\
& & \multicolumn{1}{c}{$\pm 0.007$} &   
   \\
\hline
\hline
 $ 121.3565 $ & $  1.4046 $ & $  0.031  $ & $  -0.056 \pm  0.227 $ \\
 $ 121.4868 $ & $  1.4061 $ & $  0.069  $ & $  -0.467 \pm  0.101 $ \\ 
 $ 121.6170 $ & $  1.4076 $ & $  0.193  $ & $  -0.187 \pm  0.036 $ \\ 
 $ 121.7472 $ & $  1.4091 $ & $  0.066  $ & $  -0.593 \pm  0.106 $ \\ 
 $ 121.8774 $ & $  1.4106 $ & $  0.037  $ & $  -0.340 \pm  0.189 $ \\ 
\hline
 $ 123.6345 $ & $  1.4310 $ & $  0.199  $ & $  2.615 \pm  0.035 $ \\
 $ 123.7647 $ & $  1.4325 $ & $  0.137  $ & $  2.476 \pm  0.051 $ \\ 
 $ 123.8949 $ & $  1.4340 $ & $  0.163  $ & $  2.615 \pm  0.043 $ \\ 
\hline
\hline
\end{tabular}
\label{table:4}
\end{table}

The amplitude spectra for the half-sectors S13.1 JD58660.80 and S13.2 JD58675.49 also show two triplets ($\nu_3$ and $\nu_7$) and one quintuplet ($\nu_6$). However, those do not completely pre-whiten, so there is some secular amplitude modulation. The quintuplet amplitudes are possibly different to those for S7.1 JD58497.34 and S7.2 JD58510.40 above, so the above discussion of the oblique pulsator model applied to the quadrupole mode may be over-interpreted. Also, the triplet amplitudes for $\nu_3$ give a ratio from equation\,\ref{eqn:opmdipole} that differs from that of $\nu_7$ presented about by $2.8\sigma$, another note of caution. The rapid amplitude changes in the modes of this star confound a more precise result. 

\section{Mode identification from stellar modelling}
\label{sec:modeID_models}

The ambiguity in the identification of the degree of the even modes can be addressed by looking at frequency combinations that suppress the dominant large-scale structure in the amplitude spectrum (hereafter, small separations). One such combination that includes dipole and quadrupole modes is \citep{2007A&ARv..14..217C},
\begin{equation}
\delta^1_{12}\left(\nu\right)=2\nu_{n,1}-\nu_{n-1,2}-\nu_{n,2} \, , 
\label{eq:small12}
\end{equation}
or, for dipole and radial modes,
\begin{equation}
\delta^1_{10}\left(\nu\right)=2\nu_{n,1}-\nu_{n,0}-\nu_{n+1,0} \, .
\label{eq:small10}
\end{equation}
High radial order $\ell=2$ and $\ell=0$ modes are expected to have different sensitivities to the innermost layers. Therefore, by comparing the observed small separations with those computed from model frequencies of consecutive $\ell=0,1,0$ and $\ell=2,1,2$ modes, we attempt to identify the degree of the modes observed. 

To explore this avenue for mode identification we consider three different models for HD\,60435, with properties listed in Table~\ref{tab:models_mc}. The models were selected from a set of best fits identified from grid-based modelling following the procedure and model physics described in~\cite{2021A&A...650A.125D}. Properly accounting for chemical transport in models of Ap stars is a challenge. In the upper atmosphere the strong magnetic field influences atomic diffusion, with the transport of different elements being impacted differently \citep{2002A&A...387..271A}. Moreover, if working alone, in the interior atomic diffusion would lead to a complete depletion of the photospheric helium and metals \citep{2000ApJ...529..338R}. Turbulent mixing, evoked to avoid this depletion when modelling stars more massive than the Sun, is not properly calibrated for Ap stars, whose subphotospheric layers are likely to be significantly affected by the presence of the strong magnetic field.

Acknowledging these limitations to the state-of-the-art, the grid of models constructed by~\cite{2021A&A...650A.125D}, and used here, does not include chemical transport. Since the correspondence between the current surface composition and the initial chemical composition is not possible to achieve based on this grid, no constraints are imposed on the surface metallicity for the identification of the best models. Instead, we select best models with a spread of metallicities. These are derived from exploring the grid, which has a parameter space that includes a significant range of helium and metal mass fractions, without imposing a particular enrichment law~\cite[see][for details]{2021A&A...650A.125D}. 

The constraints imposed on the forward modelling were: $T_{\rm eff} = 7900\pm 200$~K, $L/{\rm {\rm L}_\odot}=14.3\pm 0.7$, and $\Delta\nu\in  [51,58]\,\upmu$Hz. From the set of best models retrieved, we chose three representative models covering a range of metallicities. For each of these models we computed the small separations defined by equations~(\ref{eq:small12})--(\ref{eq:small10}). To account for slight differences in the mean density of the models, we further constructed the ratios of small to large frequency separations~\citep{2003A&A...411..215R} defined as
\begin{equation}
r_{12}=\frac{\delta^1_{12}}{\nu_{n,2}-\nu_{n-1,2}} \, ,
\label{eq:r010}
\end{equation}
and
\begin{equation}
r_{10}=\frac{\delta^1_{10}}{\nu_{n+1,0}-\nu_{n,0}} \, .
\label{eq:r010}
\end{equation}
These model ratios are to be compared with the observed ratios, $r_{\rm obs}$, calculated assuming that $\nu_3$, $\nu_5$, and $\nu_7$ are consecutive dipole modes and that $\nu_1$, $\nu_4$, $\nu_6$, and $\nu_8$ are consecutive modes of even degree (note that the observed ratios do not depend on whether the latter are radial or quadrupole modes).

\begin{figure}
\begin{center}
\includegraphics[width=1.\linewidth,angle=0]{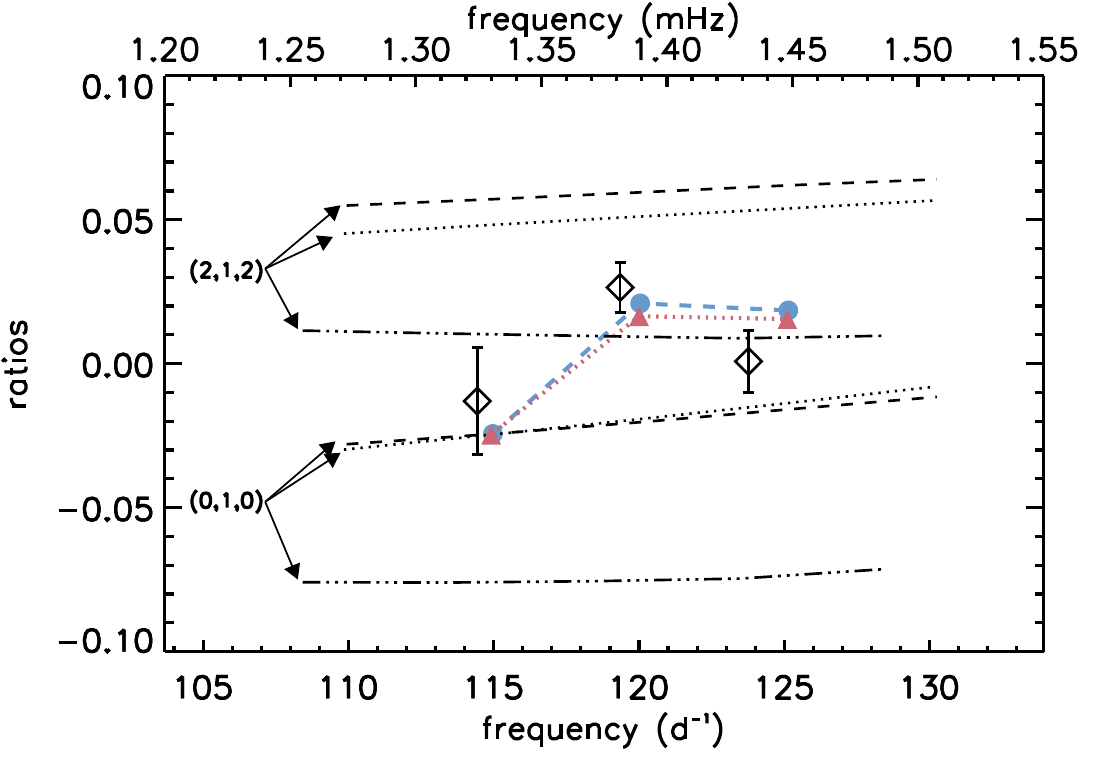}\\
\caption{Ratios of small to large frequency separations constructed from the frequencies in Table~\ref{table:3} (open diamonds). Also shown are the ratios computed for the three models listed in Table~\ref{tab:models_mc} in the region of the observed frequencies, considering a sequence of alternating even and odd degree modes: model $a$, high metallicity (black dashed lines);  model $b$, intermediate metallicity (black dotted lines);  model $c$, low metallicity (black dashed-dotted lines). For each model, we show two options for the even modes identified by the arrows: alternating $\ell=0$ and $\ell=1$ modes (0,1,0); alternating $\ell=2$ and $\ell=1$ modes (2,1,2). With filled symbols connected by lines we also identify the ratios obtained by taking the odd modes to be of $\ell=1$, $\nu_6$ to be of $\ell=2$ and $\nu_1$, $\nu_4$, and $\nu_8$ to be of $\ell=0$, for model $a$ (circles, blue dashed line) and model $b$ (triangles, red dotted line).}
\label{fig:ratios} 
\end{center}
\end{figure}

The results from the data and model comparison can be seen in Fig.~\ref{fig:ratios}, with the observed ratios and respective uncertainties shown by open symbols and the model ratios shown by black lines. For all three models, the ratios obtained when considering combinations involving quadrupole modes, hereafter (2,1,2), are positive, while those obtained for combinations involving radial modes, hereafter (0,1,0), are negative.  Moreover, the magnitude of the ratios depends on metallicity, with lower metallicity models having smaller magnitude (2,1,2) ratios and higher magnitude (0,1,0) ratios. This highlights the potential of ratios to identify the even modes, under the assumption that the odd modes are dipolar, and possibly also to constrain the initial chemical composition of roAp stars.  

In all cases,  Fig.~\ref{fig:ratios} shows that the model ratios computed for sequences of the same alternating even and odd mode degrees vary smoothly with frequency, while the observed ratios apparently do not. However,  they are still consistent with a smooth variation within 2$\sigma$, in which case the best solution would point to a set of alternating dipolar and quadrupolar modes in a star with a low initial metallicity.

This potential difference between the observations and model results may have different origins. On the one hand, the even degree modes may not all be of the same degree. If we take $\nu_6$ to be a quadrupole (cf. section~\ref{aopm}) and relax the assumption that the observed even modes are all of the same degree, the ratios of high-metallicity and intermediate metallicity models (a and b, respectively) that best represent the data are found when $\nu_6$ is a quadrupole mode and all other even degree modes (i.e.\, $\nu_1$, $\nu_4$, and $\nu_8$) are radial modes. This sequence is shown in Fig.\,\ref{fig:ratios} by filled symbols connected by lines in colour for model a (circles, dashed line) and model b (triangles, dotted line) models.

On the other hand, the  stellar interior  may have regions of sharp stratification, not accounted for in our models, that perturb the ratios away from the smooth frequency behaviour observed in Fig.\,\ref{fig:ratios}.  Regions of strong chemical gradients and transitions between  convective and radiative regions are known to induce frequency perturbations that vary cyclically with frequency \cite[see][for a review]{2020ASSP...57..185C}. In fact, \citet{2004A&A...425..179V} argued that chemical transport inside roAp stars could lead to a helium gradient below the convective envelope that, in turn, would result in a non-smooth frequency dependence of particular combinations of the observed frequencies. Depending on the location of these gradients,  the ratios computed here  could also be affected in a similar way \citep{2023A&A...673A..49D}. Hence, it is possible that the even degree  modes are all of the same degree and our models fail to capture the non-smooth dependence of the ratios on frequency because they do not include chemical transport.

\begin{table}
    \caption{Models of HD\,60435 discussed in section~\ref{sec:modeID_models}. Shown, from top to bottom, are: the model mass, effective temperature, luminosity, age, initial metal mass fraction, and initial helium mass fraction.}
    \centering
    \begin{tabular}{cccc}
    
        \hline
	& 	Model a	&	Model b	&	Model c	\\
    \hline							
    $M/{\rm M}_\odot$	&	1.9	&	1.8	&	1.6	\\
    $T_{\rm eff}$ (K)	& 	7809	&	7892	&	7942	\\
    $L/{\rm {\rm L}_\odot}$	& 	14.48	&	14.56	&	14.16	\\
    $Age$ (Myr)	& 	745	&	1072	&	1224	\\
    $Z$	& 	0.0231	&	0.0121	&	0.0071	\\
    $Y$	& 	0.29191	&	0.25192	&	0.27192	\\
    \hline
    \end{tabular}
    \label{tab:models_mc}
\end{table}

\subsection{The nature of $\nu_2$}

In the discussion above we have not considered the nature of $\nu_2$, the mode that falls off the regular frequency spacing in the amplitude  spectrum of HD\,60435 (see Fig.\,\ref{fig:as0}). Anomalies in otherwise regular frequency spacings have been seen in other multiperiodic roAp stars, such as HD\,24712 (HR\,1217), for which \cite{2001MNRAS.325..373C} interpreted the anomaly as possibly resulting from the impact of the strong magnetic field on pulsations via the Lorentz force. While generally strong magnetic fields induce frequency perturbations that vary smoothly with radial order, resulting only in a slight increase of the large frequency separation, occasionally a strong disruption of the expected asymptotic regular spacing does occur, similar to what is seen near avoided crossings in evolved solar-like pulsators  \citep{2000MNRAS.319.1020C,2004MNRAS.350..485S,2006MNRAS.365..153C}. 

In the present case, if $\nu_2$ results from such an anomaly, then we can interpret it as a dipolar mode. A similar anomaly occurring in the sequence of radial or quadrupole modes would then be expected to take place to the left of $\nu_2$  (i.e.\, at lower frequencies) where modes are either not driven or do not reach observable amplitudes, such that the sequences of even and odd modes would remain equally spaced in the observed region. 

An alternative explanation to the nature of $\nu_2$ is that this is an odd mode of higher degree.  Magnetic distortion of the eigenfunctions could give rise to lower degree spherical harmonic components in what would be a higher degree  mode, making modes of higher degree potentially easier to observe than in other stellar pulsators \citep{1996ApJ...458..338D,2000MNRAS.319.1020C,2000A&A...356..218B,2004MNRAS.350..485S}. This could also explain similar apparent anomalies in other roAp stars. 

Inspection of the frequencies of $\ell=3$ modes in the three models listed in Table~\ref{tab:models_mc} show that they are separated from the nearest $\ell=1$ mode by $\sim$$7$ to $9\,\upmu$Hz, depending on the model, thus are smaller than the nominal measured 12$\,\upmu$Hz separating $\nu_2$ from $\nu_3$. Despite this difference, considering that the individual mode frequencies were not used directly as constraints in the model selection, and that the models in Table 6 do not consider the effect of the magnetic field, some caution is called for before rejecting the possibility that $\nu_2$ is an $\ell=3$ mode.

However, the mode of $\nu_2$ presents further challenges to interpretation. It can be seen in Figs\,\ref{fig:amps}, \ref{fig:wav1}, and \ref{fig:wav2} that $\nu_2 = 113.39$\,d$^{-1}$ reaches amplitude maximum at a rotation phase of approximately 0.25, i.e., about $90^\circ$ later in rotation phase to the pulsation maxima of the dipole and quadrupole modes $\nu_7$ and $\nu_6$. To look at this more carefully, we studied the S6.1-8.1 data (see Table\,\ref{table:1b} for the dates and durations of these half-sectors) where $\nu_2$ appears to be semi-stable. 

We see clear evidence of a frequency triplet in the amplitude spectrum, and find a significant quintuplet split by exactly the rotation frequency by least-squares fitting of the frequencies. We note that the mode frequency is not fully stable over the time-span of the data set used, hence there is significant amplitude remaining after pre-whitening by the quintuplet, making the results less certain. The central frequency of the quintuplet, $113.47$\,d$^{-1}$, is the mode frequency. This is a more precise measurement of $\nu_2$ than that given in Table\,\ref{table:3} where it was estimated from the centroid of the broad peak from one half-sector when it had good visibility. We then used the frequency quintuplet to generate a plot of pulsation amplitude and pulsation phase as a function of rotational phase, similar to that seen for $\nu_6$ and $\nu_7$ in Fig.\,\ref{fig:rotation}.

\begin{figure}
\begin{center}
\includegraphics[width=1.0\linewidth,angle=0]{figs/rotation_3.pdf}\\
\includegraphics[width=1.0\linewidth,angle=0]{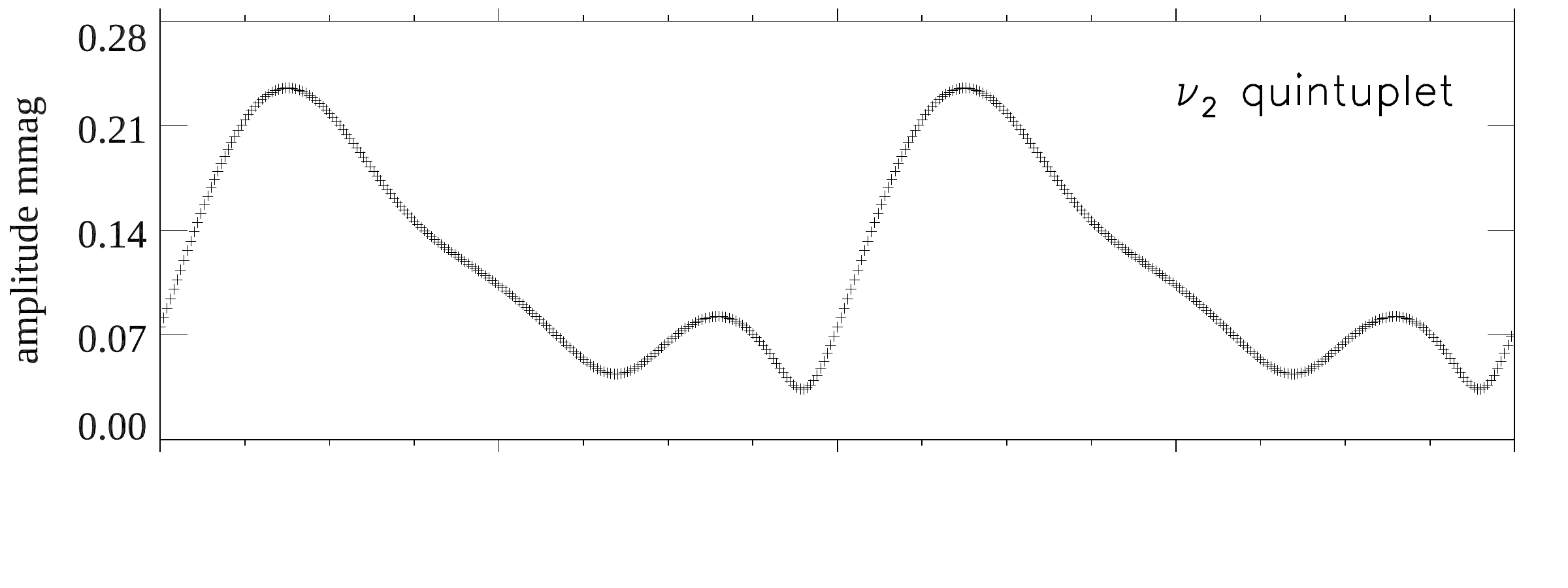}\\
\vspace{-0.7cm}
\includegraphics[width=1.0\linewidth,angle=0]{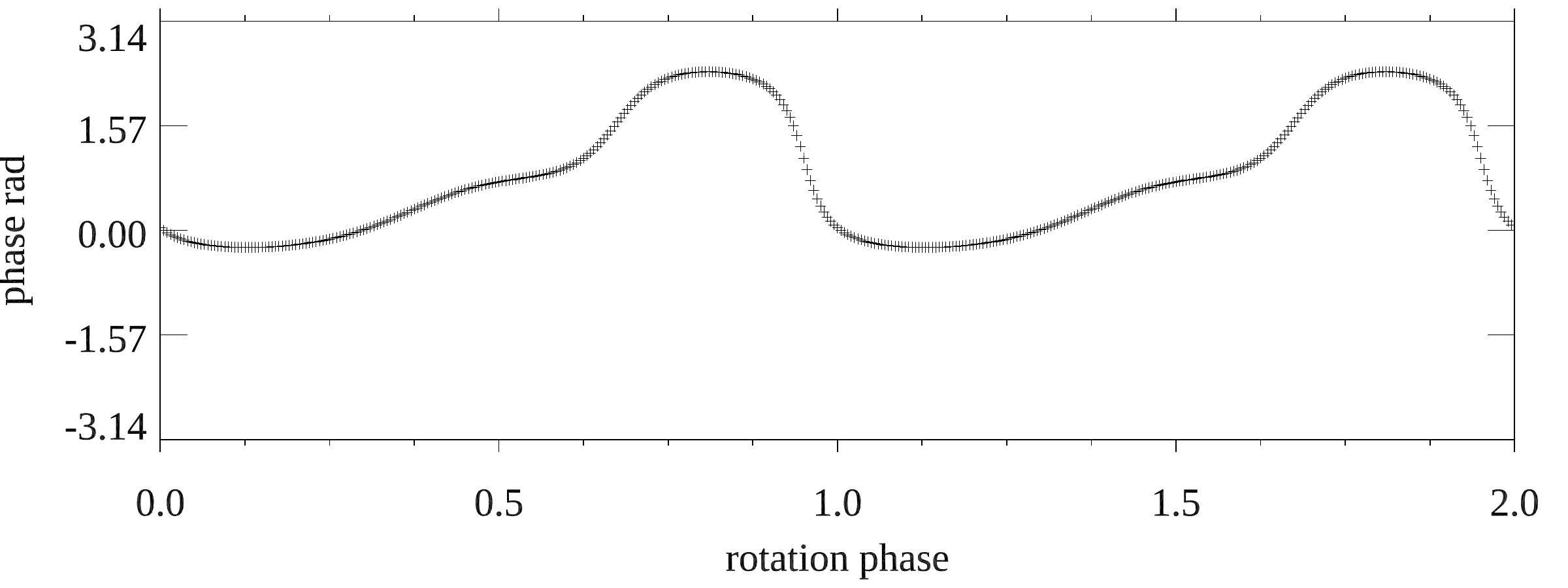}\\
\caption{Top panel: The rotation curve (as in Fig.\,\ref{fig:rotation}) for comparison. Two rotation cycles are shown for visibility across rotation phase zero. Middle and bottom panels: the pulsation amplitude and phase as a function of rotation for the $\nu_2$ mode. The pulsation phase has a zero point shift of $\uppi$\,rad for display purposes.  }
\label{fig:rotation2} 
\end{center}
\end{figure}

The result for $\nu_2$ is shown in Fig.\,\ref{fig:rotation2}, and it is perplexing. The amplitude reaches maximum near rotation phase 0.25 with a second maximum near rotation phase 0.75. The amplitude variation is not inconsistent with an $\ell =3$ octupole mode with some suppression in the equatorial region, but the rotation phase of pulsation maximum in that case would be 0, not 0.25. We note that the range of pulsation phase change is somewhat less than $\uppi$\,rad, but is much less suppressed than for the $\nu_6$ $\ell=2$ quadrupole mode that we show in models in the next section. 

\begin{figure}
\begin{center}
\includegraphics[width=1.0\linewidth,angle=0]{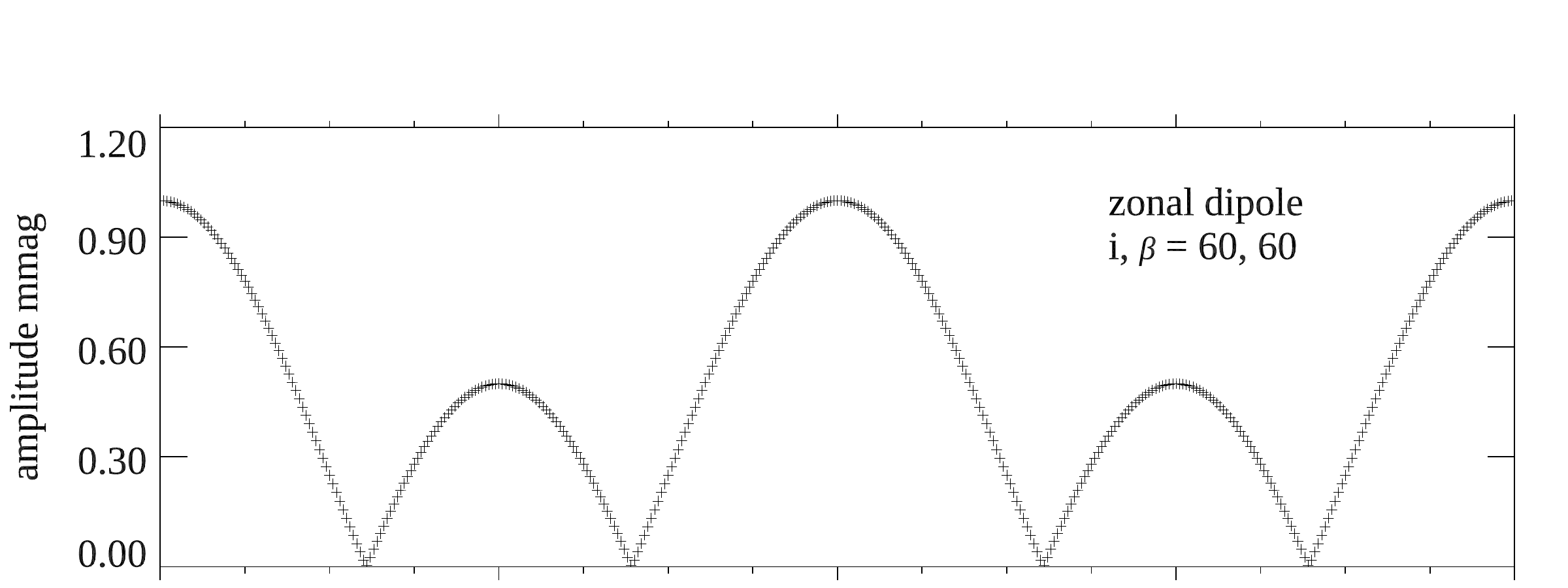}\\
\vspace{0.1cm}
\includegraphics[width=1.0\linewidth,angle=0]{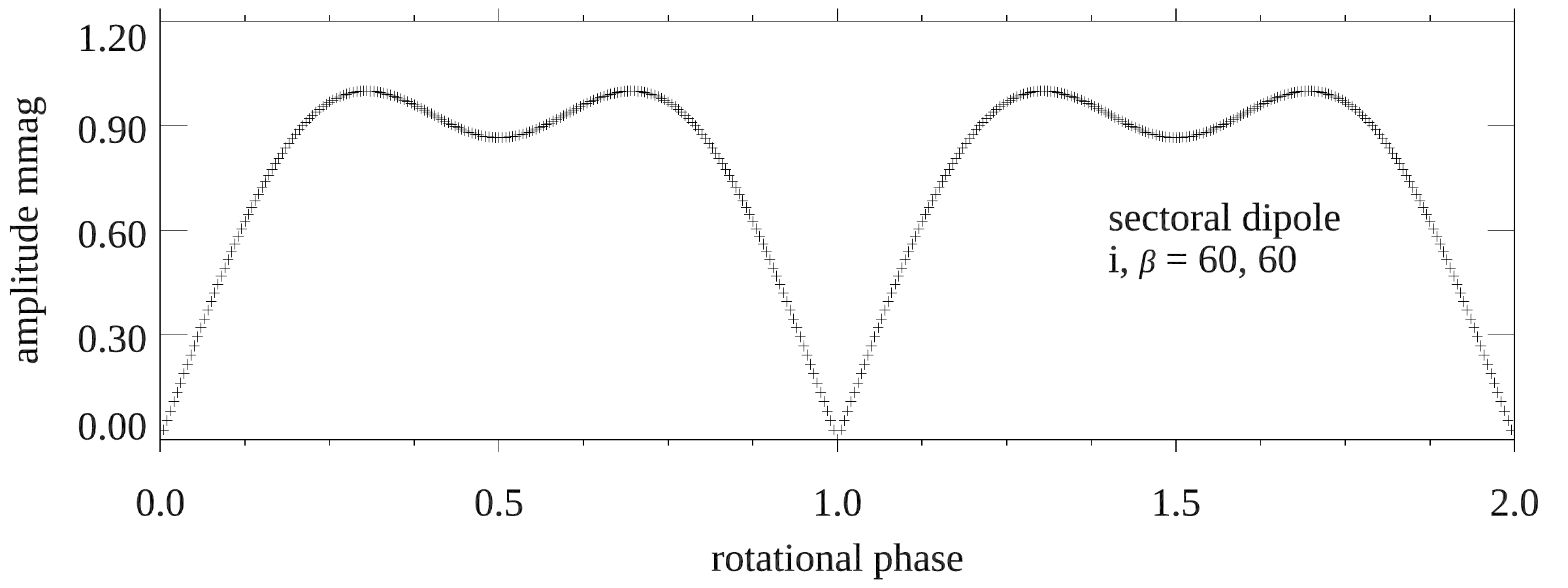}\\
\caption{The pulsation amplitude, normalised to 1.0,  as a function of rotation phase for top: a zonal dipole mode ($\ell = 1, m = 0$); bottom: a sectoral dipole mode, both with $i = 60^\circ$ and $\beta = 60^\circ$. }
\label{fig:rotation3} 
\end{center}
\end{figure}

We have also considered the possibility of a sectoral dipole mode ($\ell = 1, m=|1|$). For the values of $i = 60^\circ $ and $\beta = 60^\circ$ -- very close to the values that we derived in section\,\ref{aopm} for the $\nu_7$ dipole mode -- we calculated the amplitude modulation expected for a zonal dipole mode ($\ell = 1, m=0$) and a sectoral dipole mode ($\ell = 1, m=|1|$). We show these amplitude variations in Fig.\,\ref{fig:rotation3} where it can be seen that the pure zonal dipole mode amplitude variations are remarkably similar to those observed for $\nu_7$ (see Fig.\,\ref{fig:rotation}), supporting the identification of that mode, but the amplitude variations for the sectoral dipole mode are not similar to the observations seen in the middle panel of Fig.\,\ref{fig:rotation2}. While both observations and model show an amplitude maximum at rotation phase 0.25, after that they diverge. We dismiss the possibility that the $\nu_2$ mode could be a sectoral dipole mode with the same pulsation axis as the $\nu_7$ mode. 

The $\nu_2$ mode is reminiscent of those seen in tri-axial tidally tilted pulsators, close binary stars where pulsation modes are excited along three axes, the tidal axis, the orbital/rotation axis, and a third axis orthogonal to both of those \citep{2023arXiv231116248Z}. This is well understood theoretically (Fuller et al., 2024, in press).  The roAp stars may also show modes with different pulsation axes, as a result of the combined effect on pulsations from the magnetic and rotational effects \citep{2002A&A...391..235B}. An example of two modes with different pulsation axes has been reported by \citet{2011MNRAS.414.2550K} for the roAp star KIC\,10195926. That this is rarely observed in roAp stars is possibly the consequence of a selective driving of the mode most closely aligned with the magnetic field \citep{Balmforth01}.

The interpretation of the geometry of the modes in roAp stars is complicated further by the distortion of the modes by the magnetic field, which results in each mode being described by a depth-dependent sum of spherical harmonics of different degree, $\ell$ \citep{1996ApJ...458..338D,2000MNRAS.319.1020C,2000A&A...356..218B,2004MNRAS.350..485S,2018MNRAS.480.1676Q}. This could explain why different spectral lines, being sensitive to different atmospheric depths in roAp atmospheres, appear to show different mode geometry when the oblique pulsator model is applied, as reported by \citet{2006A&A...446.1051K} in a spectroscopic line profile study of the roAp star HR\,3831. By comparing mode geometries derived from the oblique pulsator model for roAp stars observed in Johnson $B$ and in {\it TESS} red filters, \citet{2021MNRAS.506.1073H} and  \citet{2020ASSP...57..313K} also found significant differences in the inferred geometry, depending on the atmospheric depth being sampled through different filters.

Interestingly, in a study of the roAp star HD~12098, \citet{2024MNRAS.529..556K} found that the single, stable pulsation mode in that star seems to be a dipole mode more distorted than seen in any other star. The pulsation amplitude and phase variations with rotation in HD~12098 have some similarity to those shown here for $\nu_2$ in HD\,60435; they are strongly distorted.

Firm conclusions regarding the interpretation of $\nu_2$, as well as the identification of the other observed frequencies based on the ratios, as discussed above, require more extensive modelling including the impact of the magnetic field. Such modelling is beyond the purpose of the current paper and shall be performed in future work.

\section{Models for p-mode pulsations in the presence of a dipole magnetic field}
\label{HSmodels}

Rapid oscillations in roAp stars are axisymmetric high-order p~mode pulsations with a pulsation axis that is closely aligned  to the magnetic axis which is in turn inclined to the rotation axis \citep{1982MNRAS.200..807K,2011A&A...536A..73B}.
Strictly speaking, in the presence of a magnetic field, the eigenfunction of an  axisymmetric nonradial pulsation mode cannot be represented by a single ($m=0$) spherical harmonic $Y_l^0(\theta,\phi)$, where $(\theta,\phi)$ are spherical polar co-ordinates with the axis along to the magnetic axis. For this reason, we represent the displacement vector $\boldsymbol{\xi}$ and a perturbed scalar variable $f'$ of an axisymmetric pulsation mode with a sum of terms associated with $Y_\ell^0(\theta,\phi)$ as
\begin{equation}
\boldsymbol{\xi} = \sum_\ell\left(\xi_{r\ell}Y_\ell^0\boldsymbol{e}_r+\xi_{h\ell}{\partial Y_\ell^0\over\partial\theta}\boldsymbol{e}_\theta\right),
\quad
f' = \sum_\ell f'_\ell Y_\ell^0.
\end{equation}
In the couplings with a dipole magnetic field, we have two mutually independent  series of modes; even modes (symmetric about the magnetic equator) consist of $\ell = 0, 2, 4, \ldots$, and odd modes consist of $\ell = 1, 3, 5, \ldots$. 
We also expand the magnetic perturbation $\boldsymbol{B'}$ similarly to $\boldsymbol{\xi}$.
Usually, we include twelve $\ell$ components for each variable, then we solve
$6\times 12$ differential equations with a complex eigenfrequency for non-adiabatic calculations under the Cowling approximation, where the Eulerian perturbation of gravity is neglected \citep[see][for details]{2004MNRAS.350..485S,Saio05}.  
For the representative latitudinal degree $\ell$ of a mode, we adopt $\ell$ of the component having the maximum kinetic energy. 

\begin{figure*}
\includegraphics[width=0.49\textwidth]{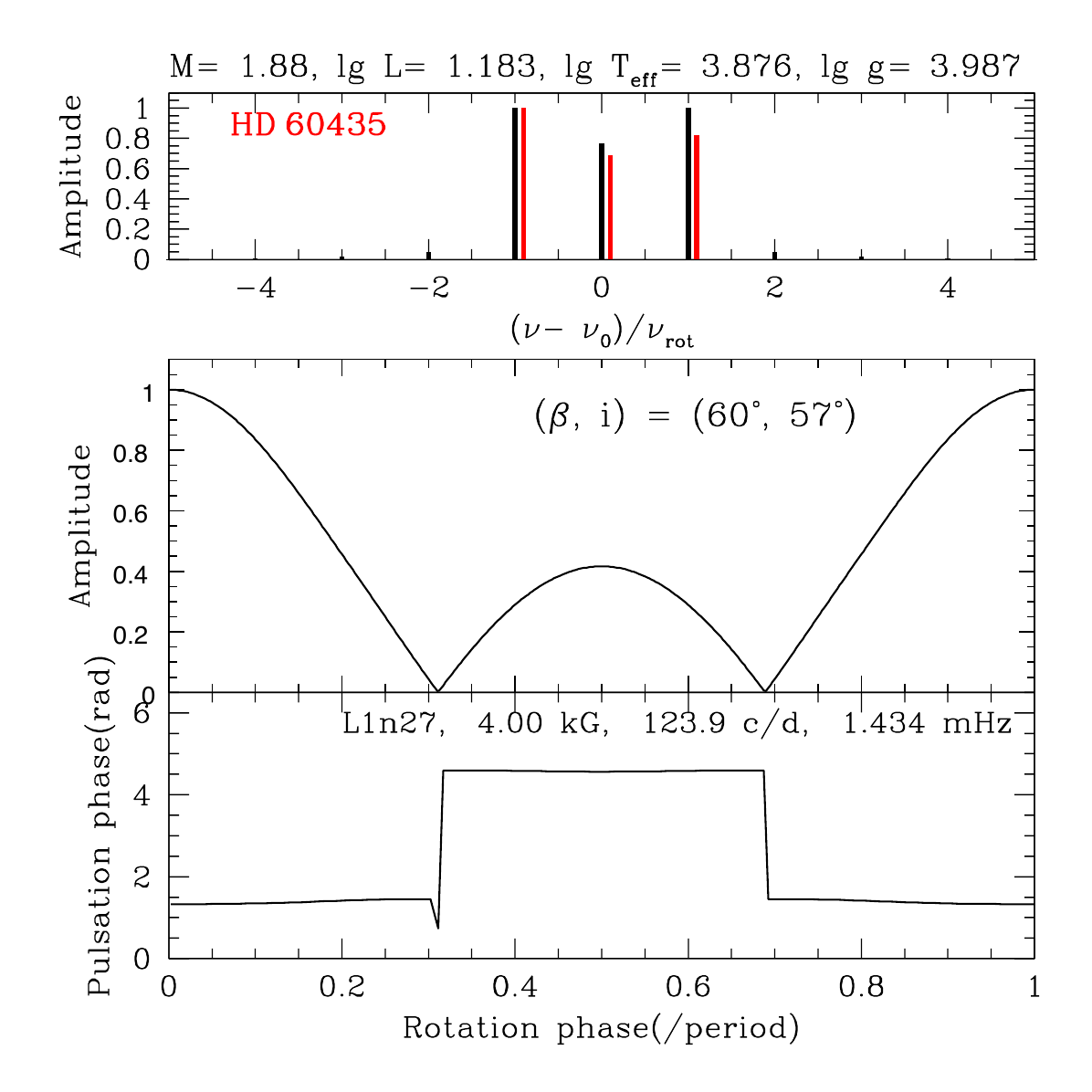}
\includegraphics[width=0.49\textwidth]{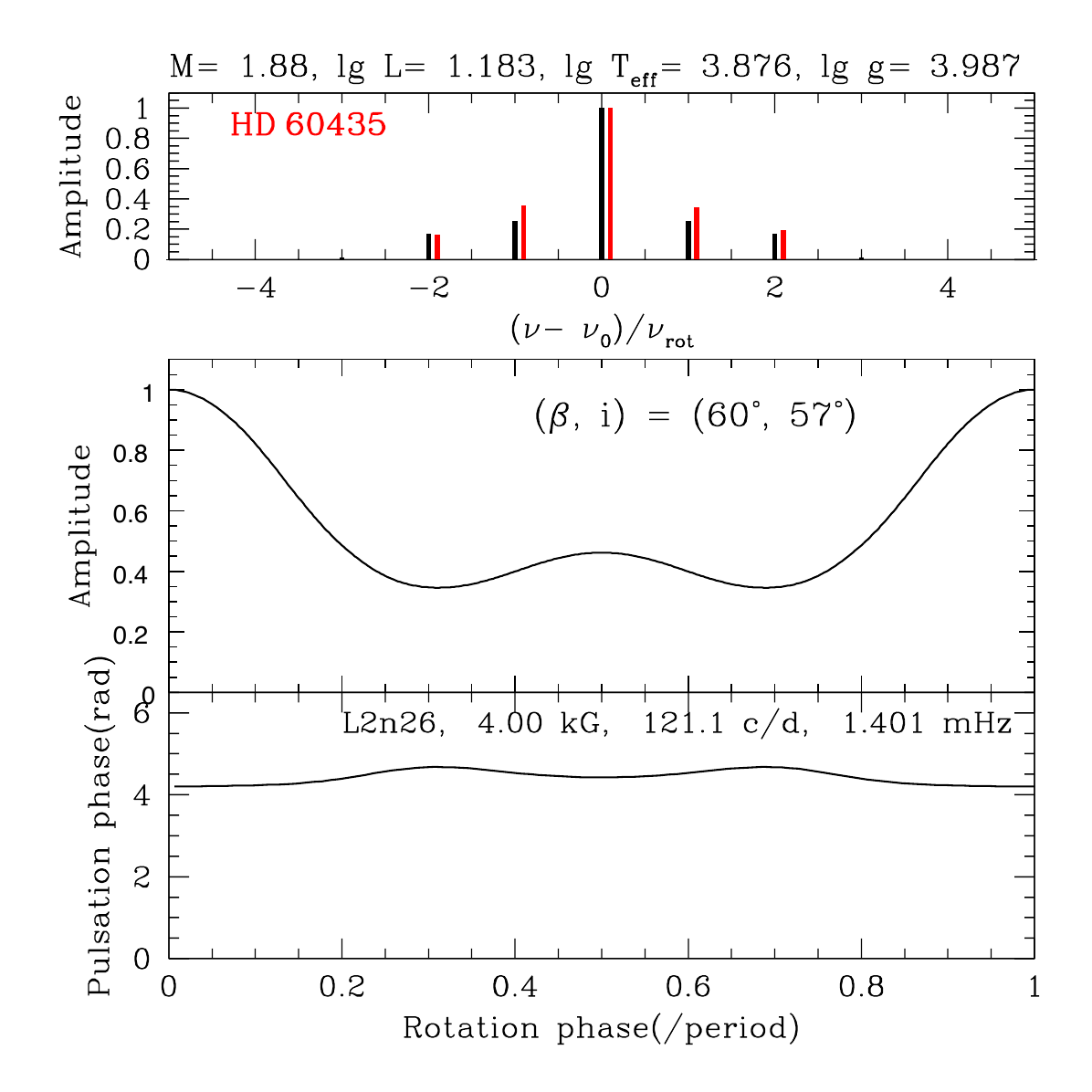}
\caption{Models for pulsation amplitude (middle panels) and phase (bottom panels) modulations for the deformed dipole mode of $\nu_7$ (left panel) and  the deformed quadrupole mode of $\nu_6$ (right panel). The top panels show the  amplitudes of the pulsation and rotation sidelobes (normalised to unity at maximum) for the corresponding pulsations; red and black lines are, respectively, observed amplitudes of HD\,60435 and model predictions at $B_{\rm p} = 4$\,kG.  }
\label{fig:AMmodel}
\end{figure*}

The equilibrium stellar model adopted has parameters of $M=1.88$\,M$_\odot$, $L=15.25$\,L$_\odot$, $T_{\rm eff} = 7525$\,K, and  $R=2.30$\,R$_\odot$. The model was taken from the evolution model with a central hydrogen abundance of 0.28, and was chosen for the large frequency separation to be $\approx$$51$\,$\upmu$Hz at $B_{\rm p}=0$ and global parameters not far from the estimates discussed in section 6. This model has an abundance of heavy elements $Z=0.02$, but helium is assumed to be depleted to $Y_{\rm s}=0.01$ above the He\,{\sc{ii}} ionisation zone, and convection in the envelope to be suppressed, assuming a strong magnetic field to stabilize the outer layers \citep[polar model of][]{Balmforth01}.  

The eight frequencies observed for HD\,60435 approximately agree with three dipole modes, four $\ell=2$ modes, and an $\ell=3$ mode of the model at $B_{\rm p}\approx 4$\,kG. The magnetic field strength is consistent with the estimate discussed in section 4 based on the longitudinal magnetic field observed by \citet{2006AN....327..289H}.  We have obtained amplitude and phase modulations using the method discussed in  \citet{2004MNRAS.350..485S} for $\ell=1$ and 2 modes close to $\nu_7$ and $\nu_6$ at $B_{\rm p}=4$\,kG to compare with the observed ones shown in Fig.\,\ref{fig:as6}. The results of our model are shown in Fig.\,\ref{fig:AMmodel}, where we adopted a rotational inclination angle $i=57^{\circ}$ derived from the radius of our model, $2.30$\,R$_\odot$, with the rotation period and $v\sin i$ of HD\,60435 determined in this paper and $\beta=60^\circ$ then derived from equation \ref{eqn:opmdipole}. We note that the adopted $i$ and $\beta$ are within the uncertainties discussed in section 7.

The phase modulation of the distorted $\ell=2$ mode $\nu_6$ is strongly suppressed in accordance with our model. Such suppressions are known to occur in other roAp stars \citep[e.g.,][]{Dan2018J1640,Dan2018J1940}.

\section{Discussion and Conclusions}

It is known that modes in stochastically driven pulsators -- solar-like and red giant oscillators -- have short lifetimes, hence show strong amplitude modulation. This is unsurprising, given the non-periodic nature of the driving. It is also known that heat-engine pulsators -- those driven by the $\kappa$-mechanism -- often show amplitude modulation on a variety of time scales. Examples of this for Polaris, Deneb, Spica, and white dwarfs were discussed in the introduction, section 1.  While the physical conditions and processes governing pulsation amplitude stability in these stars are expected to be different to those in main sequence pulsators, the phenomenological similarity between these objects is striking. In some heat engine driven pulsators some modes have been observed to appear, and to disappear. But in no previous case has a pulsating star been observed to cease all pulsations, as has the roAp star HD\,60435. 

We know that HD\,60435 was pulsating when it was observed in 1983 \citep[][Johnson $B$]{1984MNRAS.209..841K}, 1984  \citep[][Johnson $B$]{1986ApJ...300..348M}, 1985 \citep[][Johnson $B$]{1987ApJ...313..782M}, 1991 \citep[][HST UV]{1993ApJ...413L.125T}, 2014 (this paper, radial velocities), and from 2018 September to 2020 October ({\it TESS} red bandpass), after which the pulsation stopped, at least up to the last {\it TESS} (as of this writing) observations in 2023 September, and the LCOGT ground-based observations in 2024 January. We do not know if it had observable pulsations at the times when it was not observed; the largest gap in the observing record is 13\,yr. But other roAp stars have shown  nearly stable pulsations over decades. There has never been any reason to suspect that an roAp star -- or any other pulsator -- could cease pulsating on a time scale far shorter than an evolutionary time. Thus, the primary result of this paper is that HD\,60435 stopped pulsating. It is not understood why, so we now make some conjectures on the implications of this. 

Figs\,\ref{fig:as}, \ref{fig:amps}, and \ref{fig:wav1} and \ref{fig:as2} and \ref{fig:wav2} suggest that when modes near $\nu_4$ are strong, modes $\nu_2$, $\nu_6$ and $\nu_7$ are weak, and vice versa. These modes look to be coupled and exchanging energy. Nonlinear mode coupling among observed modes and other modes that are of too high a degree to be observed has been proposed  by \citet{1985AcA....35....5D} and discussed by \citet{2016MNRAS.460.1970B} as a possible explanation for amplitude variability among 983 $\delta$~Sct stars observed by {\it Kepler}. This requires a tight resonance among `parent' and `child' mode frequencies, which seems possible for $\delta$~Sct stars where only a few of many modes show amplitude modulation, and the frequency spectrum is densely populated with mode frequencies in models.  

This could also be an explanation for the disappearance of observable pulsation in HD\,60435, i.e., the observed modes coupled with, and transferred their pulsation energy to, higher degree modes that do not produce observable light variations. It would require that all observed mode frequencies -- not just some as in $\delta$~Sct stars -- coupled resonantly with higher degree, unobservable mode frequencies. That is more challenging to accept than the case for $\delta$~Sct stars with only a few resonances generating amplitude modulation. We see no way at present to test this conjecture. 

An alternative is that in the balance between the pulsation driving and the global damping of the modes, damping won (i.e., the linear growths rates became negative). This is what must happen when stars evolve out of instability regions, so can we conjecture that we have just caught HD\,60435 evolving out of the roAp instability region? Given its $T_{\rm eff}$ and $L$, that is unlikely, but if that conjecture were correct, then we can predict that HD\,60435 will not be observed to pulsate in future observations. That is falsifiable if the pulsations return. Future observations are planned to test this conjecture. Alternatively, can pulsations just stop for reasons other than evolutionary changes in the stellar structure on a time scale of weeks for HD\,60435? That would require a very fine balance between the driving and damping, and then a subtle change in the star to tip the balance. At present, we have no models with sufficient confidence to address this question. Note, for example, that the modes in HD\,60435 are theoretically not driven at all in the model we presented in section 9. 

There is the further question of why we do not see in the {\it TESS} data the long series of frequencies between $0.7 - 1.3$\,mHz found by \citet{1987ApJ...313..782M} and listed in Table\,\ref{table:3}. Were those frequencies actually present in the 1984--1985 data? If so, those modes had stopped pulsating by the time of the {\it TESS} observations. Or, were the challenges of signal-to-noise and complex spectral windows in the ground-based data such that those frequency identifications were not secure? Those ground-based data were obtained from Cerro Tololo InterAmerican Observatory (CTIO), Las Campanas Observatory (LCO) and the South African Astronomical Observatory (SAAO) contemporaneously, but the lower frequencies identified are only visible in the LCO data, hence perhaps were an artefact. Should they reappear in future observations, then their reality will be proved. But should they not be observed again, no further conclusion can be drawn, since we know HD\,60435 is capable of stopping pulsating. 

The ceasing of pulsations in HD\,60435 has an impact on the question of why only 5.5\,per\,cent  of magnetic Ap stars are roAp stars \citep{2024MNRAS.527.9548H}, the rest being labelled as noAp (non-oscillating Ap) stars. Since we now have this one case of an roAp star becoming an noAp star in a matter of weeks, the distinction between roAp and noAp is blurred. 

This problem is more general. For example,  \citet{2024ApJ...972..137G} examined millions of light curves in {\it TESS} 30-min cadence data and found that out of nearly 16000 $\delta$~Sct stars the pulsator fraction peaks at $50 - 70$\,per\,cent in the centre of the $\delta$~Sct instability strip. They found a correlation between pulsator fraction and rotation, hence deduced that rotation plays a role in driving in $\delta$~Sct stars. \citet{2024arXiv240913135M} showed that the $\delta$~Sct pulsator fraction is also dependent on age, even for stars that remain inside the instability strip. The general question is why do some stars in an instability strip pulsate while other do not, when the stars may share $T_{\rm eff}$, $L$, age, [Fe/H], binarity. What are the additional factors that govern this?  For the $\delta$~Sct stars \citeauthor{2024ApJ...972..137G} showed rotation is one such factor. But for the roAp stars we do not know what the factors are that select between roAp and noAp, although it does not appear to be rotation, as the vast majority of Ap stars are relatively slow rotators with periods between $1-10$\,d \citep{2023A&A...676A..55L,2012MNRAS.420..757W}, and some are super-slowly rotating with periods up to centuries \citep{2024A&A...683A.227M}. The cessation of pulsation in HD\,60435 highlights that this important question needs more work.

In addition to the major conclusion of this paper that HD\,60435 stopped pulsating, with all of the implication of that, we have also learned more from this star. Since it showed stable pulsations for more than a rotation period, we were able to apply the oblique pulsator model and extract constraints on the geometry of the pulsation modes, i.e. constrain the rotational inclination, $i$, and the pulsation axis (hence the magnetic axis) obliquity, $\beta$.  We showed that the magnetic pulsation model of \citet{Saio05} inferred a polar magnetic field strength of 4\,kG, in good agreement with the one magnetic observation that has been published. This reinforces confidence in the theory for that model, which explains the suppression of pulsation amplitude in the pulsational equatorial zone for quadrupole  modes, thus the small changes in pulsation phase for a quadrupole mode with the rotation of the star. Thus HD\,60435 joins a small number of roAp stars for which this phase suppression in obliquely pulsating quadrupole modes has been observed \citep[e.g.,][]{Dan2018J1640,Dan2018J1940}.

Using the models of \citet{2021A&A...650A.125D} we examined the frequency spacing in HD\,60435 to address two issues: the degree of the even modes, and the degree and frequency spacing of the  $\nu_2$ mode. We found that $\nu_6$ is a quadrupole mode, but that the other even degree  modes could be either quadrupole or radial modes, depending on the metallicity of the star. This highlights the potential of the frequency spacings in roAp stars to constrain the global metallicity, something that cannot be done by spectroscopy because of the extreme atmospheric abundance anomalies in Ap stars. 

While identification of the degree of the $\nu_2$ mode awaits future modelling, from the frequency spacing it is potentially a highly distorted octupole ($\ell = 3$) mode with strong suppression of the pulsation in the equatorial band making the mode visible. If so, HD\,60435 highlights opportunities for asteroseismic constraints with higher degree modes than  the $\ell = 1, 2$ that have previously been identified and modelled in roAp stars.

\section*{acknowledgements}

We thank Morgan Deal for providing the information on the models considered in section\,\ref{sec:modeID_models}. We thank Barry Smalley for providing a determination of $T_{\rm eff}$ from the IRFM. We thank Guy Davies and David Mkrtichian for useful discussions on HD 60435. This work has been partially supported by Funda\c c\~ao pa a Ci\^encia e Tecnologia FCT-MCTES, Portugal, through national funds by these grants: DOI: 10.54499/UIDB/04434/2020; DOI: 10.54499/UIDP/04434/2020 and DOI:10.54499/2022.03993.PTDC. MC is funded by FCT-MCTES by the contract with reference CEECIND/02619/2017. GH thanks the Polish National Center for Science (NCN) for supporting this study through grant 2021/43/B/ST9/02972. SJM is supported by the Australian Research Council through Future Fellowship FT210100485. Co-funded by the European Union (ERC, MAGNIFY, Project 101126182). Views and opinions expressed are however those of the authors only and do not necessarily reflect those of the European Union or the European Research Council. Neither the European Union nor the granting authority can be held responsible for them. This paper includes data collected by the {\it TESS} mission. Funding for {\it TESS} is provided by NASA's Science Mission Directorate. Resources used in this work were provided by the NASA High End Computing (HEC) Program through the NASA Advanced Supercomputing (NAS) Division at Ames Research Center for the production of the SPOC data products.Based in part on data acquired at the Anglo-Australian Telescope, under programs A/2024A/016 and A/2024B/011. We acknowledge the traditional custodians of the land on which the AAT stands, the Gamilaraay people, and pay our respects to elders past and present. Some of the observations reported in this paper were obtained with the Southern African Large Telescope (SALT). 
This work made use of observations from the Las Cumbres Observatory global telescope network. This work has made use of data from the European Space Agency (ESA) mission {\it Gaia} (\url{https://www.cosmos.esa.int/gaia}), processed by the {\it Gaia} Data Processing and Analysis Consortium (DPAC, \url{https://www.cosmos.esa.int/web/gaia/dpac/consortium}). Funding for the DPAChas been provided by national institutions, in particular the institutionsparticipating in the {\it Gaia} Multilateral Agreement. This work made use of Astropy, a community-developed core Python package and an ecosystem of tools and resources for astronomy.

\section*{data availability}

The {\it TESS} data used in this study are available on MAST. The SALT spectra are available from the SALT data archive: {\url https://ssda.saao.ac.za}. The ground-based data from \citet{1987ApJ...313..782M} are available from DWK on request. The AAT spectra are available from the AAT archive\footnote{https://archives.datacentral.org.au/query}, or from SJM on request.

\bibliography{hd60435.bib}{}
\bibliographystyle{aasjournal}
 
 \clearpage
 
\appendix

\section{The half-sector information and plots}

\begin{table*}
\centering
\caption{Times and durations of the {\it TESS} half-sectors and full sectors. The data half-sectors are also named with the mid-time of each half-sector for further identification of the individual amplitude spectra in Figs\,\ref{fig:as} and \ref{fig:as2}, and to judge the time distributions. Columns 5 and 6 give, respectively, the duration of the half-sectors and their combined sector in days. All the data are 120-s cadence, except S61, 62, 63, and 67, which are 200-s cadence.} 
\begin{tabular}{ccrrrr}
\hline 
\multicolumn{1}{c}{{\it TESS} Sector} & \multicolumn{1}{c}{data set name} & \multicolumn{1}{c}{BJD time start} & \multicolumn{1}{c}{BJD time end} & \multicolumn{2}{c}{duration days}  \\
\multicolumn{1}{c}{} & \multicolumn{1}{c}{} & \multicolumn{2}{c}{$2400000+$} & \multicolumn{1}{r}{half sector} &  \multicolumn{1}{r}{full sector}  \\
\hline
\hline
3.1 & JD58390.03 & 58385.934815 & 58394.154367 & 8.22 &   \\ 
3.2 & JD58401.42 & 58396.639116 & 58406.212895 & 9.57 & 20.28  \\ 
6.1 & JD58472.65 & 58468.272565 & 58477.021355 & 8.75 &   \\ 
6.2 & JD58484.18 & 58478.243596 & 58490.045204 & 11.80 & 21.77  \\ 
7.1 & JD58497.34 & 58491.634114 & 58503.038448 & 11.40 &   \\ 
7.2 & JD58510.40 & 58504.710685 & 58516.087190 & 11.38 & 24.45  \\ 
8.1 & JD58523.56 & 58517.984416 & 58529.066416 & 11.08 &   \\ 
8.2 & JD58539.01 & 58536.019205 & 58542.001135 & 5.98 & 24.02  \\ 
9.1 & JD58549.94 & 58544.327512 & 58555.542723 & 11.22 &   \\ 
9.2 & JD58563.02 & 58557.571871 & 58568.475921 & 10.90 & 24.15  \\ 
10.1 & JD58576.50 & 58571.218945 & 58581.785451 & 10.57 &   \\ 
10.2 & JD58590.40 & 58585.104836 & 58595.681025 & 10.58 & 24.46  \\ 
13.1 & JD58660.80 & 58653.919997 & 58667.689174 & 13.77 &   \\ 
13.2 & JD58675.49 & 58668.623876 & 58682.355603 & 13.73 & 28.44  \\ 
27.1 & JD59041.26 & 59036.277847 & 59046.545744 & 10.27 &   \\ 
27.2 & JD59053.87 & 59049.152647 & 59058.588651 & 9.44 & 22.31  \\ 
30.1 & JD59121.35 & 59115.884497 & 59126.794366 & 10.91 &   \\ 
30.2 & JD59135.83 & 59130.208301 & 59141.441833 & 11.23 & 25.56  \\ 
33.1 & JD59207.80 & 59201.733424 & 59213.864218 & 12.13 &   \\ 
33.2 & JD59221.50 & 59215.435079 & 59227.572775 & 12.14 & 25.84  \\ 
34.1 & JD59234.95 & 59229.000568 & 59240.906264 & 11.91 &   \\ 
34.2 & JD59248.20 & 59242.357667 & 59254.066081 & 11.71 & 25.07  \\ 
35.1 & JD59260.96 & 59255.814700 & 59266.110553 & 10.30 &   \\ 
35.2 & JD59276.00 & 59272.013326 & 59279.979952 & 7.97 & 24.17  \\ 
37.1 & JD59313.64 & 59308.254647 & 59319.030847 & 10.78 &   \\ 
37.2 & JD59327.00 & 59321.412748 & 59332.579186 & 11.17 & 24.32  \\ 
61.1 & JD59969.00 & 59962.796133 & 59975.196726 & 12.40 &  \\ 
61.2 & JD59981.82 & 59975.412006 & 59988.224578 & 12.81 & 25.43 \\ 
62.1 & JD59994.68 & 59988.439857 & 60000.923658 & 12.48 &  \\ 
62.2 & JD60007.65 & 60001.136621 & 60014.152763 & 13.02 & 25.71 \\ 
63.1 & JD60020.88 & 60014.368039 & 60027.391059 & 13.02 &  \\ 
63.2 & JD60034.25 & 60027.604019 & 60040.897809 & 13.29 & 26.53 \\ 
67.1 & JD60133.44 & 60126.639041 & 60140.247607 & 13.61 &  \\ 
67.2 & JD60147.43 & 60140.460567 & 60154.393260 & 13.93 & 27.75 \\ 
68.1 & JD60160.23 & 60154.910362 & 60165.607499 & 10.70 &   \\ 
68.2 & JD60173.98 & 60168.485256 & 60179.510225 & 11.02 & 24.60  \\ 
69.1 & JD60187.58 & 60182.353276 & 60192.596356 & 10.24 &   \\ 
69.2 & JD60200.70 & 60195.396365 & 60206.035332 & 10.64 & 23.68  \\ 
\hline
\hline
\end{tabular}
\label{table:1b}
\end{table*}

\clearpage
\newpage

\begin{figure}
\begin{center}
\includegraphics[width=1.0\linewidth,angle=0]{figs/JD58390_2.pdf}\\
\vspace{-0.6cm}
\includegraphics[width=1.0\linewidth,angle=0]{figs/JD58401_2.pdf}\\
\vspace{-0.6cm}
\includegraphics[width=1.0\linewidth,angle=0]{figs/JD58472_2.pdf}\\
\vspace{-0.6cm}
\includegraphics[width=1.0\linewidth,angle=0]{figs/JD58484_2.pdf}\\
\vspace{-0.6cm}
\includegraphics[width=1.0\linewidth,angle=0]{figs/JD58497_2.pdf}\\
\vspace{-0.6cm}
\includegraphics[width=1.0\linewidth,angle=0]{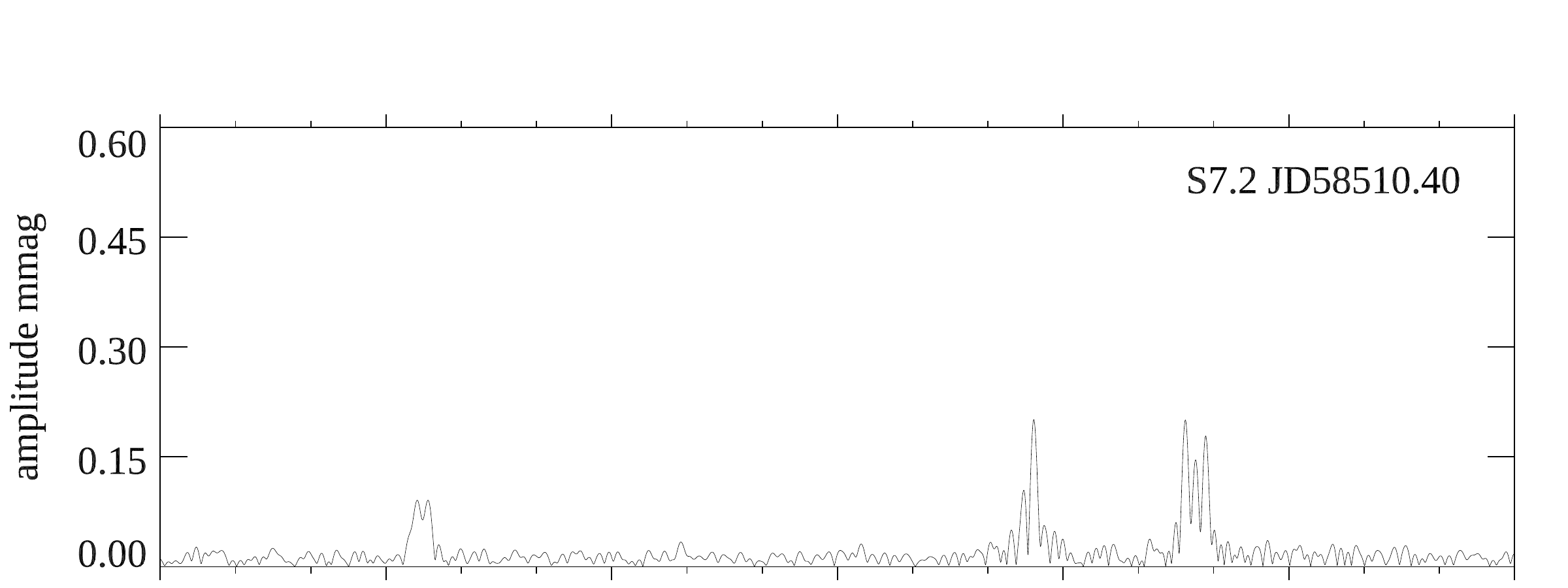}\\
\vspace{-0.6cm}
\includegraphics[width=1.0\linewidth,angle=0]{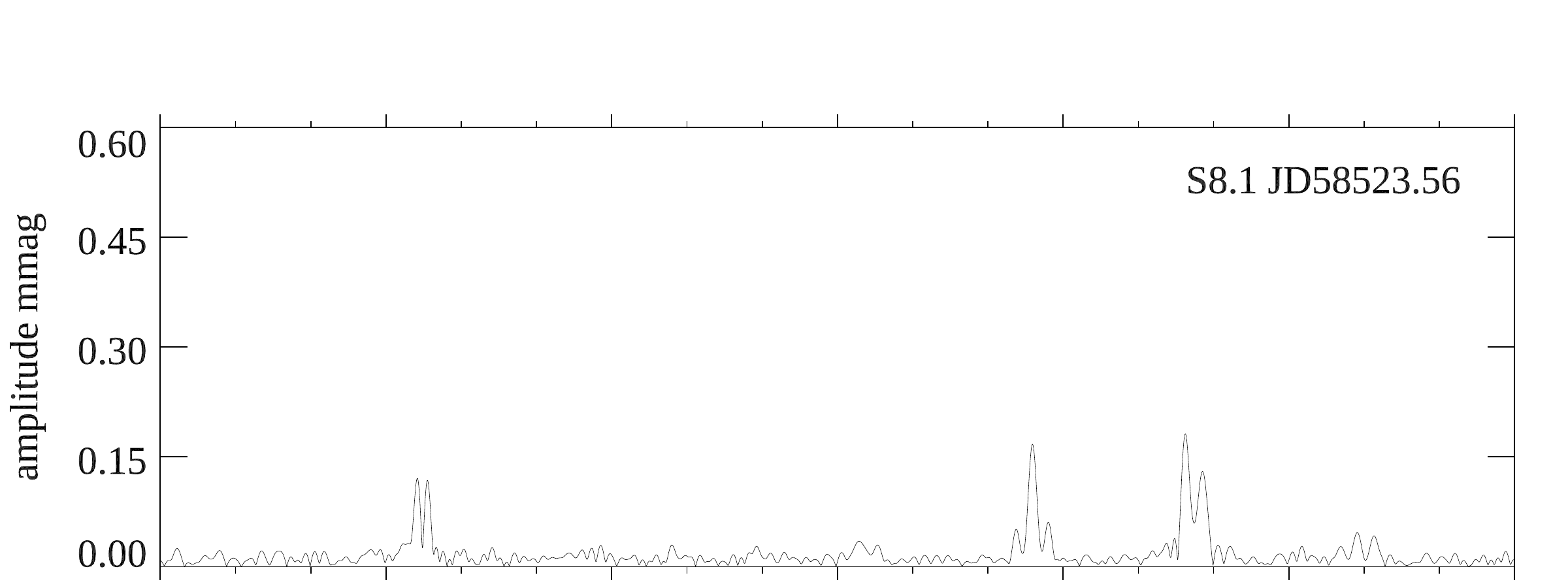}\\
\includegraphics[width=1.0\linewidth,angle=0]{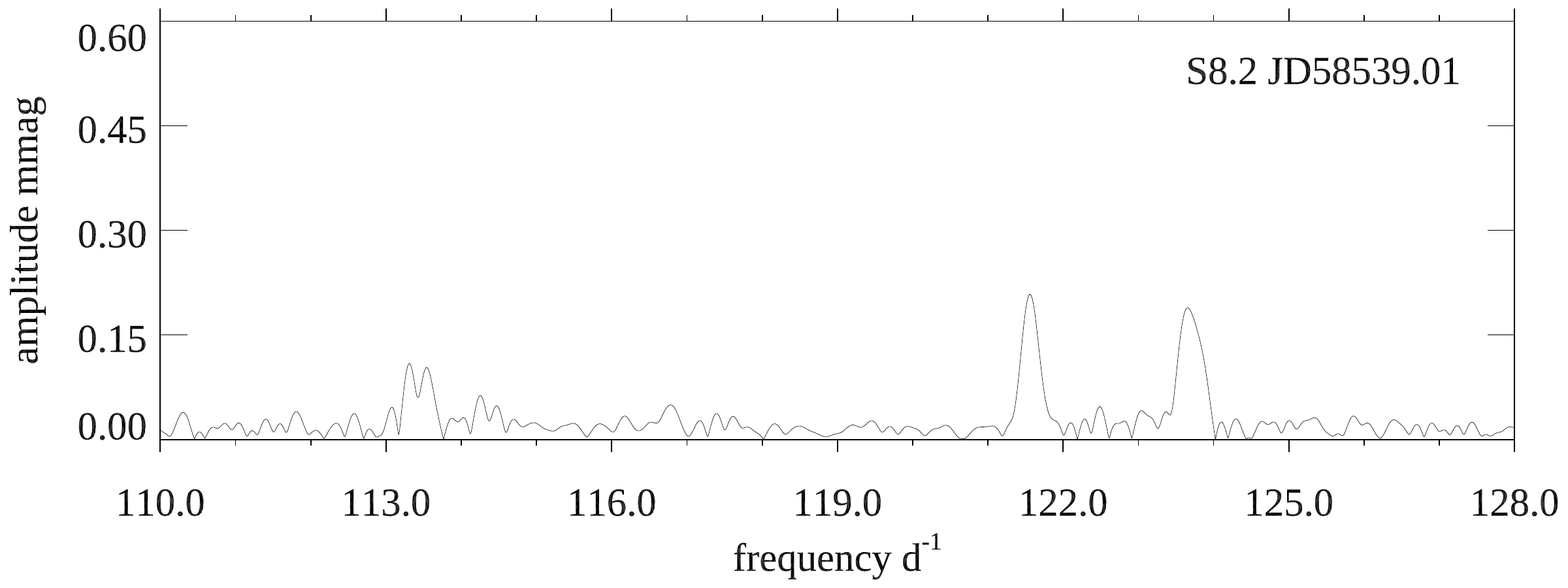}\\
\caption{Amplitude spectra of the half-sectors showing the amplitude modulation and mode changes.}
\label{fig:as2} 
\end{center}
\end{figure}

\addtocounter{figure}{-1}
\begin{figure}
\begin{center}
\includegraphics[width=1.0\linewidth,angle=0]{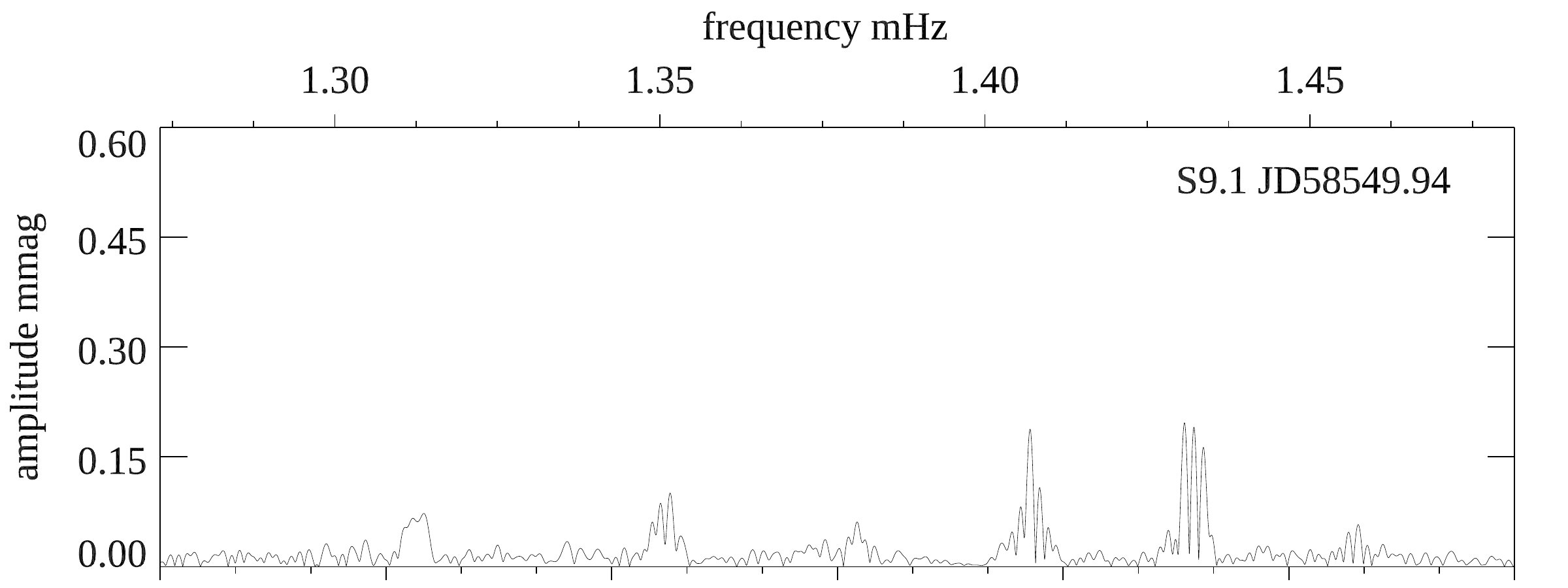}\\
\vspace{-0.6cm}
\includegraphics[width=1.0\linewidth,angle=0]{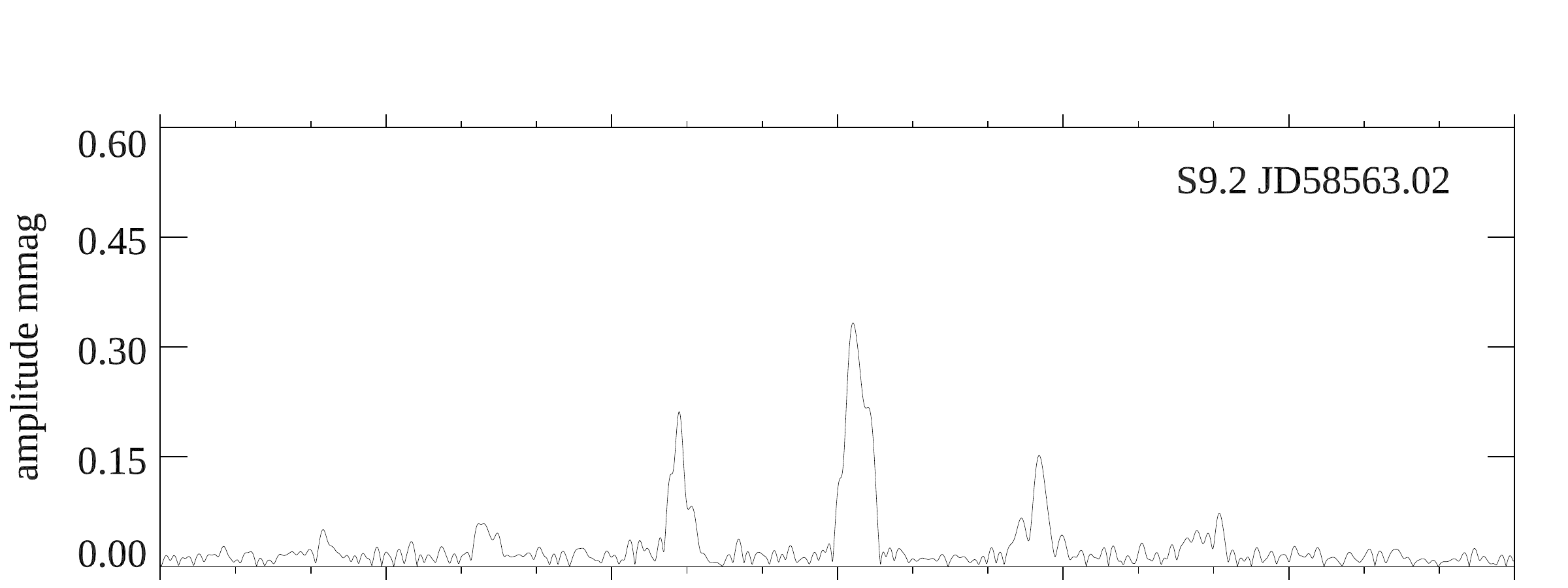}\\
\vspace{-0.6cm}
\includegraphics[width=1.0\linewidth,angle=0]{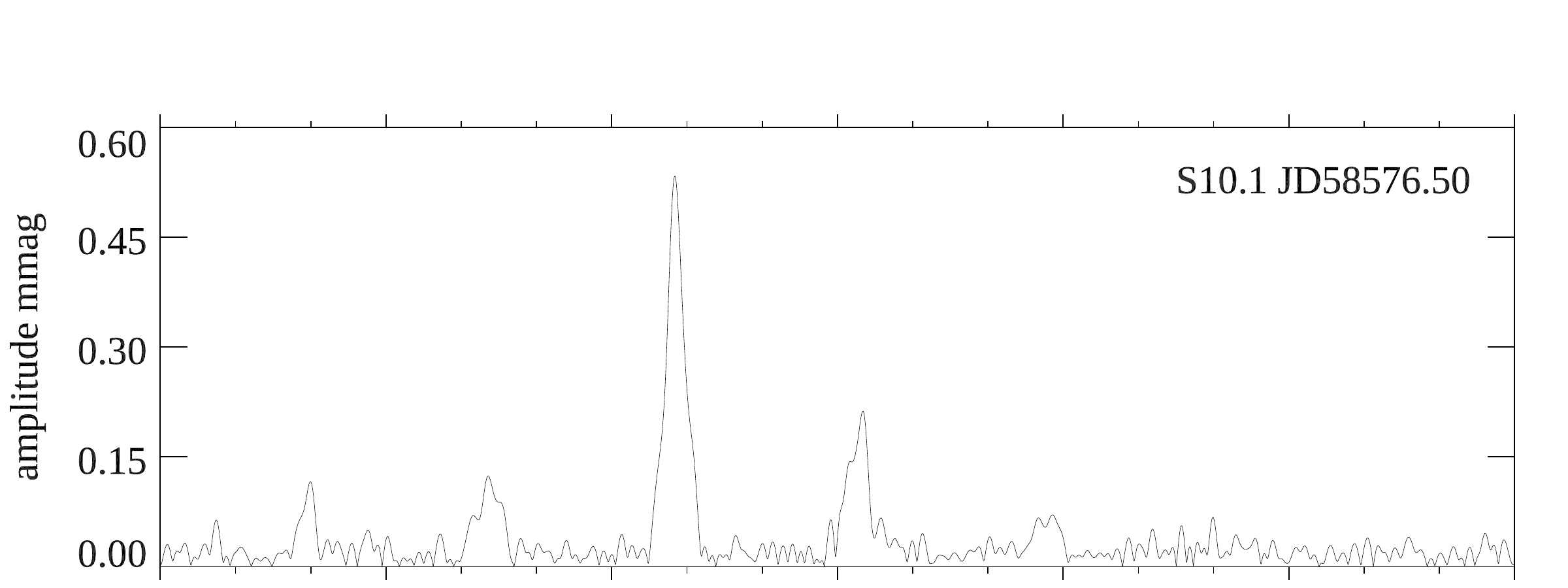}\\
\vspace{-0.6cm}
\includegraphics[width=1.0\linewidth,angle=0]{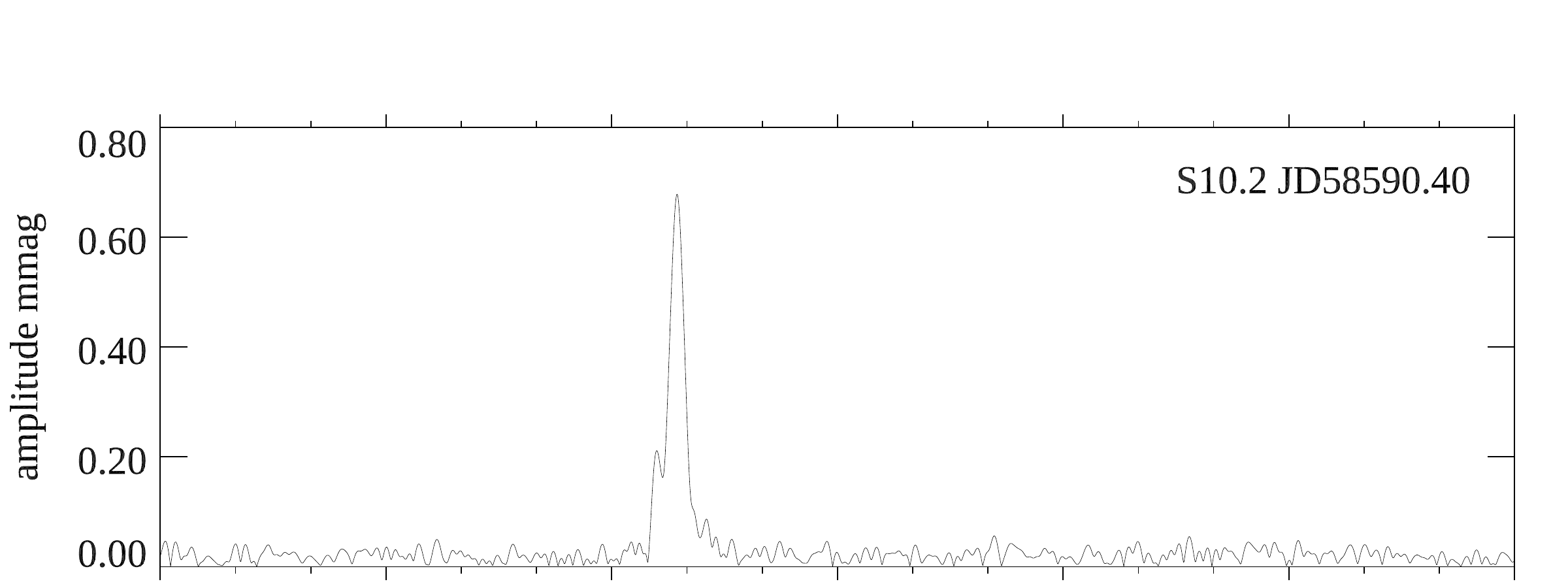}\\
\vspace{-0.6cm}
\includegraphics[width=1.0\linewidth,angle=0]{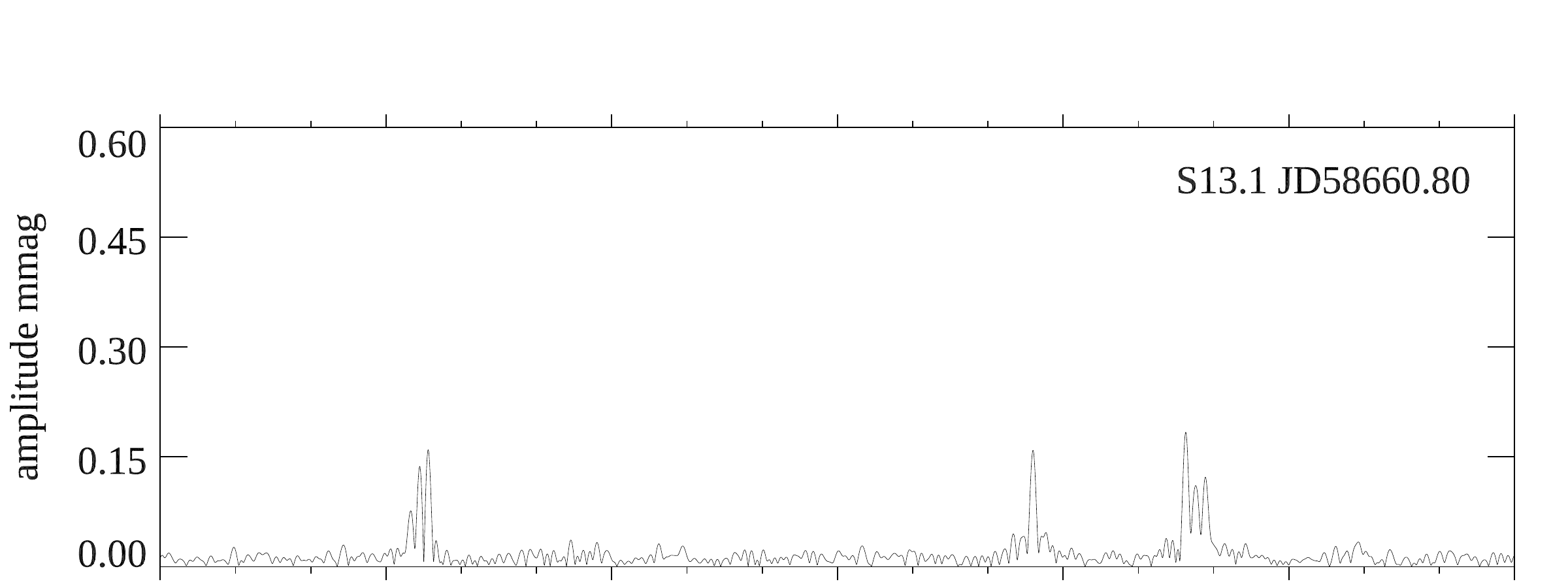}\\
\vspace{-0.6cm}
\includegraphics[width=1.0\linewidth,angle=0]{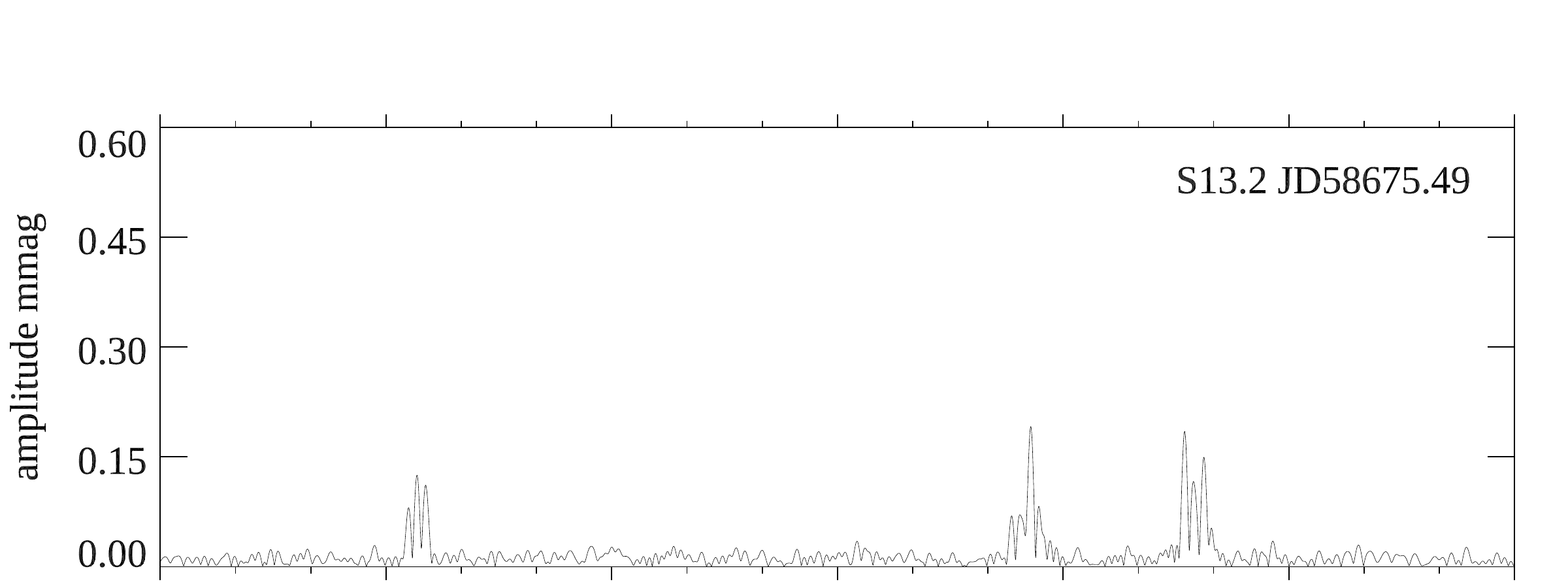}\\
\vspace{-0.6cm}
\includegraphics[width=1.0\linewidth,angle=0]{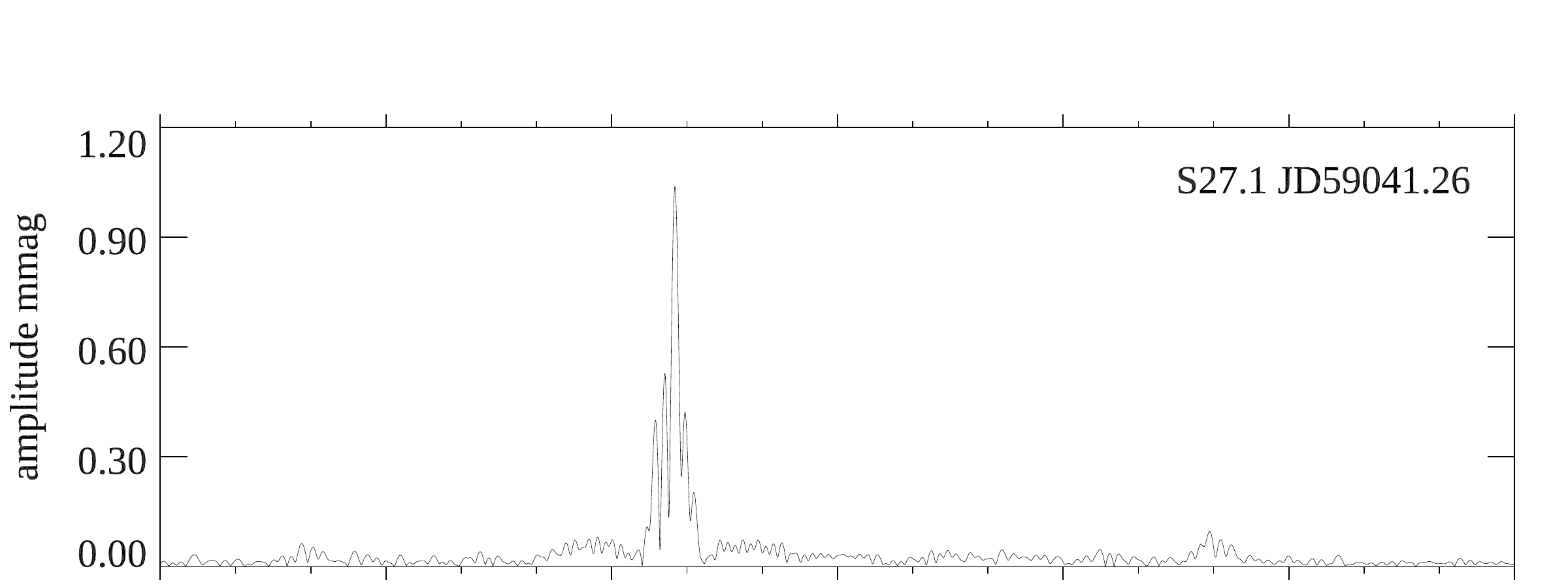}\\
\includegraphics[width=1.0\linewidth,angle=0]{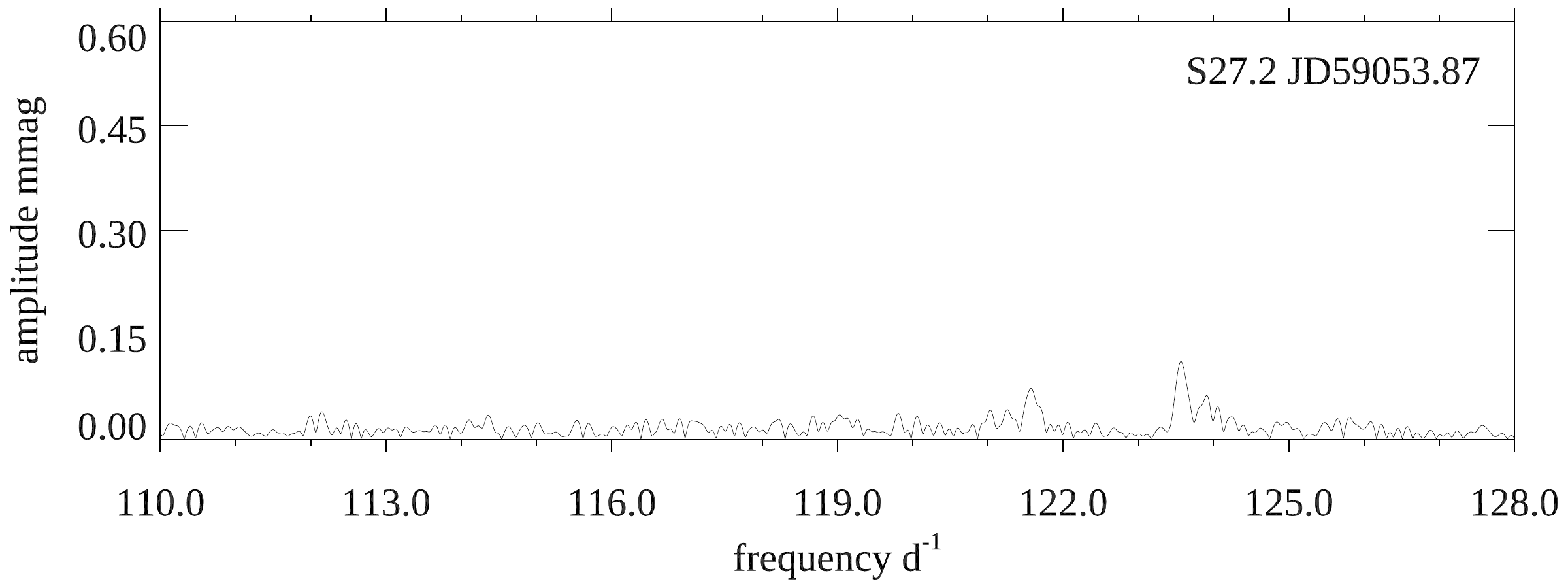}\\
\caption{Amplitude spectra continued. Note the ordinate scale changes for S10.2 JD58590.40 and S27.1 JD59041.26.}
\label{fig:as2} 
\end{center}
\end{figure}

\addtocounter{figure}{-1}
\begin{figure}
\begin{center}
\includegraphics[width=1.0\linewidth,angle=0]{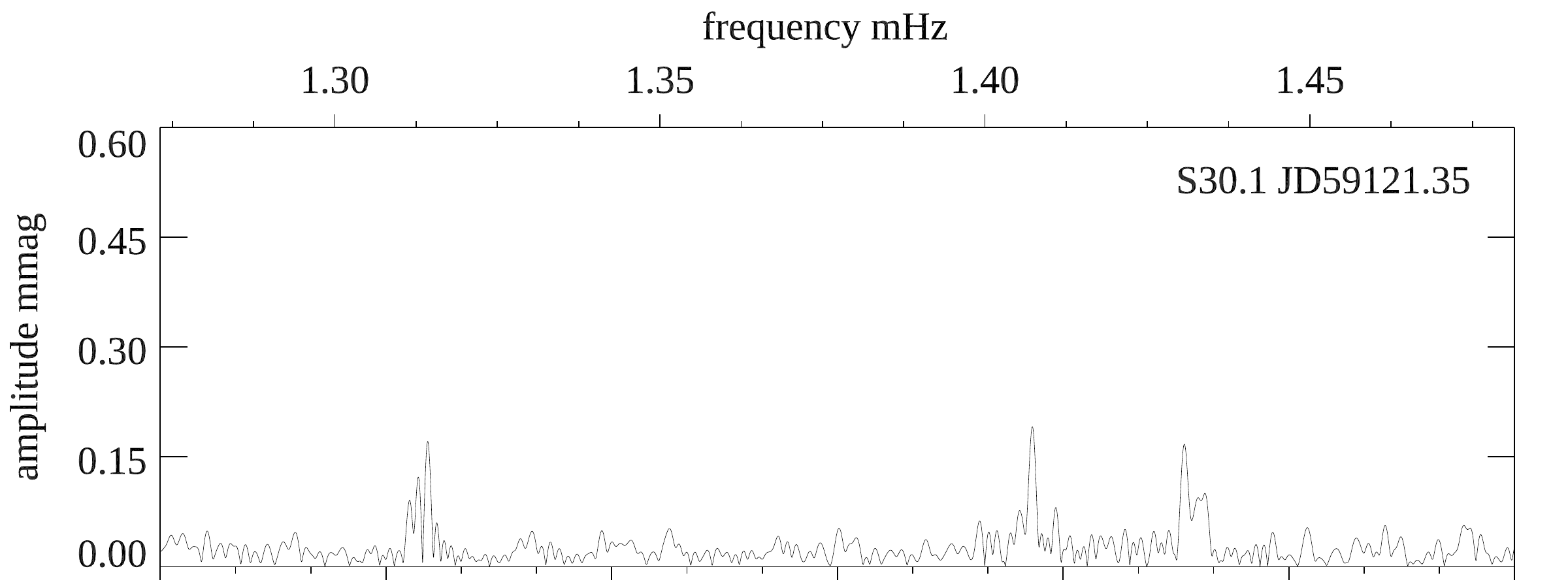}\\
\vspace{-0.6cm}
\includegraphics[width=1.0\linewidth,angle=0]{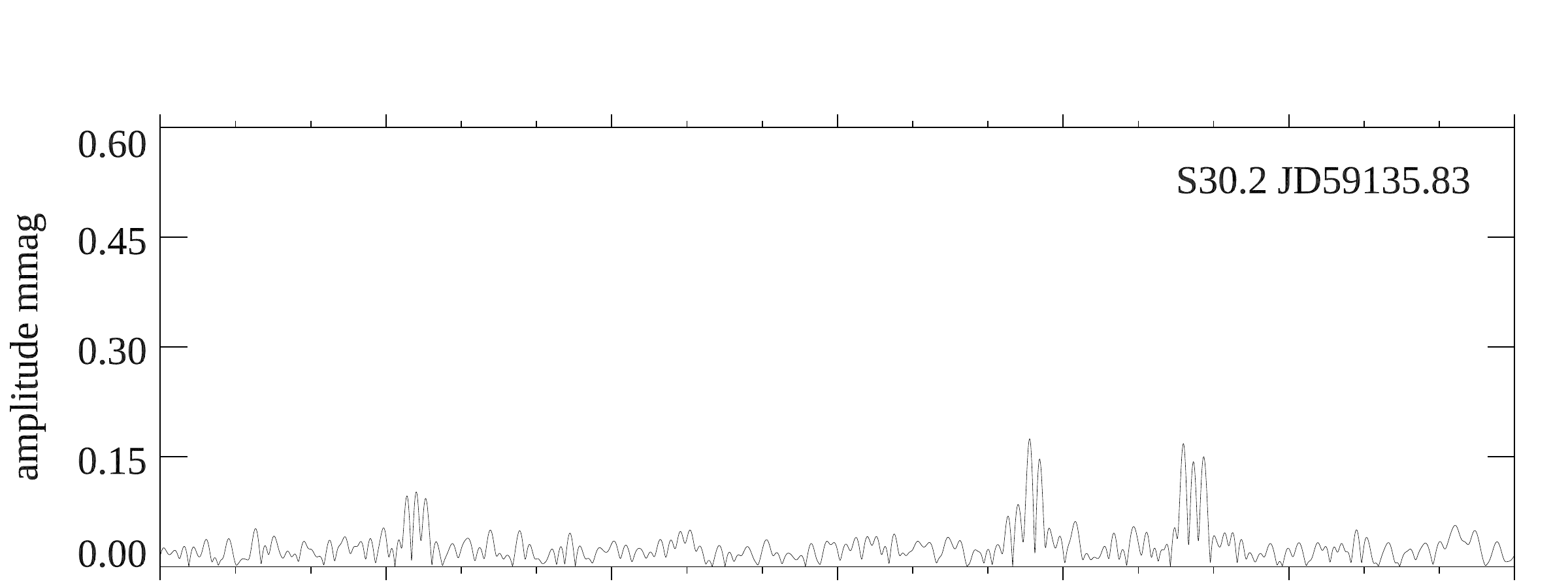}\\
\vspace{-0.6cm}
\includegraphics[width=1.0\linewidth,angle=0]{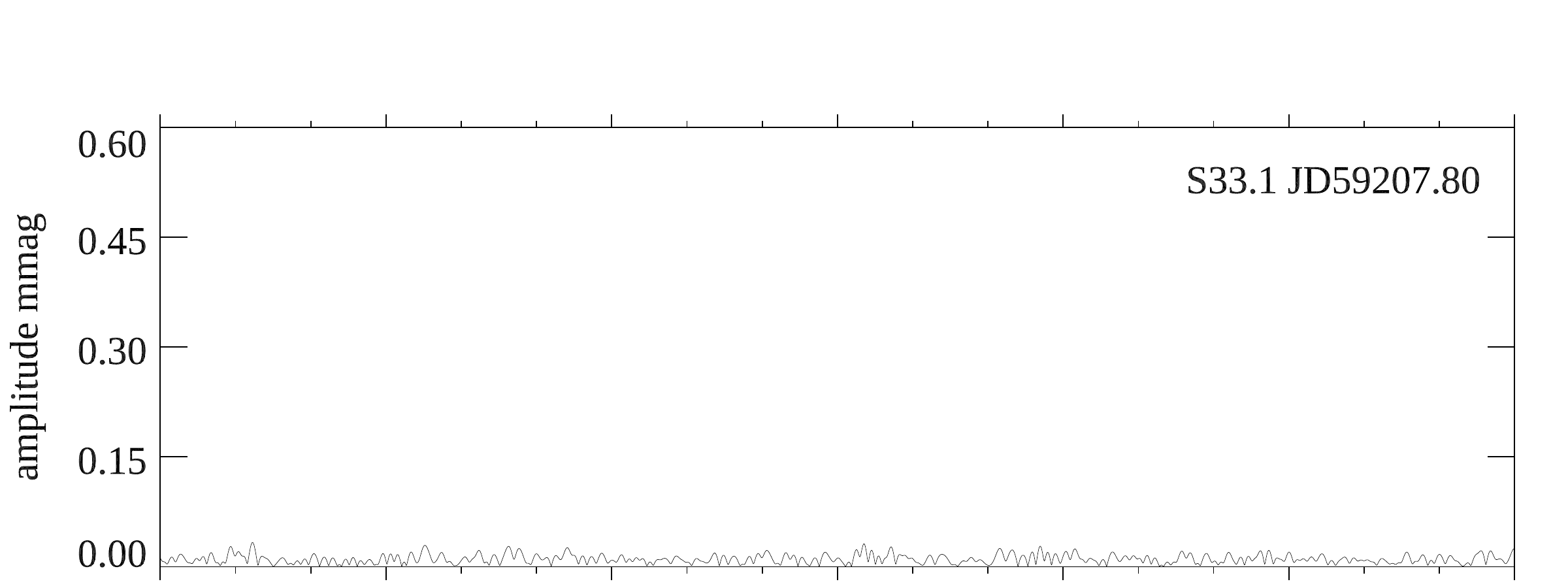}\\
\vspace{-0.6cm}
\includegraphics[width=1.0\linewidth,angle=0]{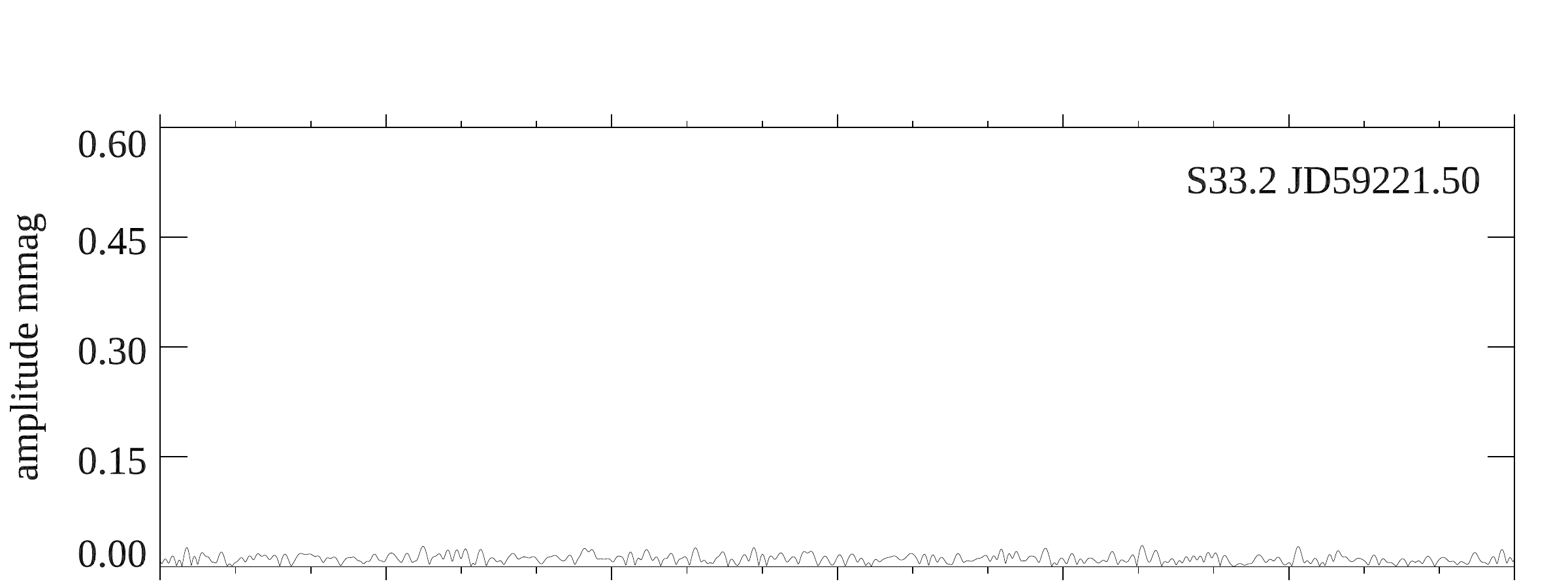}\\
\vspace{-0.6cm}
\includegraphics[width=1.0\linewidth,angle=0]{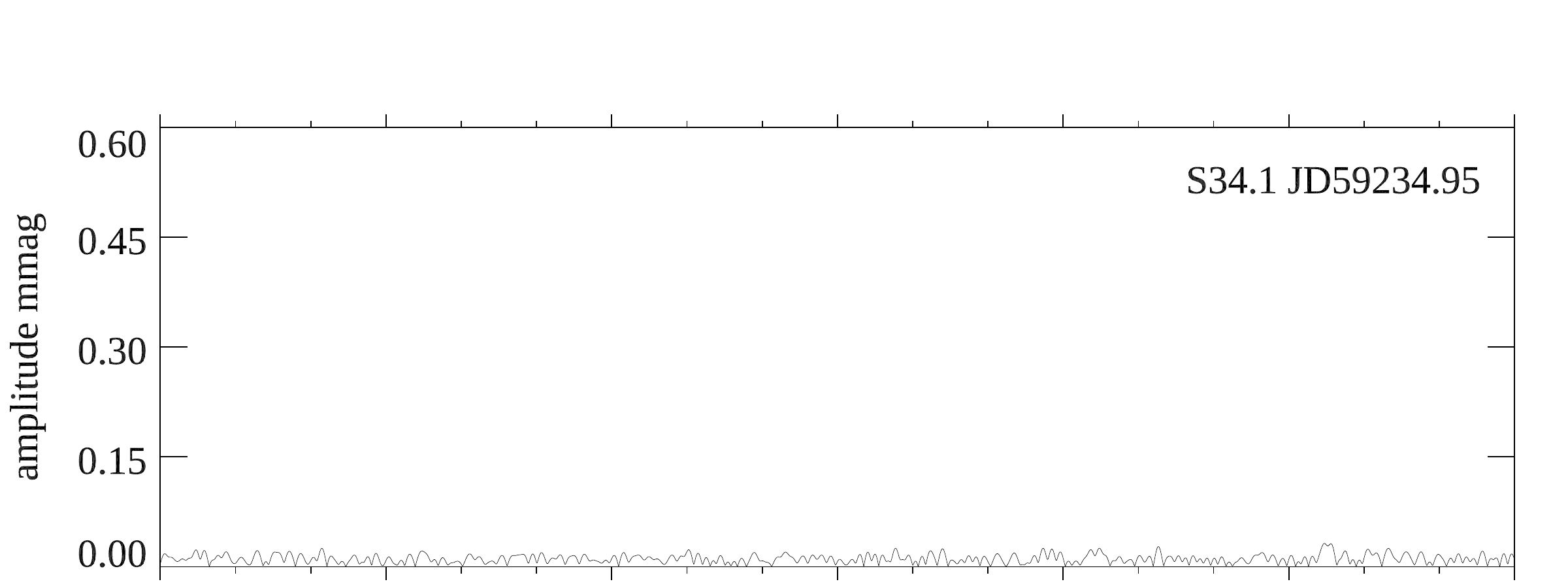}\\
\vspace{-0.6cm}
\includegraphics[width=1.0\linewidth,angle=0]{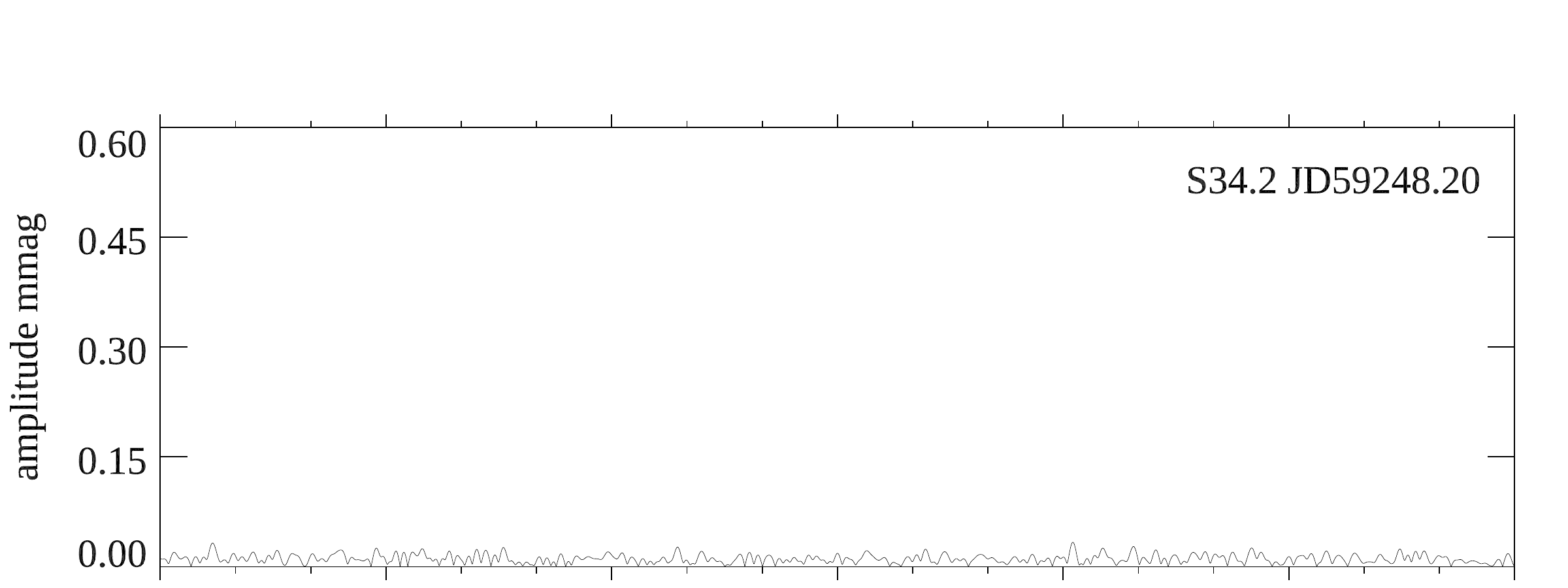}\\
\vspace{-0.6cm}
\includegraphics[width=1.0\linewidth,angle=0]{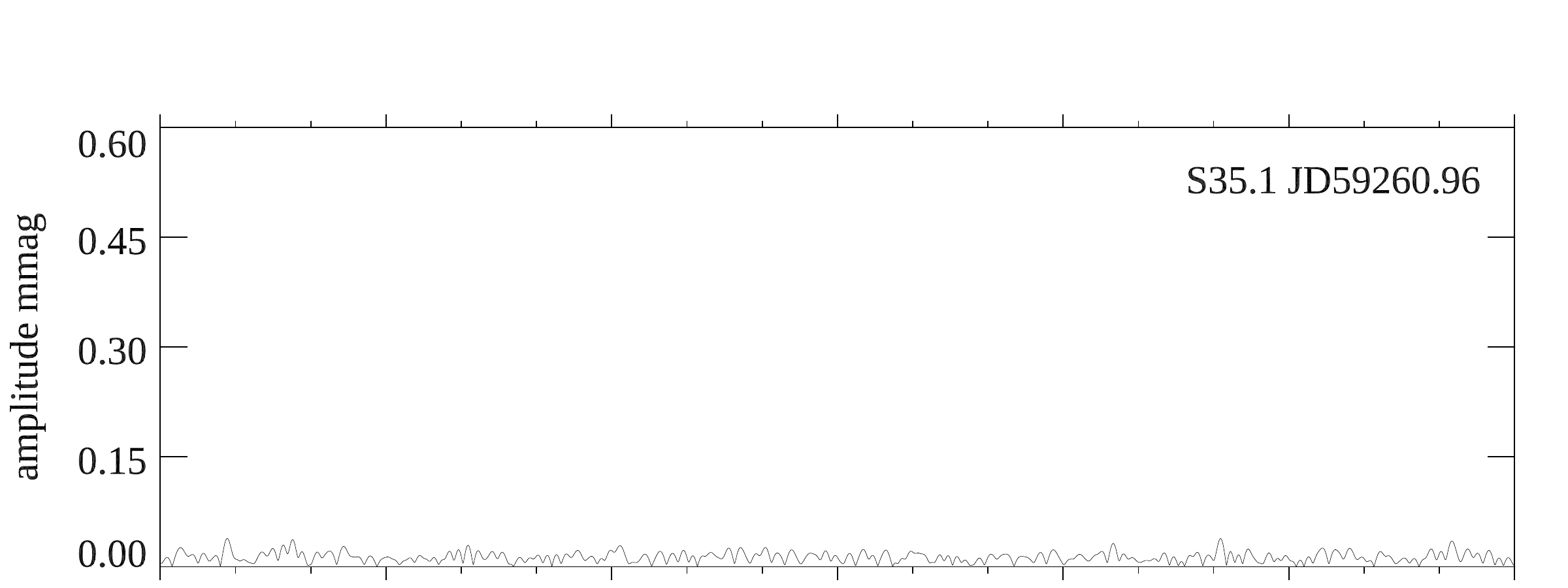}\\
\includegraphics[width=1.0\linewidth,angle=0]{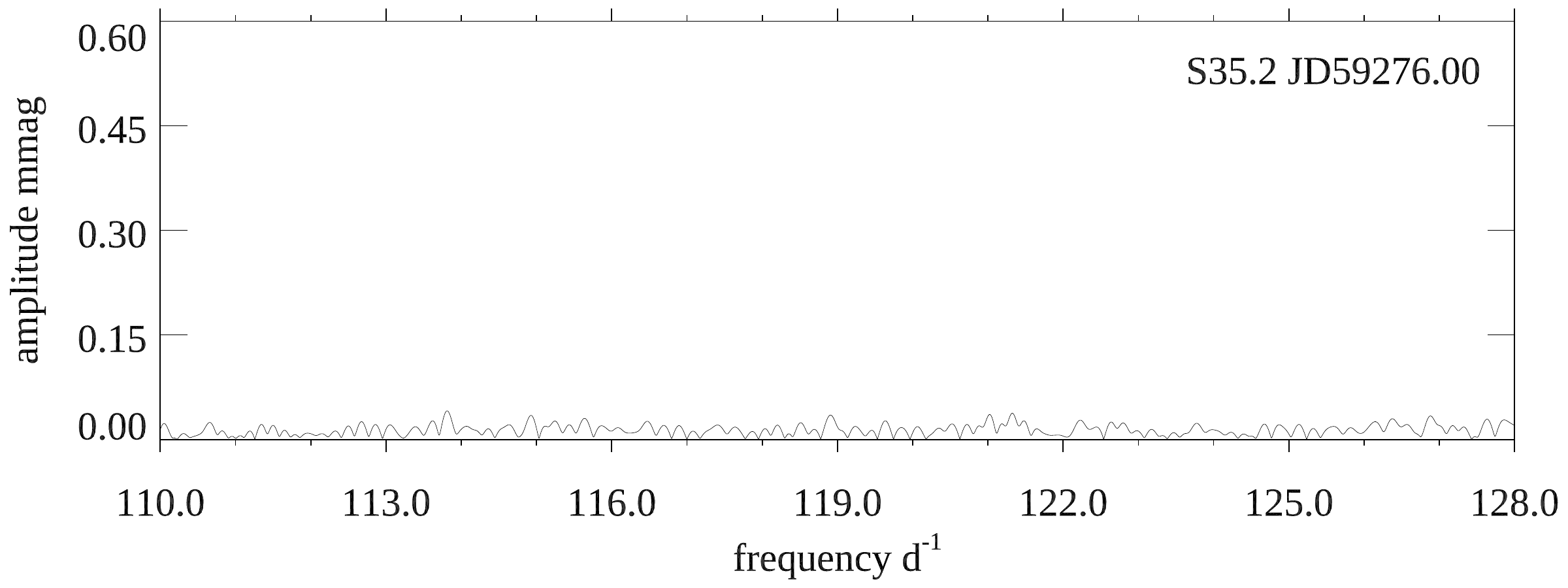}\\
\caption{Amplitude spectra continued. }
\label{fig:as2} 
\end{center}
\end{figure}

\addtocounter{figure}{-1}
\begin{figure}
\begin{center}
\includegraphics[width=1.0\linewidth,angle=0]{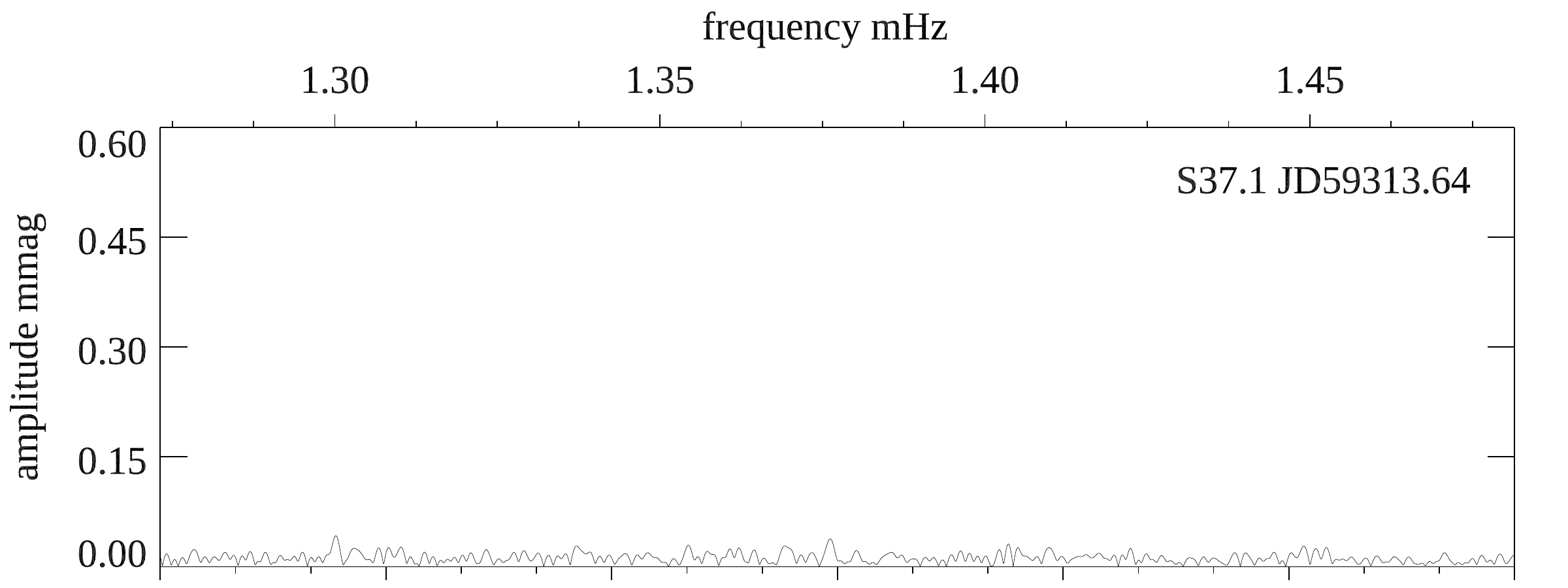}\\
\vspace{-0.6cm}
\includegraphics[width=1.0\linewidth,angle=0]{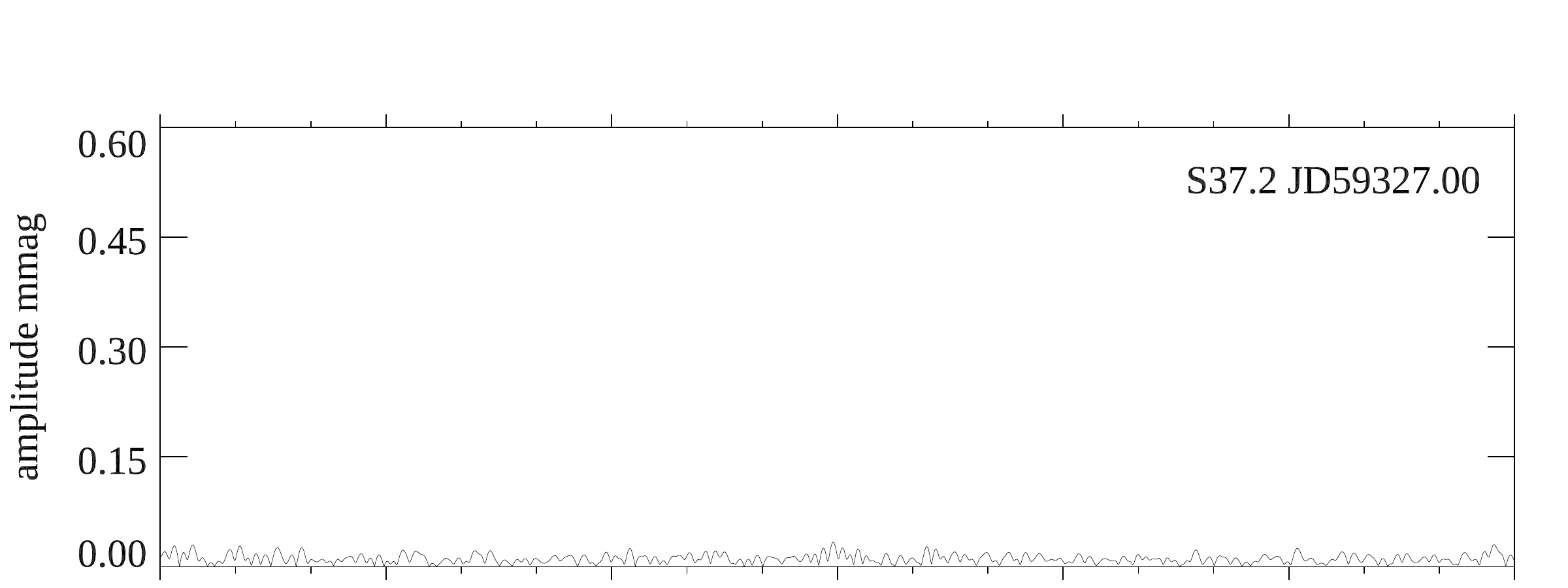}\\
\vspace{-0.6cm}
\includegraphics[width=1.0\linewidth,angle=0]{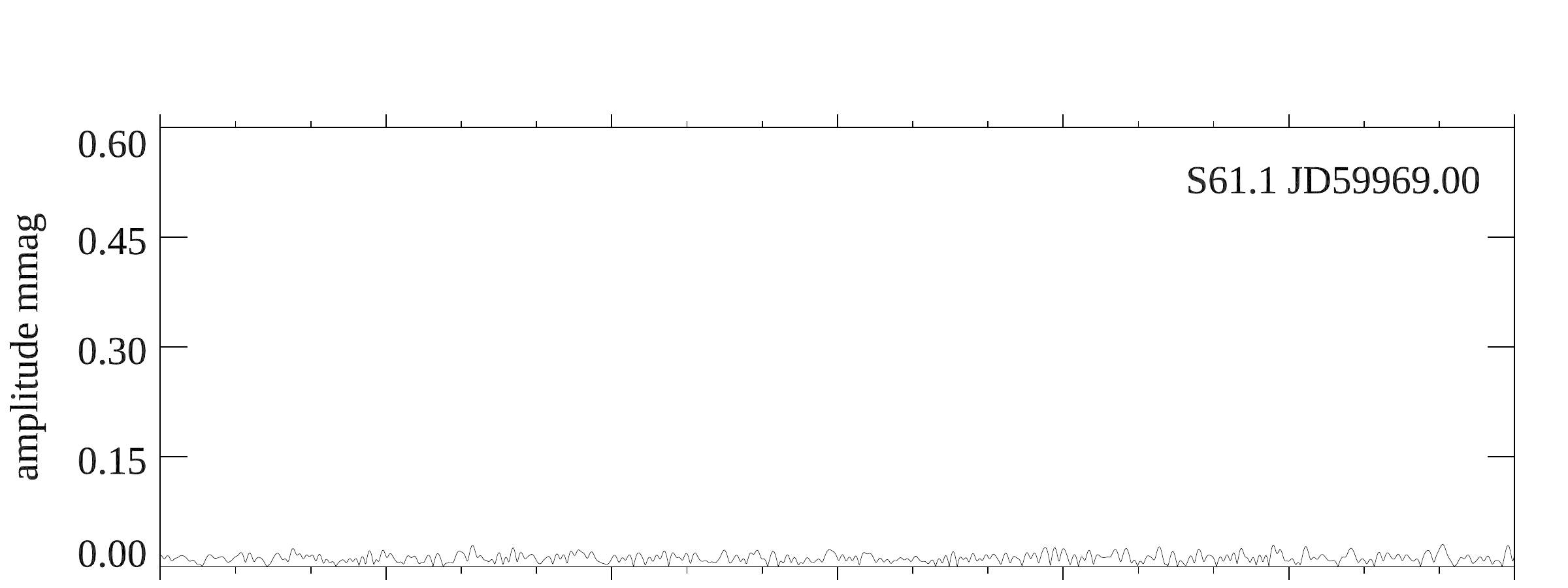}\\
\vspace{-0.6cm}
\includegraphics[width=1.0\linewidth,angle=0]{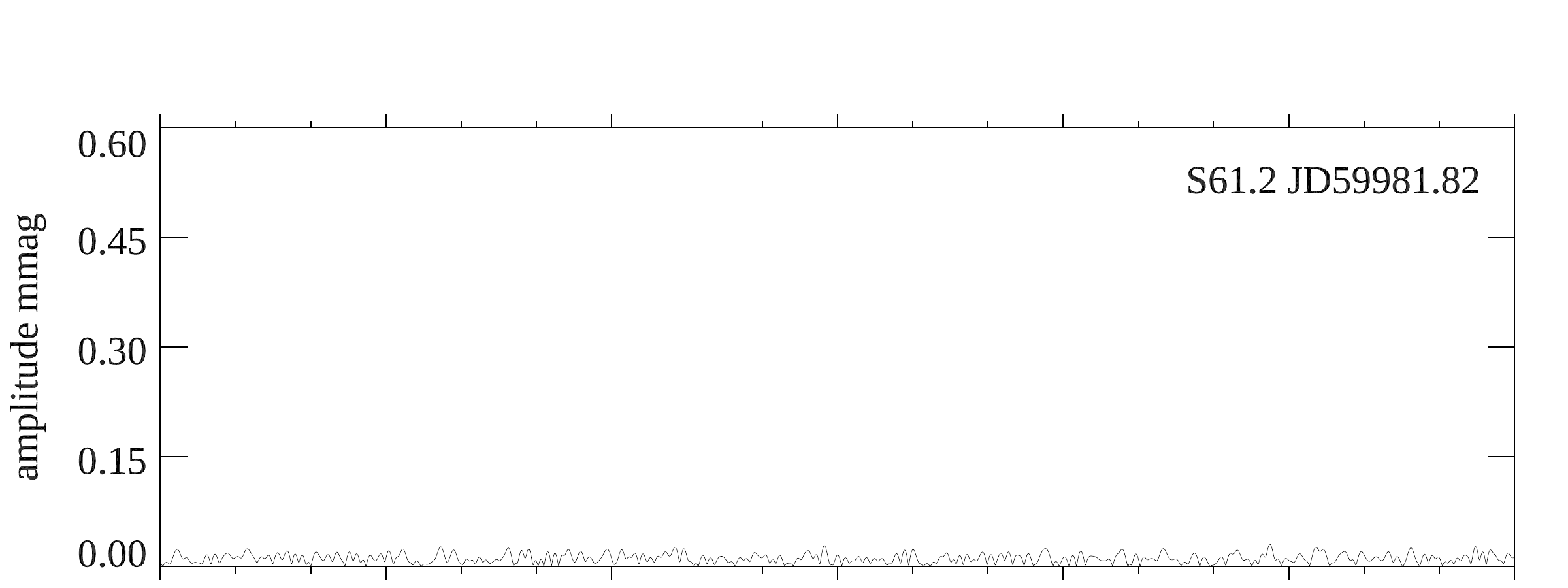}\\
\vspace{-0.6cm}
\includegraphics[width=1.0\linewidth,angle=0]{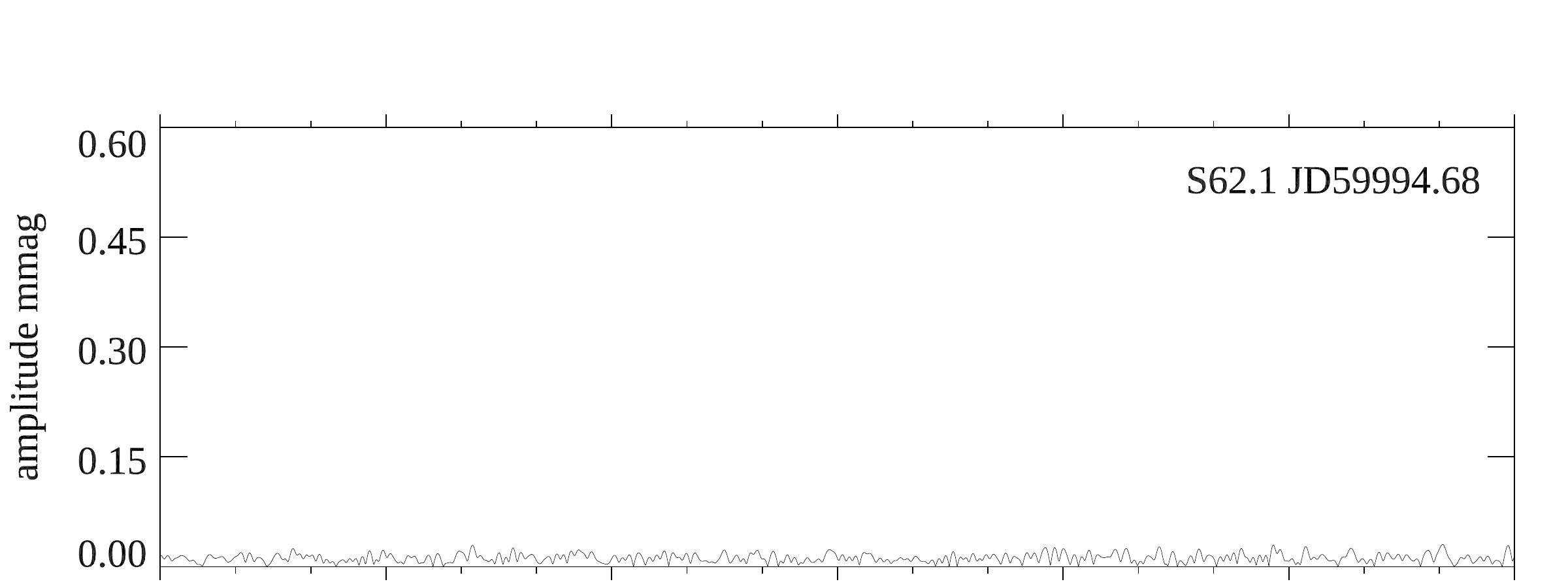}\\
\vspace{-0.6cm}
\includegraphics[width=1.0\linewidth,angle=0]{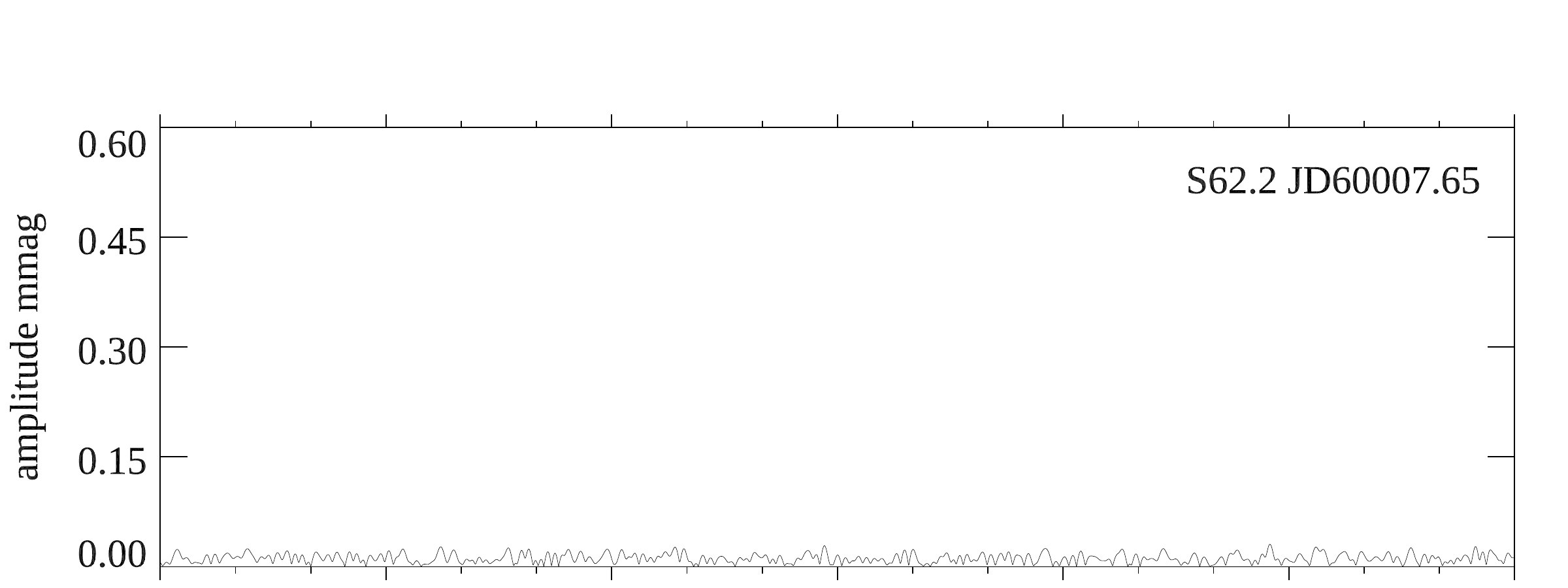}\\
\vspace{-0.6cm}
\includegraphics[width=1.0\linewidth,angle=0]{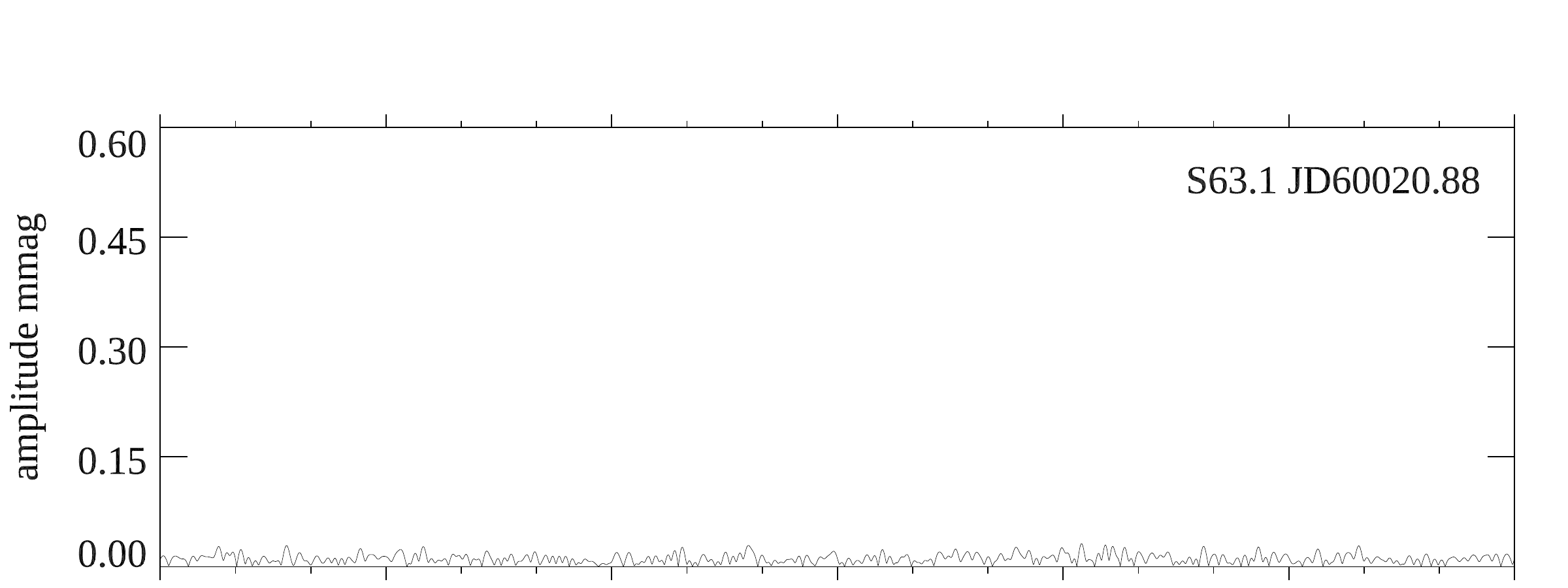}\\
\includegraphics[width=1.0\linewidth,angle=0]{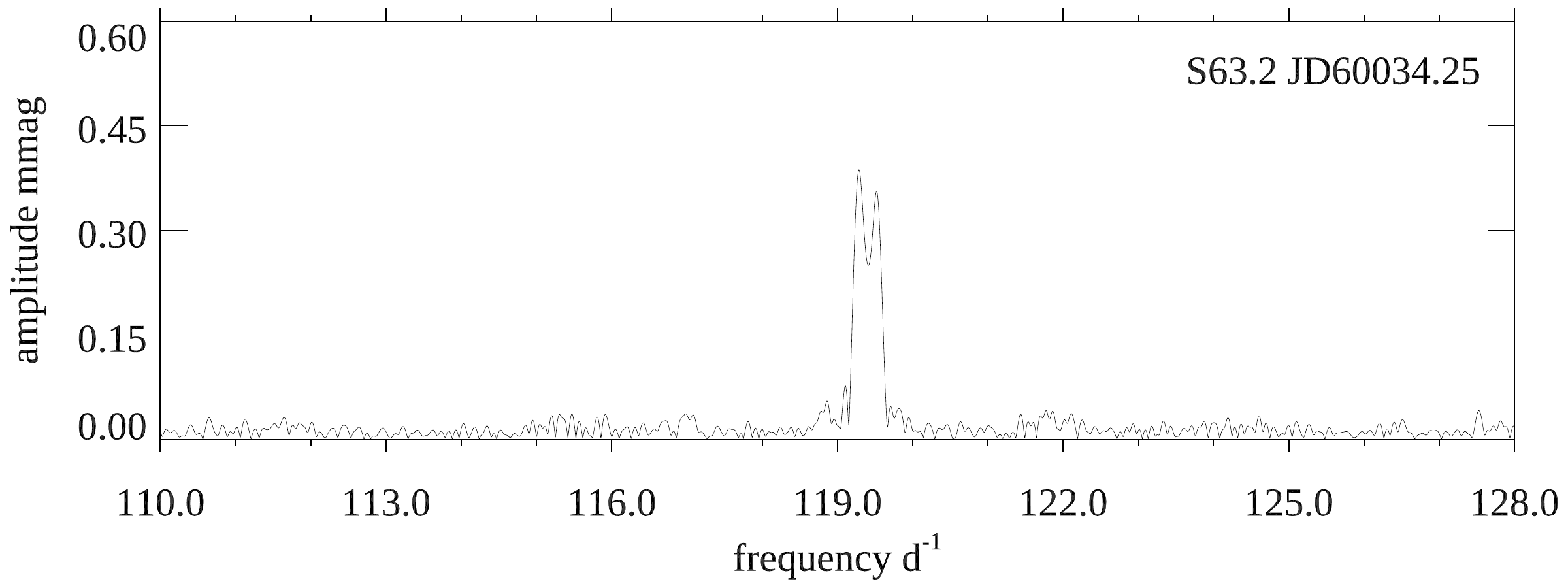}\\
\caption{Amplitude spectra continued. }
\label{fig:as2} 
\end{center}
\end{figure}

\addtocounter{figure}{-1}
\begin{figure}
\begin{center}
\includegraphics[width=1.0\linewidth,angle=0]{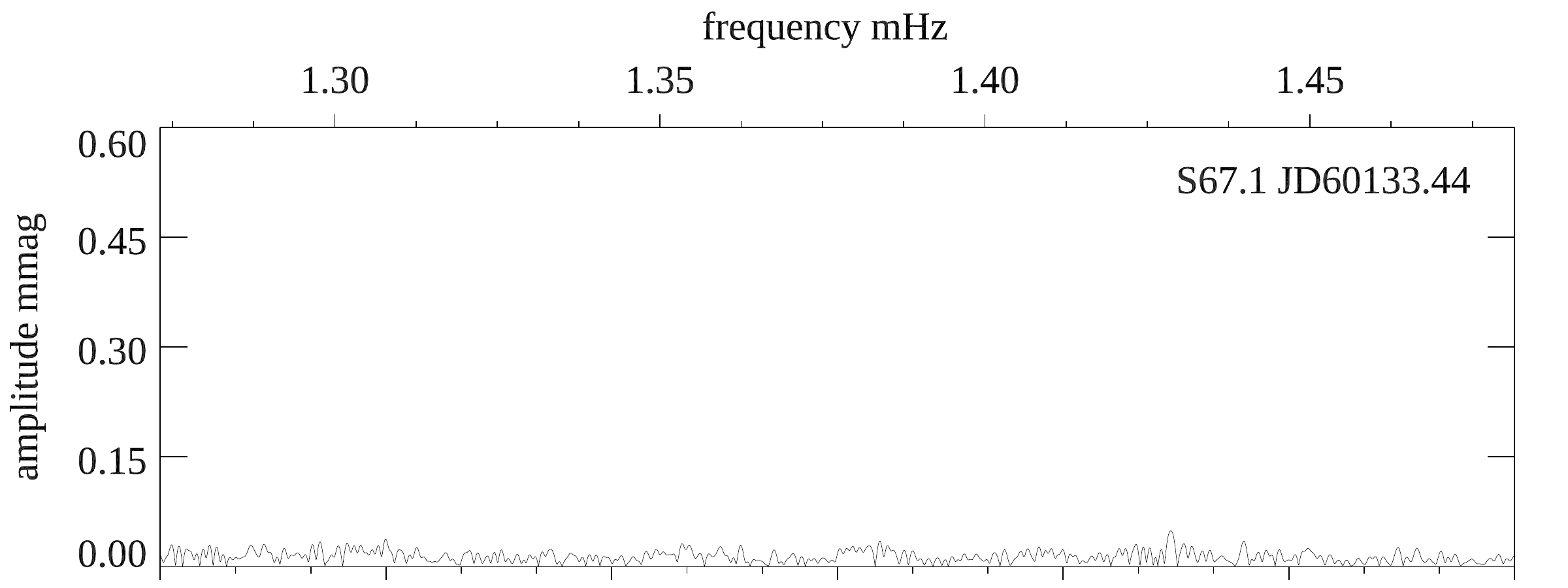}\\
\vspace{-0.6cm}
\includegraphics[width=1.0\linewidth,angle=0]{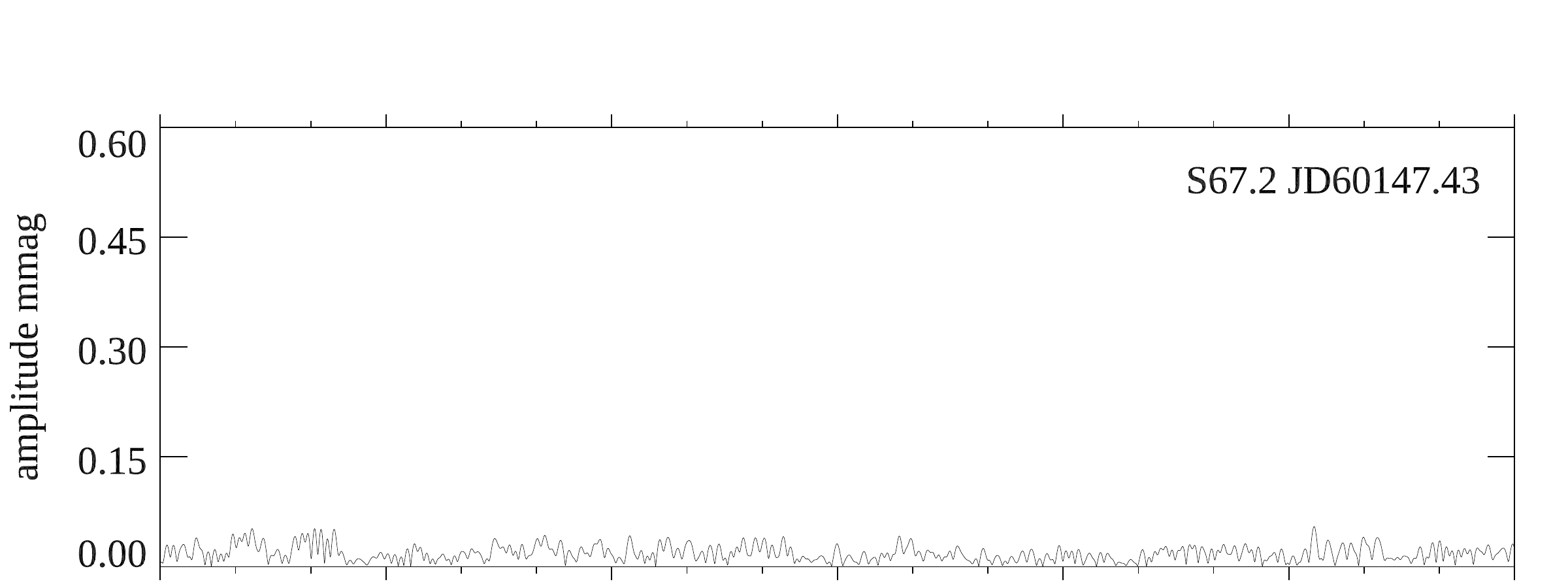}\\
\vspace{-0.6cm}
\includegraphics[width=1.0\linewidth,angle=0]{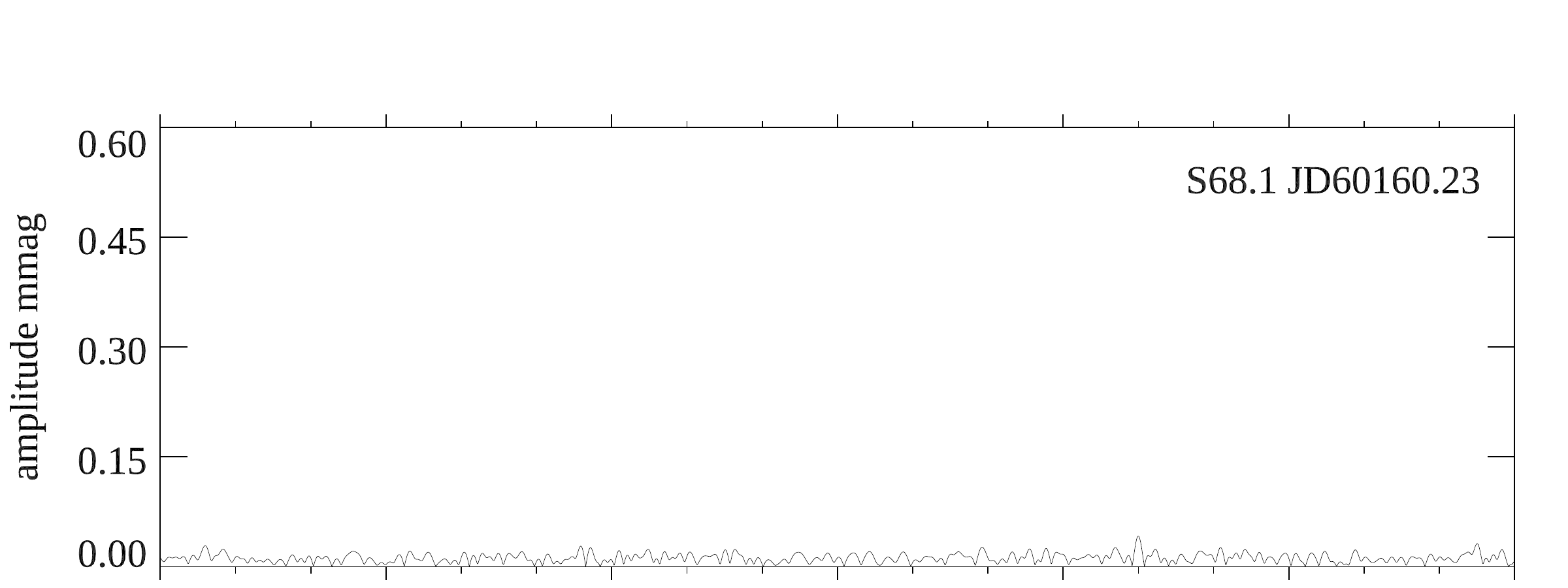}\\
\vspace{-0.6cm}
\includegraphics[width=1.0\linewidth,angle=0]{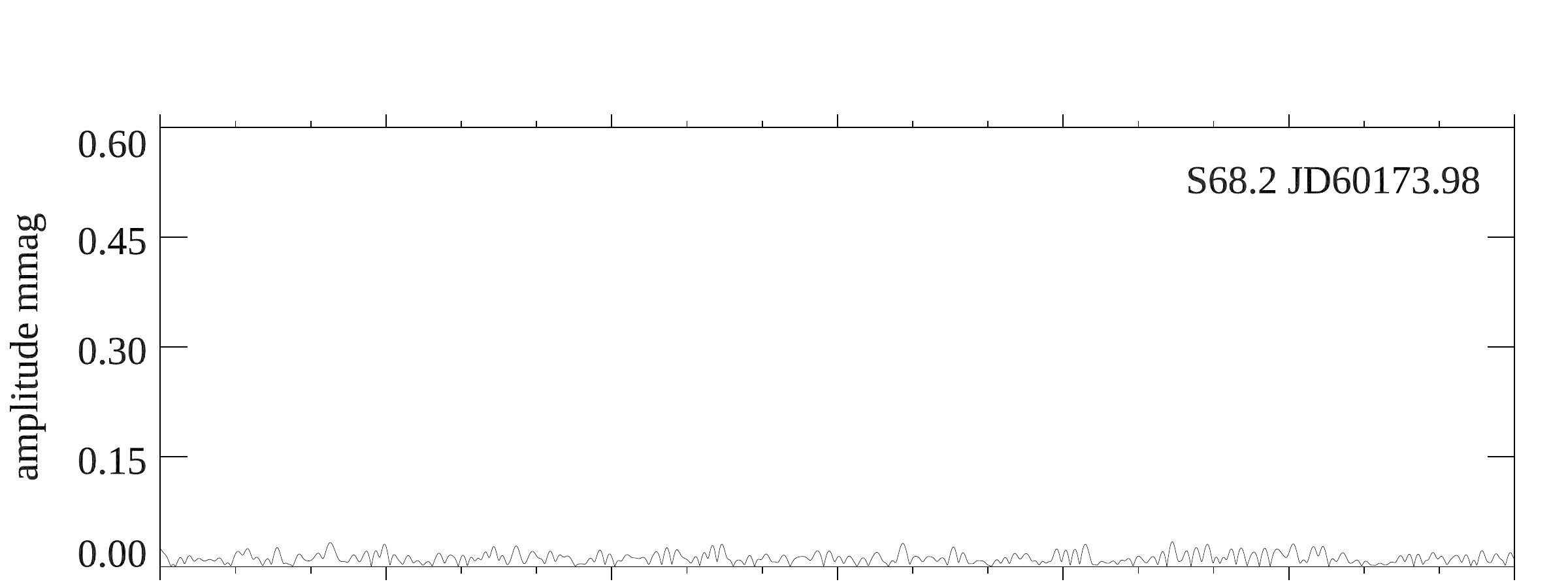}\\
\vspace{-0.6cm}
\includegraphics[width=1.0\linewidth,angle=0]{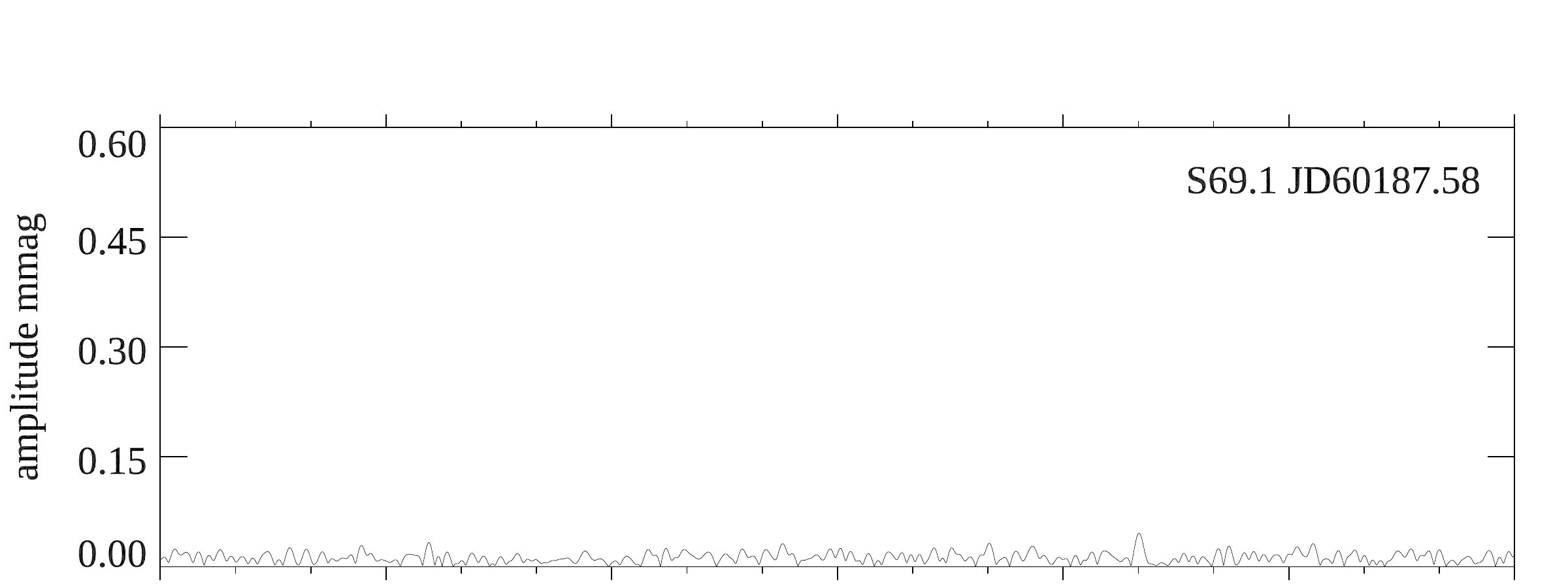}\\
\includegraphics[width=1.0\linewidth,angle=0]{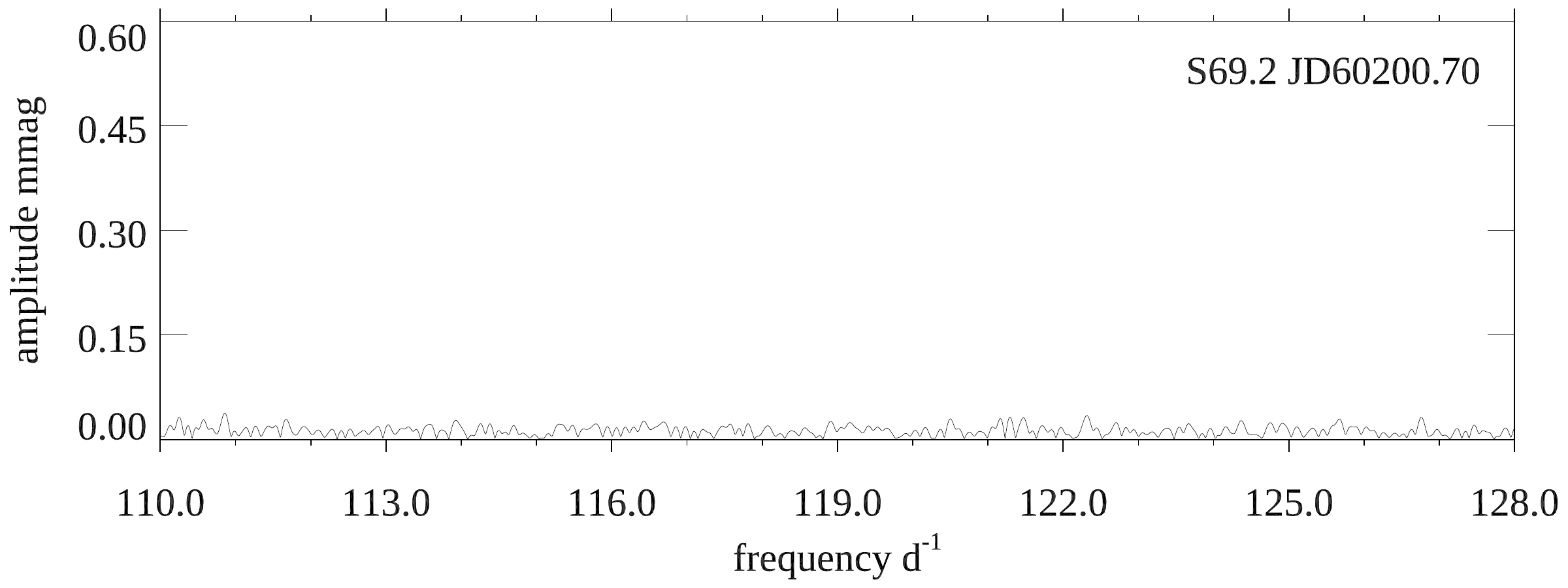}
\caption{Amplitude spectra continued. }
\label{fig:as2} 
\end{center}
\end{figure}

\clearpage

\begin{figure}
\begin{center}
\includegraphics[width=1.0\linewidth,angle=0]{figs/3.pdf}\\
\includegraphics[width=1.0\linewidth,angle=0]{figs/6.pdf}\\
\includegraphics[width=1.0\linewidth,angle=0]{figs/7.pdf}\\
\includegraphics[width=1.0\linewidth,angle=0]{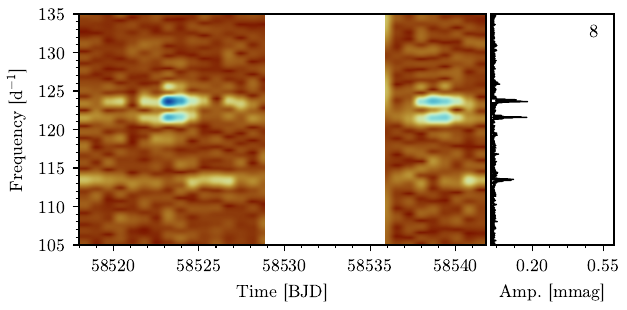}\\
\caption{Wavelet plots by sector. The time is ${\rm BJD} - 2400000.$  The amplitude modulations from both the changing view of oblique pulsation with the $7.679696$-d rotation and the secular changes can be seen. The colours range linearly from amplitude zero (red) to amplitude 0.6\,mmag (blue). The sectors are numbered in the top right corner of each plot. The plots are for 120-s cadence data, unless otherwise noted. }
\label{fig:wav2} 
\end{center}
\end{figure}

\addtocounter{figure}{-1}
\begin{figure}
\begin{center}
\includegraphics[width=1.0\linewidth,angle=0]{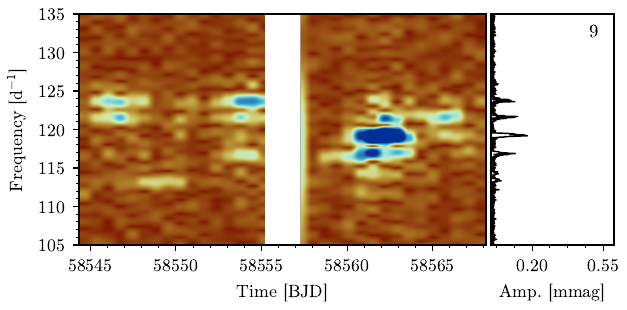}\\
\includegraphics[width=1.0\linewidth,angle=0]{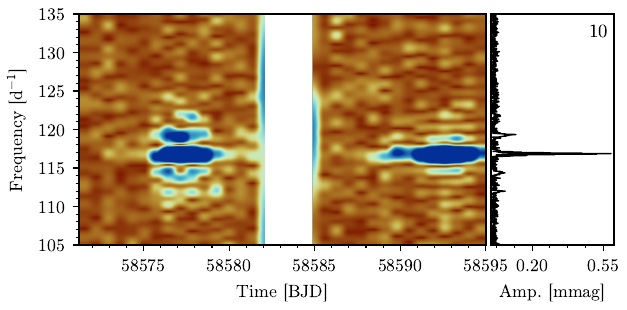}\\
\includegraphics[width=1.0\linewidth,angle=0]{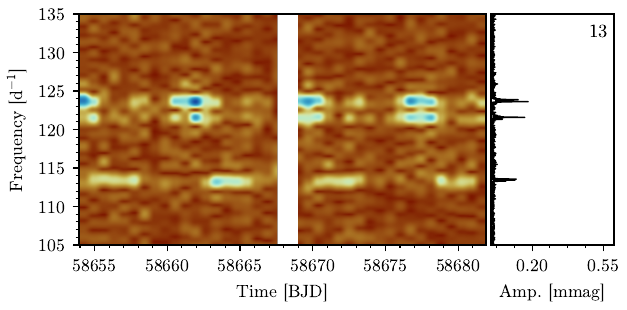}\\
\includegraphics[width=1.0\linewidth,angle=0]{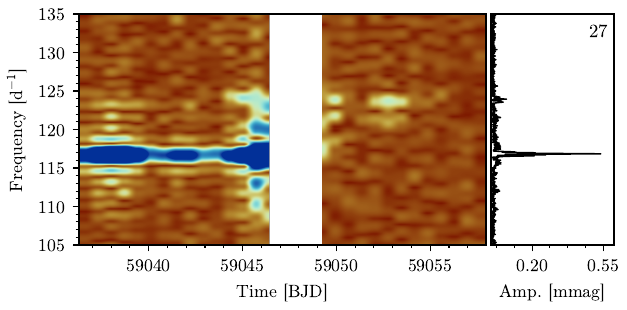}\\
\caption{Wavelets continued.}
\label{fig:wav2} 
\end{center}
\end{figure}

\begin{figure}
\begin{center}
\includegraphics[width=1.0\linewidth,angle=0]{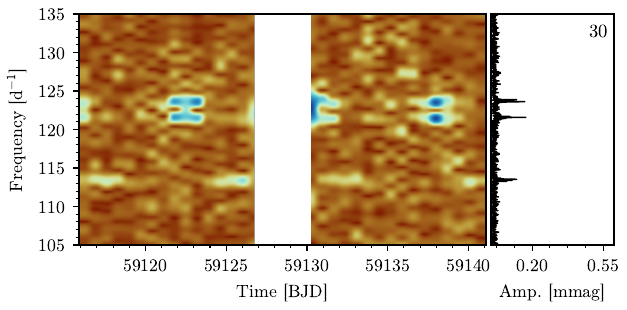}\\
\includegraphics[width=1.0\linewidth,angle=0]{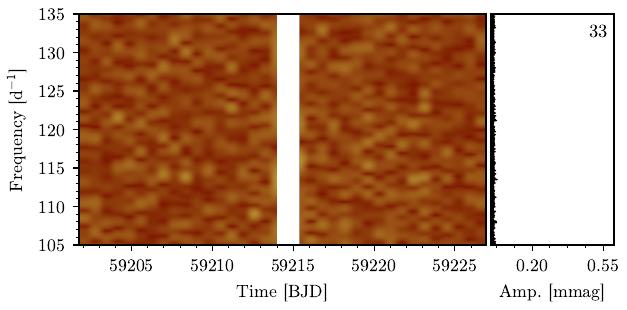}\\
\includegraphics[width=1.0\linewidth,angle=0]{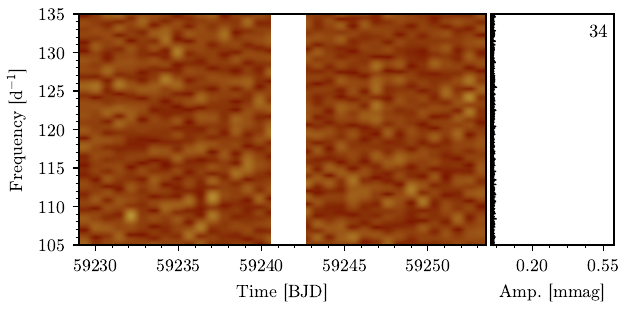}\\
\includegraphics[width=1.0\linewidth,angle=0]{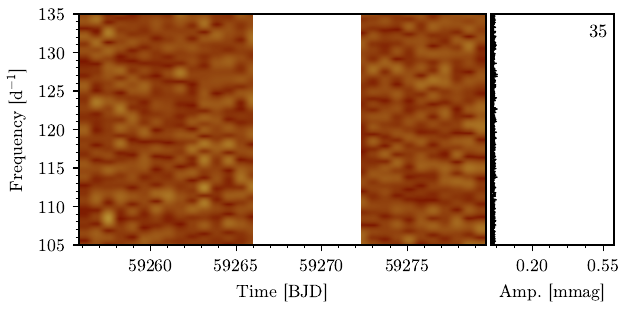}\\
\includegraphics[width=1.0\linewidth,angle=0]{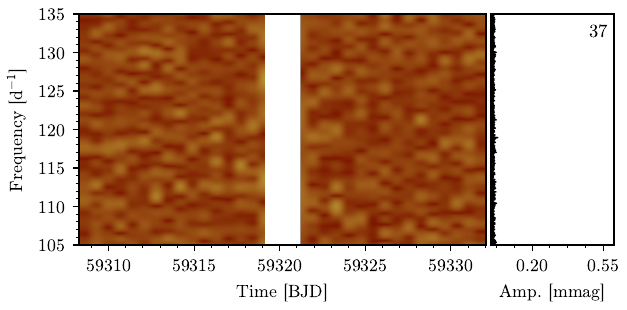}\\
\caption{Wavelets continued. }
\label{fig:wav2} 
\end{center}
\end{figure}

\addtocounter{figure}{-1}
\begin{figure}
\begin{center}
\includegraphics[width=1.0\linewidth,angle=0]{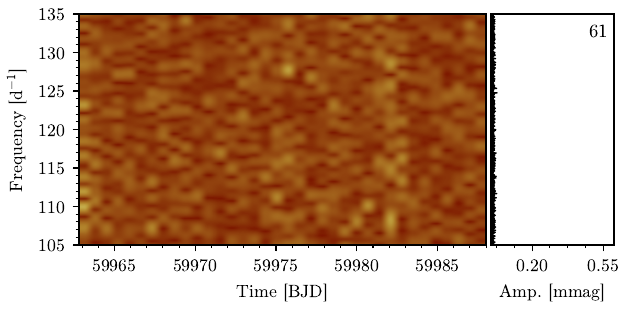}\\
\includegraphics[width=1.0\linewidth,angle=0]{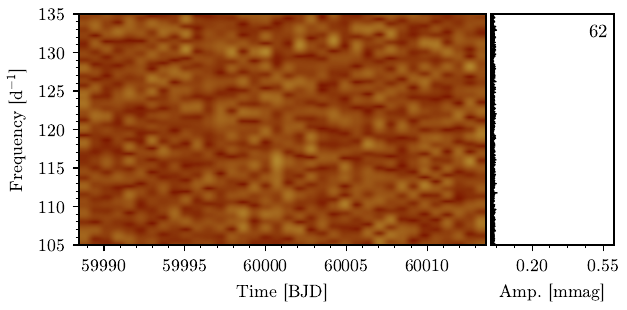}\\
\includegraphics[width=1.0\linewidth,angle=0]{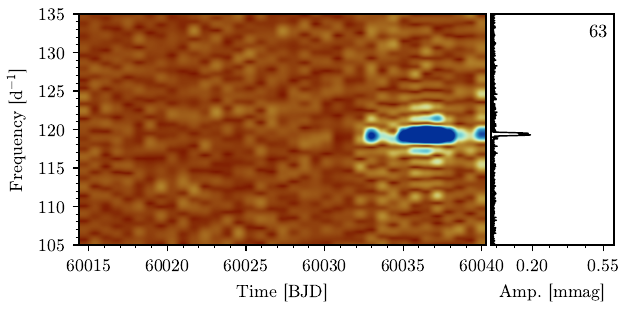}\\
\includegraphics[width=1.0\linewidth,angle=0]{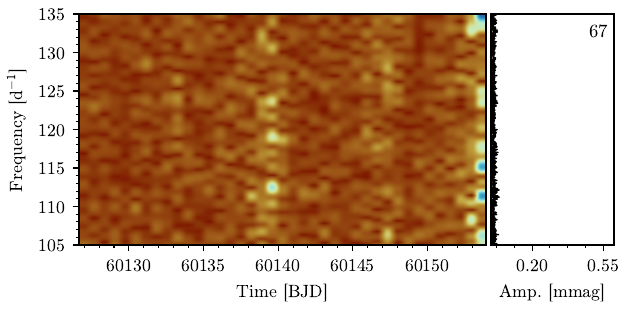}\\
\caption{Wavelets continued. These four sectors are for 200-s cadence data.}
\label{fig:wav2} 
\end{center}
\end{figure}\addtocounter{figure}{-1}

\begin{figure}
\begin{center}
\includegraphics[width=1.0\linewidth,angle=0]{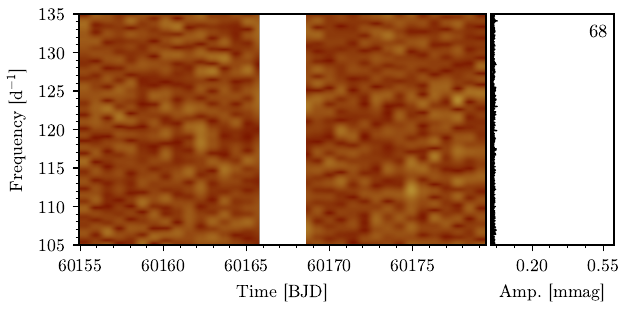}\\
\includegraphics[width=1.0\linewidth,angle=0]{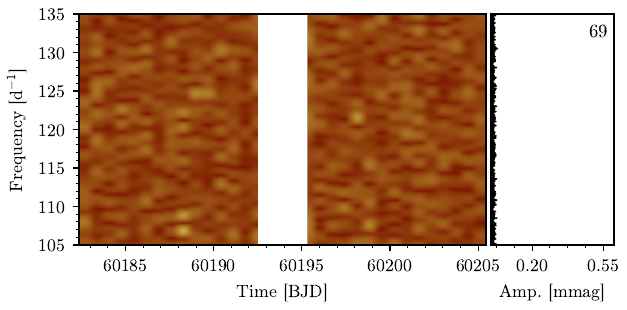}\\
\caption{Wavelets continued.}
\label{fig:wav2} 
\end{center}
\end{figure}

\end{document}